\documentclass[a4paper,accepted=2025-08-22]{quantumarticle}
\pdfoutput=1
\usepackage[utf8]{inputenc}
\usepackage[english]{babel}
\usepackage[T1]{fontenc}

\newcommand{\curlE}{\mathcal{E}}
\newcommand{\ii}{\text{i}}

\usepackage{graphicx}
\usepackage{tikz}
\usepackage{amsmath}
\usepackage{mathtools}
\usepackage{amssymb}
\usepackage{xcolor}
\usepackage{nicefrac}
\usepackage{siunitx}
\usepackage{bbold}
\usepackage{braket}
\usepackage{hhline}
\usepackage{lipsum}
\usepackage{multirow}
\usepackage{placeins}

\usepackage[numbers,sort&compress]{natbib}
\usepackage{hyperref}
\hypersetup{
     pdftitle={Efficient fault-tolerant code switching via one-way transversal CNOT gates},
}

\begin{document}

\title{Efficient fault-tolerant code switching \hspace{10cm} \linebreak via one-way transversal CNOT gates}

\author{Sascha Heu\ss en}
\email{s.heussen@neqxt.org}
\affiliation{\href{https://www.neqxt.org/}{neQxt, 50670 Cologne, Germany}}
\orcid{0000-0002-7581-2148}

\author{Janine Hilder}
\affiliation{\href{https://www.neqxt.org/}{neQxt, 63906 Erlenbach am Main, Germany}}
\affiliation{QUANTUM, Institut für Physik, Universität Mainz, 55128 Mainz, Germany \hspace{5cm}September 20, 2024}
\orcid{0000-0001-9129-1314}

\maketitle 

\begin{abstract}

Code switching is an established technique that facilitates a universal set of fault-tolerant (FT) quantum gate operations by combining two quantum error correcting (QEC) codes with complementary sets of gates, which each by themselves are easy to implement fault-tolerantly. In this work, we present a code switching scheme that respects the constraints of FT circuit design by only making use of transversal gates. These gates are intrinsically FT without additional qubit overhead. 
We analyze application of the scheme to low-distance color codes, which are suitable for operation in existing quantum processors, for instance based on trapped ions or neutral atoms. We also briefly discuss connectivity constraints that arise for architectures based on superconducting qubits. 
Numerical simulations of circuit-level noise indicate that a logical $T$-gate, facilitated by our scheme, could outperform both flag-FT magic state injection protocols and a physical $T$-gate at low physical error rates. Transversal code switching naturally scales to code pairs of arbitrary code distance. We observe improved performance of a distance-5 protocol compared to both the distance-3 implementation and the physical gate for realistic error rates. We discuss how the scheme can be implemented with a large degree of parallelization, provided that logical auxiliary qubits can be prepared reliably enough. Our logical $T$-gate circumvents the need for potentially costly magic state factories.
The requirements to perform QEC and to achieve an FT universal gate set are then essentially the same: Prepare logical auxiliary qubits offline, execute transversal gates (ideally in parallel) and perform fast-enough measurements. 
Transversal code switching thus serves to enable more practical hardware realizations of FT universal quantum computation. The scheme alleviates resource requirements for experimental demonstrations of quantum algorithms run on logical qubits.

\end{abstract}

\section{Introduction}

Universal quantum computation offers the potential to efficiently solve computational tasks that classical computers require an exponentially growing time or memory for. Crucially, a discrete universal set of quantum gate operations should be implemented on error-corrected logical qubits in a fault-tolerant (FT) manner \cite{campbell2017roads}. Quantum error correction (QEC) allows one to systematically suppress noise such that the quantum computation can in principle be upheld for arbitrarily long times, provided the physical noise level is below some threshold value \cite{terhal2015quantum}. The formal requirements of quantum fault tolerance add an additional gate and/or qubit overhead to any quantum circuit, which -- however -- has been shown to not scale unfavorably with the size of such circuit \cite{gottesman2014faulttolerant}. Remarkably, these added components have demonstrated improvements of overall circuit performance in real devices compared to a non-FT implementation, despite being noisy themselves \cite{ryan2021realization, ryan2022implementing, postler2022demonstration, google2023suppressing, dasilva2024demonstrationlogicalqubitsrepeated, mayer2024benchmarkinglogicalthreequbitquantum, ryananderson2024highfidelity, acharya2024quantumerrorcorrectionsurface}. 

A prominent modern technique to render quantum circuits FT with little qubit and gate overhead is the flag qubit paradigm \cite{goto2016minimizing, yoder2017surface, chao2018fault, chamberland2018flag, reichardt2020fault, prabhu2021fault, Bhatnagar_2023, du2024parallel, liou2024reducingquantumerrorcorrection}. Faults within the circuit that may cause uncorrectable errors on the logical qubit are heralded by controlled propagation onto extra auxiliary qubits via additional CNOT gates, which are interleaved into the quantum circuit in an intricate way.
When such a flag qubit is measured in a particular state, one may employ further circuity to treat such faults or repeat the circuit altogether until the flag is clear. 
On the one hand, flag circuits may be favorable due to their conceptual simplicity and typically small qubit overhead. On the other hand, dynamical circuit branching with relatively long circuit sequences might pose a challenge for practical applications of QEC with flag circuits. Also, recent developments of quantum computing hardware suggest that the number of qubits may no longer be the most critical limiting factor \cite{bravyi2022future, bluvstein2022quantum, Pause_2024, bluvstein2024logical, manetsch2024tweezerarray6100highly}. 

\begin{figure*}[!th]
	\centering
	\includegraphics[width=0.99\linewidth]{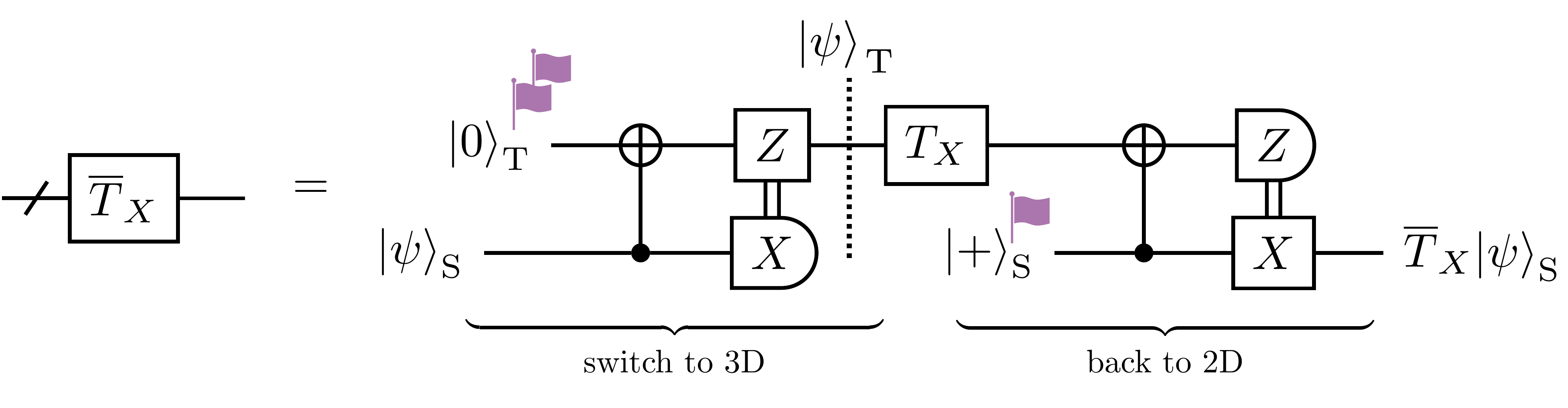}
	\caption{Implementation of the logical $T$-gate on a self-dual code via transversal code switching between the self-dual code (S) on the lower wire and a doubled triorthogonal code (T) on the upper wire. An arbitrary input state $\ket{\psi}$ is moved to the upper wire with a teleportation subroutine that only requires transversal logical operations and preparation of an auxiliary qubit in the logical zero state. The triorthogonal code implements the non-Clifford $T$-gate transversally. A second teleportation subroutine moves the logical state $T_X\!\ket{\psi}$ back to the lower wire where, in the meantime, a logical plus state has been prepared. The state preparation subroutines might require flag qubits or another type of verification to comply with fault tolerance. A concrete application of the scheme to the 2D/3D color codes of distance $d = 3$, the Steane code and the Tetrahedral code, is described in Sec.~\ref{sec:colors}.}
	\label{fig:scheme}
\end{figure*}

Although possibly requiring a larger number of physical qubits,
using \emph{transversal} gate operations could offer numerous operational advantages. Not only are they inherently FT but also straight-forwardly scale up to larger code distances and can directly be parallelized if the physical qubit architecture permits. Other FT QEC algorithms, like Steane-type \cite{steane1997active, postler2024demonstration} or Knill-type \cite{knill2004scalable, knill2005quantum, ryananderson2024highfidelity} error correction, make use of transversal gates. For instance, Steane-type EC has been shown to outperform flag EC if the limiting factor is the entangling gate error rate $p_2$ in the limit of small $p_2~(\lesssim 3\%)$ even though using more physical qubits \cite{postler2024demonstration}. 

For the controlled manipulation of encoded information, FT gate operations must be performed on the logical qubits. Transversal CNOT gates are appealing because they need not induce a time-overhead to respect fault tolerance such as, e.g., repeated operator measurements in lattice surgery \cite{horsman2012surface, cohen2022low, cowtan2024css}. It is known that no non-trivial QEC code offers a transversal universal gate set \cite{eastin2009restrictions}. Therefore, given any code with a set of transversal gates, additional strategies are imperative to establish universality. There exist several procedures such as gate teleportation \cite{gottesman1999demonstrating} or code conversion \cite{fowler2012surface, anderson2014fault} that vary in intricacy from both a theoretical and experimental point of view. Code switching in particular refers to the idea of employing \emph{two} different QEC codes, which each on their own have a different set of transversal gates but in conjunction offer an FT universal gate set that is simple to perform on the logical level \cite{bombin2015gauge, kubica2015universal}. The difficulty of implementing a particular gate is traded in for transferring the logical information between the two codes at will in an FT fashion. 

All Clifford gates have a transversal implementation in two-dimensional (2D) color codes \cite{bombin2006distillation}. To unlock universality, an efficient FT construction of a non-Clifford gate, such as the $T$-gate, is required. The $T$-gate is transversal in three-dimensional (3D) color codes (as well as the CNOT gate but the Hadamard gate $H$ is not) \cite{bombin2007exact}. An FT code switching technique based on measuring potentially long sequences of stabilizers with flag qubits has been introduced recently for distance-3 color codes, which serve to correct arbitrary single Pauli errors \cite{butt2024fault}. It has subsequently been demonstrated in practice on an ion-trap quantum processor \cite{pogorelov2024experimental}. 

In this work, we investigate a new approach to implement the non-Clifford $T$-gate fault-tolerantly on the 2D color code via FT code switching to the 3D color code that is based on code doubling \cite{betsumiya2012triply, bravyi2015doubledcolorcodes, shi2024triorthogonal, sullivan2024code}. A notable strength of the scheme is that it can be viewed as \emph{deterministic} in practice: Only QEC codes of distance $d \geq 3$ are employed, instead of mere quantum error \emph{detecting} codes with $d=2$, and no post-selection is required. The scheme is considered efficient because it relies on a transversal construction of the logical CNOT gate, which works as long as one restricts the connection of the gate to be one-directional. Transversal code switching is performed via logical state teleportation. The logical $T$-gate implementation is then built on two switching steps ``2D $\rightarrow$ 3D $\rightarrow$ 2D'', depicted with logical building blocks in Fig.~\ref{fig:scheme}. Logical auxiliary qubits in both codes can be prepared with Clifford circuits only, so that no complicated magic state subroutines are needed. We find that transversal code switching can outperform other universality schemes in terms of logical failure rate, entangling gate count and circuit depth while only slightly increasing the number of required physical qubits.

This manuscript is organized as follows: The efficient FT $T$-gate construction is discussed in detail in Sec.~\ref{sec:T}. We offer insights on parallelization and embedding into a reduced-connectivity hardware architecture. In Sec.~\ref{sec:noise}, we present simulations of incoherent circuit-level Pauli noise with both a simple single-parameter depolarizing noise model and a more elaborate multi-parameter noise model that can be used to capture the essential noise processes of a trapped-ion quantum computer. A comparison to other strategies for establishing an FT universal gate set is provided in Sec.~\ref{sec:comp}. In Sec.~\ref{sec:scaleup}, we consider scaling up our scheme to QEC codes of arbitrary distance and lastly draw conclusions and provide an outlook on future work in Sec.~\ref{sec:outlook}.

\section{Transversal code switching}\label{sec:T}

Code switching has originally been deployed as a method relying on two topological stabilizer codes each manifesting as a particular gauge of the same, common subsystem code \cite{bombin2015gauge}. The switching is performed by gauge fixing in this framework and a simplified switching scheme has been developed soon after \cite{kubica2015universal}.

Essential for any distance-$d$ QEC code, FT code switching must ensure that no more than $t = \lfloor \frac{d-1}{2} \rfloor$ \emph{errors} of any Pauli type $X$ or $Z$ can result from up to $t$ \emph{fault} events in a circuit-level noise model. Such faults must be assumed to happen during the execution of the actual quantum circuits of the switching algorithm where every physical operation acting on $q$ qubits is subjected to a $q$-qubit depolarizing noise channel. A logical gate is deemed transversal if it can be implemented by applying bitwise physical gate operations to all physical qubits and, crucially, no interactions between physical qubits happen within the same logical block. Therefore, faults are confined to the qubits on which they initially occur and never spread uncontrollably within a code block.

The Hadamard gate is not directly transversal in the 3D color code because the $X$- and $Z$-type stabilizers are not invariant under the exchange of $X$ and $Z$, i.e., it is not self-dual, as opposed to the 2D color code. The $T$-gate has a transversal implementation for the class of triorthogonal codes, to which the 3D color code belongs but the 2D color code does not. By using more elaborate transversal operations and an auxiliary system, it was shown how to perform the FT Hadamard gate on triorthogonal codes \cite{paetznick2013universal}. The FT $T$-gate can be implemented on a small 2D color code via flag-assisted code switching, which requires an intermediate (half-)cycle of QEC \cite{butt2024fault}.

In the following, we describe a novel FT $T$-gate implementation for 2D color codes based on transversal code switching. We first review the construction of so-called doubled codes, which are appropriate triorthogonal codes to enable the transversal $T$-gate and efficient switching procedures. Then, we focus on FT code switching between 2D and 3D color codes. Usage of the smallest code representatives of distance $d=3$ is explained in detail on the level of physical qubit quantum circuits. Practical aspects of the scheme such as logical auxiliary state preparation, parallelized gate operations and limited qubit-connectivity are discussed subsequently.

\subsection{Doubled-code construction}
In Ref.~\cite{sullivan2024code}, a code construction is described that allows one to perform FT code switching with the help of logical auxiliary qubits and one-way transversal CNOT gates. The central ingredients are a $[[n, k=1, d]]$ self-dual Calderbank-Shor-Steane (CSS) code\footnote{This need not be a color code.} and a triorthogonal $[[\Tilde{n}, 1, \Tilde{d}]]$ code, from which a so-called \emph{doubled} code is constructed, which is itself again a triorthogonal code with parameters \mbox{$[[n'=2n+\Tilde{n}, 1, d'=\text{min}(d, \Tilde{d}+2)]]$}. The stabilizers and logical operators of the two ingredient codes are recombined to a new, bigger triorthogonal matrix in the process of code doubling. The initial self-dual code is complemented with additional $Z$-type stabilizers to yield the new triorthogonal code, which may have non-local stabilizers. Importantly, the original $Z$-type stabilizers of the self-dual code remain intact and are incorporated into the doubled code. As a consequence, the one-way logical CNOT gate can be executed by connecting $n$ physical CNOT gates in a pair-wise fashion between the $n$ physical qubits of the self-dual code and a subset of $n$ physical qubits of the doubled code. This enables bi-directional switching between the two codes via standard quantum state teleportation circuits at the logical level. Transversality ensures that fault tolerance is respected for a code switching step. Additionally, it must be guaranteed that no uncorrectable error can ever propagate between the two codes from the preparation of the logical auxiliary qubits. While code doubling has been investigated before \cite{betsumiya2012triply, bravyi2015doubledcolorcodes, shi2024triorthogonal}, the one-way transversal CNOT gate is a new ingredient identified in Ref.~\cite{sullivan2024code}.

\subsection{Reduction to 2D/3D color code combination}\label{sec:colors}

\begin{figure}
    \centering
	\includegraphics[width=0.99\linewidth]{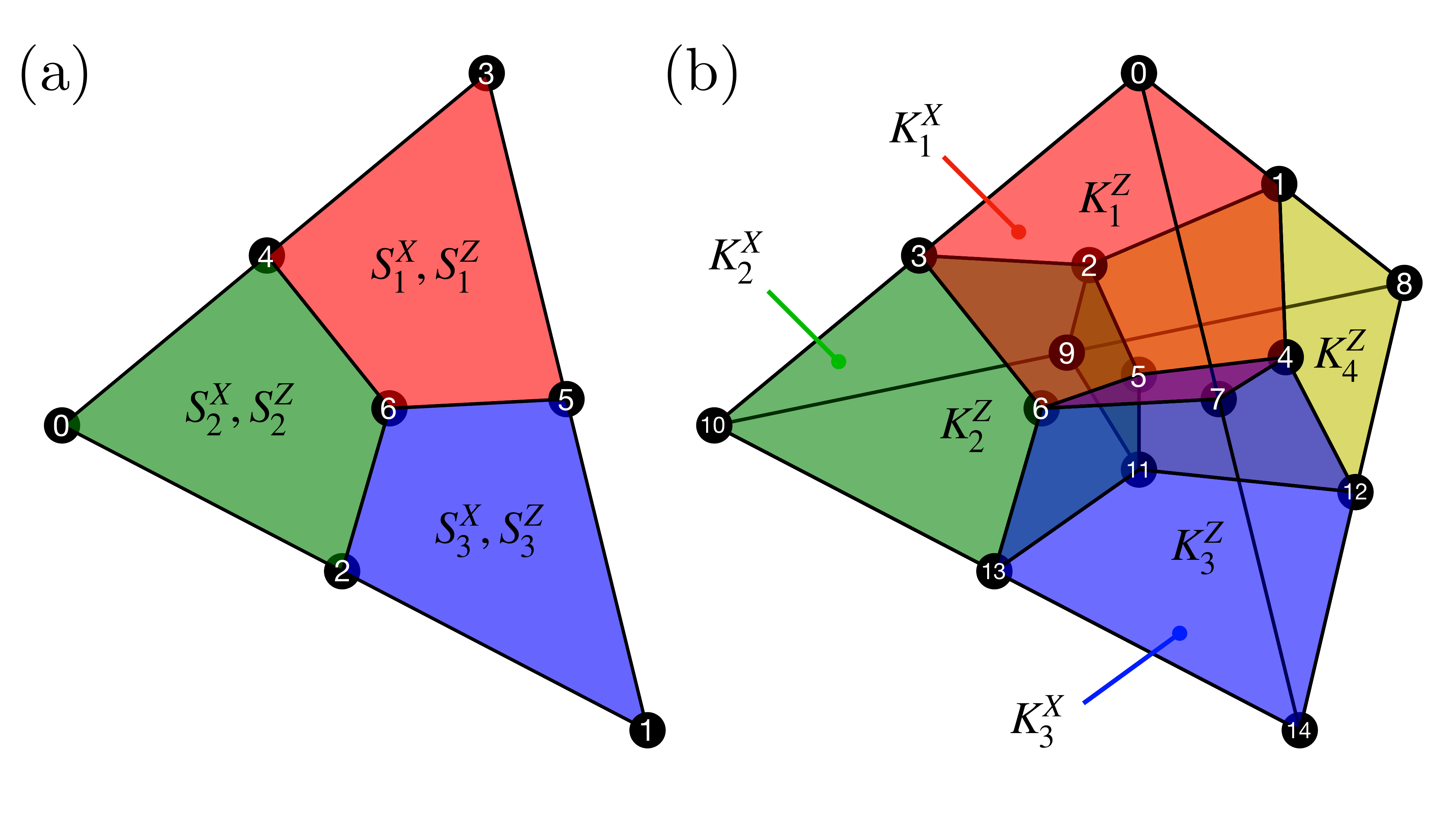}
	\caption{The smallest representatives of the code families of (a) two-dimensional and (b) three-dimensional CSS color codes. These distance-3 codes are commonly referred to as the Steane code and the Tetrahedral code respectively. The Clifford gates $\{H, S, \text{CNOT}\}$ are transversal in the Steane code and the gate set $\{X, T, \text{CNOT}\}$ is transversal in the Tetrahedral code. The weight-8 cell operators are defined as the $Z$-type stabilizer generators of the Tetrahedral code; three out of the 10 weight-4 face $X$-type stabilizer generators are labelled explicitly. They are inherited from the $X$-plaquettes of the Steane code.}
	\label{fig:codes}
\end{figure}

The logical CNOT gate can be performed directly between a 2D and a 3D color code at the price of restricting it to only one direction. The reason for this is that the 3D color code already is a doubled code. For the code definitions used in Ref.~\cite{sullivan2024code}, this means that the control of the logical CNOT may only be encoded in the 3D color code and the target of the logical CNOT must be encoded in the 2D color code. It can be ensured that all stabilizer operators of each code propagate through the logical CNOT onto stabilizer operators of the respective other code, using only $n$ physical CNOT gates. The product of the stabilizer groups of both codes is therefore left invariant. Simultaneously, logical operators of both $X$- and $Z$-type are mapped between the two codes as prescribed by the action of a logical CNOT gate.

The remainder of this section is dedicated to applying the aforementioned code switching protocol to the smallest instances of the 2D and 3D color code respectively depicted in Fig.~\ref{fig:codes}, i.e., the seven-qubit Steane code and the fifteen-qubit Tetrahedral code. Both QEC codes have distance $d = 3$ and thus correct $t = 1$ arbitrary Pauli error. FT state preparation of the logical auxiliary qubit states will be discussed in Sec.~\ref{sec:ftstateprep}. After providing a detailed discussion of this small-scale implementation, we broaden the horizon again and discuss scaling up to large distances in Sec.~\ref{sec:scaleup}. 

\subsubsection{Distance-3 transversal code switching}\label{sec:steanetetraswitch}

\begin{figure*}
    \centering
	\includegraphics[width=0.99\linewidth]{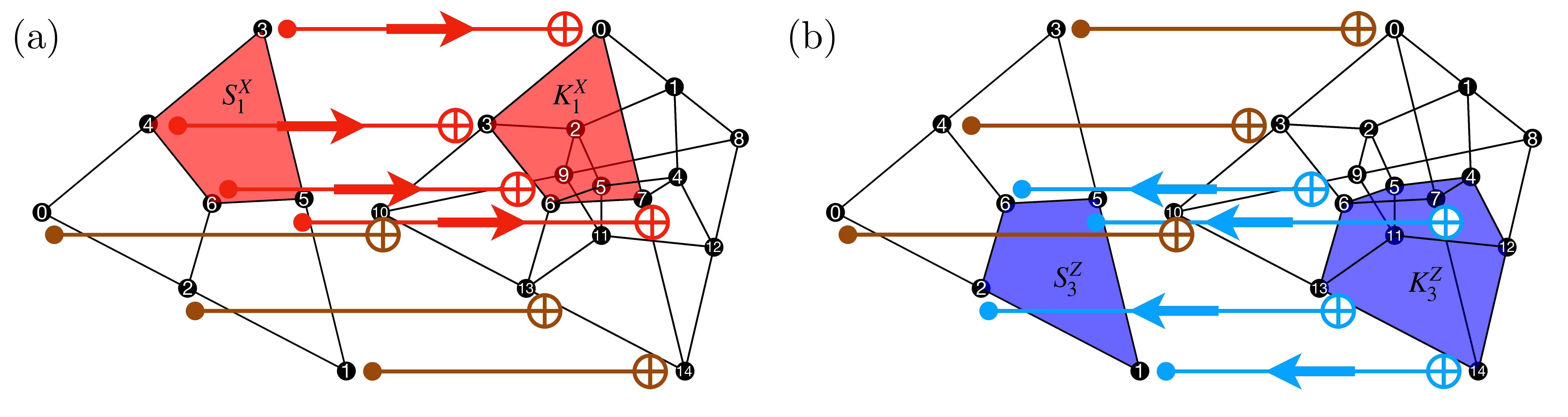} \caption{Propagation of stabilizers through the one-way transversal CNOT gate. (a) The red $X$-plaquette of the Steane code $S_1^X$ is mapped to the red face of the Tetrahedral code $K_1^X$. (b) The blue cell of the Tetrahedral code $K_3^Z$ propagates partly to the Steane code such that it corresponds to the blue $Z$-plaquette $S_3^Z$. }
	\label{fig:stabprop}
\end{figure*}

We depict the scheme with logical building blocks in Fig.~\ref{fig:scheme}. The workhorse of the code switching procedure is the reduced-support transversal CNOT gate that can be applied between the Steane code and the Tetrahedral code \cite{sullivan2024code}. The $[[7,1,3]]$ Steane code \cite{steane1996multiple} is defined by the stabilizer generators
\begin{align}
    S_1^X &= X_3X_4X_5X_6, ~~~~~~ S_1^Z = Z_3Z_4Z_5Z_6, \notag \\
    S_2^X &= X_0X_2X_4X_6, ~~~~~~ S_2^Z = Z_0Z_2Z_4Z_6, \\
    S_3^X &= X_1X_2X_5X_6, ~~~~~~ S_3^Z = Z_1Z_2Z_5Z_6. \notag
\end{align}
The logical operators have minimal weight 3. For example, the edges of the Steane code triangle, such as $\overline{X}_\mathrm{S} = X_0X_3X_4$, correspond to logical operators. We use a definition of the Tetrahedral code, where all $X$- and $Z$-type stabilizers are interchanged compared to the conventional definition \cite{steane1999quantum}. This way, we achieve protection against $t = 3$ $Z$-errors since the $[[15,1,3]]$ code defined by the stabilizer generators
\begin{align}
    K_1^X &= X_0X_3X_6X_7, & K_2^X &= X_3X_6X_{10}X_{13}, \notag \\
    K_3^X &= X_6X_7X_{13}X_{14}, & K_4^X &= X_8X_9X_{11}X_{12}, \notag \\
    K_5^X &= X_1X_2X_4X_5, & K_6^X &= X_4X_5X_6X_7, \label{eq:tetraxstabs} \\
    K_7^X &= X_2X_3X_5X_6, & K_8^X &= X_4X_5X_{11}X_{12}, \notag \\
    K_9^X &= X_2X_5X_9X_{11}, & K_{10}^X &= X_5X_6X_{11}X_{13} \notag
\end{align}
and
\begin{align}
    K_1^Z &= Z_0Z_1Z_2Z_3Z_4Z_5Z_6Z_7, \notag \\
    K_2^Z &= Z_2Z_3Z_5Z_6Z_9Z_{10}Z_{11}Z_{13}, \notag \\
    K_3^Z &= Z_4Z_5Z_6Z_7Z_{11}Z_{12}Z_{13}Z_{14}, \notag \\ 
    K_4^Z &= Z_1Z_2Z_4Z_5Z_8Z_9Z_{11}Z_{12} \label{eq:tetrazstabs}
\end{align}
has distance $d_Z = 7$ and distance $d_X = 3$. This reflects the fact that the 3D color code is not self-dual. The ten $X$-type stabilizers are faces of the tetrahedron of weight 4 in Fig.~\ref{fig:codes} and the four $Z$-type stabilizers correspond to weight-8 cells. The interchange of $X$- and $Z$-type stabilizers has the consequence that, for our implementation, the one-way transversal CNOT may only connect the logical control to the Steane code and the logical target to the Tetrahedral code. Logical $X$($Z$)- operators can be chosen to correspond to any edge (side) of the tetrahedron. For example, $\overline{X}_\mathrm{T} = X_0X_3X_{10}$ and $\overline{Z}_\mathrm{T} = Z_0Z_1Z_4Z_7Z_8Z_{12}Z_{14}$ are logical operators. Other representatives can be obtained by multiplication with stabilizers. As noted in Ref.~\cite{sullivan2024code}, the Tetrahedral code can be constructed as a doubled code from the Steane code and the trivial triorthogonal code with parameters $[[\Tilde{n}=1,1,\Tilde{d}=1]]$. This is why the $X$-generators of the Tetrahedral code explicitly contain the Steane code's $X$-plaquettes (also see Fig.~\ref{fig:codes}).

\begin{figure*}\centering
	\includegraphics[width=0.99\linewidth]{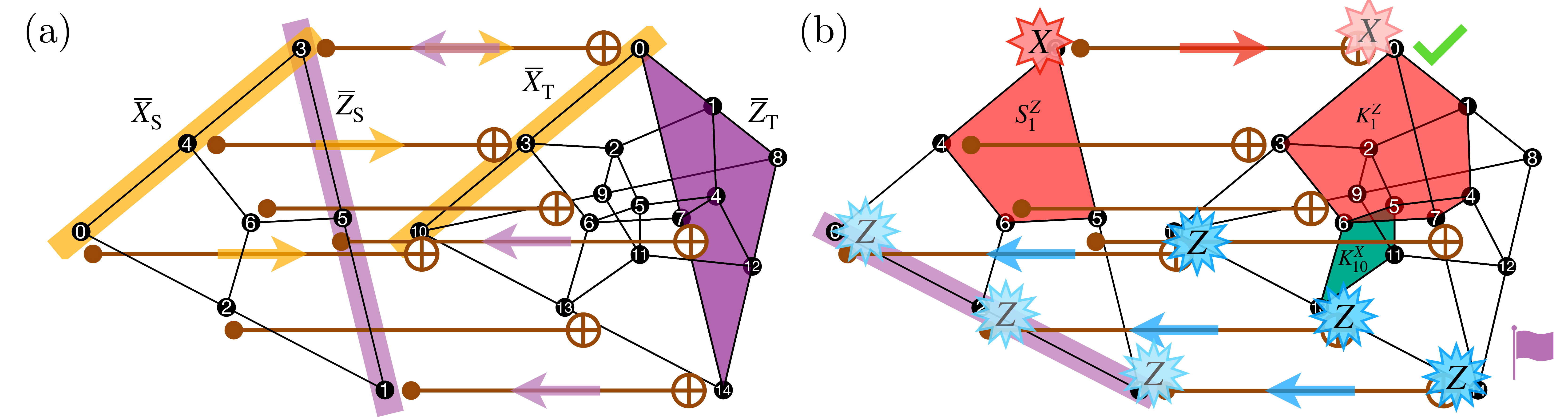}
	\caption{Propagation of logical operators and errors through the one-way transversal CNOT gate. (a) Propagating logical operators (see Eq.~\eqref{eq:cnotmap}) match the codes' geometries: The logical $X$-operator $\overline{X}_\mathrm{S} = X_0X_3X_4$ on the Steane code's left edge is mapped to an edge of the tetrahedron (both yellow), reflecting the fact that it has $d_X = 3$. Three Pauli operators of the logical $Z$-operator $Z_0Z_1Z_4Z_7Z_8Z_{12}Z_{14}$ on the right side of the Tetrahedral code (purple) propagate to the right edge of the Steane code (light purple) The logical operator $Z_0Z_3Z_6Z_7Z_{10}Z_{13}Z_{14}$ on the front side of the Tetrahedral code (not marked) maps to the weight-7 operator $Z^{\otimes 7}$ on the Steane code, which is also a logical operator. Every logical $Z$-operator on the Tetrahedral code has overlap 3 or 7 with the physical CNOT gates. It is important that a weight-3 logical $X$-operator on the Tetrahedral code, which may only have overlap 1 with the physical CNOT gates, such as $X_0X_1X_8$ (not marked), does \emph{not} propagate to the Steane code. (b) A single $X$-error on the Steane code (red star) propagates to a single correctable $X$-error on the Tetrahedral code (light red star). A weight-3 $Z$-error (blue stars), which is correctable in the $d_Z = 7$-version of the Tetrahedral code, propagates directly to a logical $Z$-error on the Steane code (light blue stars) and must therefore be prevented; for instance by utilizing a flag qubit that heralds such errors. Stabilizers that anticommute with the given correctable errors are colored in red and turquoise respectively.}
	\label{fig:logerrprop}
\end{figure*}

Let us now explain why this one-way transversal CNOT gate works; Figs.~\ref{fig:stabprop} and \ref{fig:logerrprop} show the propagation of stabilizers and logical operators through the seven physical CNOTs wired between the seven qubits of the Steane code and one side of the Tetrahedron. With the given qubit labels, the logical CNOT is implemented as
\begin{align}
    \overline{\text{CNOT}} =\, &\text{CNOT}(0, 10)\,\text{CNOT}(1, 14)\,\text{CNOT}(2, 13) \notag \\ &\text{CNOT}(3, 0)\,\text{CNOT}(4, 3)\,\text{CNOT}(5, 7)\notag \\ &\text{CNOT}(6, 6) \label{eq:logcnot}.
\end{align}
The labels of the physical controls refer to the Steane code and the labels of the physical targets refer to the Tetrahedral code (see also the circuit diagram in Fig.~\ref{fig:tv}). A physical CNOT gate propagates a single $X$ from the control to its target. A single $Z$ is propagated from the target to the control qubit. A Pauli $X$($Z$) on the target (control) does not propagate. Consequentially, the one-way logical CNOT in Eq.~\eqref{eq:logcnot} leaves the respective stabilizers of the Steane code and the Tetrahedral code invariant. The $X$-type stabilizers can only propagate from the Steane code to the Tetrahedral code but not vice versa; plaquettes of the Steane code are mapped to faces of the Tetrahedral code, which indeed are stabilizers by construction \cite{sullivan2024code}. As an example, the propagation 
\begin{align}
    S_1^X \otimes I \rightarrow S_1^X \otimes K_1^X
\end{align}
is depicted in Fig.~\ref{fig:stabprop}a. Since four physical CNOT gates are connected to three cells each, the weight-8 $Z$-type stabilizers of the Tetrahedral code propagate to only weight-4 plaquettes of the Steane code. Consider, for instance, the propagation
\begin{align}
    I \otimes K_3^Z \rightarrow S_3^Z \otimes K_3^Z
\end{align}
as shown in Fig.~\ref{fig:stabprop}b. 

A weight-3 logical $X$-operator on one edge of the Steane code is correctly mapped to a weight-3 logical $X$-operator on one edge of the Tetrahedral code. Logical $Z$-operators have minimal weight 7 and correspond to any side of the tetrahedron. They propagate to weight-3 logical $Z$-operators on the Steane code. These two propagation directions are illustrated with example operators in Fig.~\ref{fig:logerrprop}a. Since \emph{all} logical operators propagate like this up to stabilizer-equivalences (or do not propagate at all), Eq.~\eqref{eq:logcnot} implements the map
\begin{align}
    \overline{X}_{\mathrm{S}} &\rightarrow \overline{X}_{\mathrm{S}}\overline{X}_{\mathrm{T}} \notag \\
    \overline{X}_{\mathrm{T}} &\rightarrow \overline{X}_{\mathrm{T}} \notag \\
    \overline{Z}_{\mathrm{S}} &\rightarrow \overline{Z}_{\mathrm{S}} \notag \\
    \overline{Z}_{\mathrm{T}} &\rightarrow \overline{Z}_{\mathrm{S}}\overline{Z}_{\mathrm{T}}, \label{eq:cnotmap}
\end{align}
which is the action of a logical C$_{\mathrm{S}}$NOT$_{\mathrm{T}}$ gate between a Steane code (S) and a Tetrahedral code (T) state. 

Equipped with this logical CNOT gate, we now consider the FT $T$-gate implementation based on two FT code switching steps, depicted in Fig.~\ref{fig:scheme}. First, a logical zero state $\ket{0}_{\mathrm{T}}$ is prepared in the Tetrahedral code. Then, a logical state teleportation step moves the arbitrary input state $\ket{\psi}_{\mathrm{S}}$, encoded in the Steane code, onto the upper logical wire by the transversal logical CNOT gate, transversal measurements and transversal correction operations that are classically-conditioned on the logical measurement outcome. On the Tetrahedral code, the logical $T$-gate can simply be executed by performing fifteen physical $X$-rotations $R^X(\theta)$ of angles $\theta = \pm \pi/4$ since $\overline{T}_X = R^X_{\{0,2,4,6,8,10,11,14\}}(\pi/4)R^X_{\{1,3,5,7,9,12,13\}}(-\pi/4)$ \cite{bombin2006distillation}. After preparing a logical auxiliary qubit in the $\ket{+}_{\mathrm{S}}$ state of the Steane code, the second code switching step moves the logical state back to the lower logical wire. Note that the logical CNOT is connected in the same direction as for the first step. Transversal measurements of the Tetrahedral code state are performed and a conditional logical $X$-correction may be applied in the teleportation subroutine. As a result, the logical $T$-gate
\begin{align}
    \overline{T}_X &= \cos(\pi/8) - \ii \sin(\pi/8)\,\overline{X}_\mathrm{S} \end{align}
has been applied to the arbitrary Steane code state $\ket{\psi}_{\mathrm{S}} = \alpha \ket{0}_{\mathrm{S}} + \beta \ket{1}_{\mathrm{S}}$. 

We now also consider the propagation of errors. It is obvious that a single correctable error in the Steane code can, due to transversality, at most propagate to a single error on the Tetrahedral code, which is therein also correctable. Correctable $Z$-errors of weight-3 in the Tetrahedral code may either propagate to correctable weight-1 errors in the Steane code through a single physical CNOT gate but can also, and more severely, directly propagate to uncorrectable weight-3 $Z$-errors on the Steane code, as shown in Fig.~\ref{fig:logerrprop}b. Their appearance must therefore be prevented by an appropriate FT state preparation circuit. It is not enough to only fault-tolerantly prepare the 3D color code state but the state preparation routine must be FT with respect to the 2D color code as well in order to render the whole code switching scheme FT. 

It comes in handy that some correctable errors can be physically removed within a single code switching step at no extra cost: Instead of just conditionally applying logical operators during the teleportation part, one may also infer syndromes from the classical measurement result to adapt the feed-forward logical operation to include error correction. For instance, a single $Z_0$($X_0$)-error after preparation of $\ket{0}_{\mathrm{T}}$ ($\ket{+}_{\mathrm{S}}$) propagates down (up) through the subsequent CNOT gate and will cause the non-trivial $X$($Z$)-syndrome $\{+1, -1, +1\}$ ($\{-1, +1, +1, +1\}$) on the Steane (Tetrahedral) code, which can be obtained in classical post-processing from the $7$($15$)-bit string output by the transversal $X$($Z$)-basis measurement. Instead of, say, the logical operators $Z^{\otimes 15}$($X^{\otimes 7}$), in this particular case we would apply $Z^{\otimes 15}/Z_0 = Z_1Z_2\cdots Z_{14}$($X^{\otimes 7}/X_0 = X_1X_2\cdots X_6$) in case the logical measurement yields 1 and $Z_0$($X_0$) in case the logical measurement yields 0. 

Not \emph{all} correctable errors can be removed this way but only the ones that are detectable by the respective measurements. One example is the single error $Z_2$. Since the logical CNOT gate has no support on qubit 2 of the Tetrahedral code, see Eq.~\eqref{eq:logcnot}, the error $Z_2$ after preparing $\ket{0}_\mathrm{T}$ does not propagate to the Steane code at all and hence can not be detected and corrected via the transversal $X$-measurement on the Steane code qubit. Advantageously however, an error $Z_2$ after the transversal $T$-gate also does not propagate to the Steane code in the second switching step (but a $Z_0$ error would propagate, undetected by the transversal $Z$-measurements). 

\subsubsection{FT state preparation}\label{sec:ftstateprep}

As stressed in Ref.~\cite{butt2024fault}, FT state preparation of the 3D color code state must be reviewed in the light of switching to the 2D color code: While the 2D color code corrects $t = \lfloor \frac{d-1}{2} \rfloor$ errors of each Pauli type $X$ or $Z$, the corresponding 3D color code corrects $t$ errors of one Pauli type and a larger number of errors for the respective other type. This implies that one type of high-weight errors -- correctable in the 3D color code -- might propagate onto the 2D color code where they are uncorrectable but cause logical failure. Such high-weight errors should be avoided by a more restrictive FT state preparation of the 3D color code auxiliary qubit \cite{gottesman2009introductionquantumerrorcorrection, chamberland2018flag, zhou2024algorithmicfaulttolerancefast}.

Regarding the Tetrahedral code, the switching scheme only requires preparation of the logical zero state (see Fig.~\ref{fig:scheme}). A unitary encoding circuit, constructed via the Latin rectangle method, amended by a single flag qubit that catches one type of Pauli faults has been given before \cite{butt2024fault}. While this construction would suffice to fault-tolerantly prepare the state $\ket{0}_{\mathrm{T}}$, it is not enough to ensure fault tolerance of the full code switching scheme. Three kinds of dangerous faults must be prevented in our scheme and we now describe these in detail. 

\textbf{1) High-weight $X$-errors:} Since the Tetrahedral code can only correct a single $X$-error, all uncorrectable $X$-errors of weight 2 or higher must be sorted out. This can be done by an appropriate flag qubit that measures the logical $Z$-operator $Z_1Z_3Z_5Z_7Z_{9}Z_{12}Z_{13}$ \cite{butt2024fault}. 

\textbf{2) High-weight $Z$-errors:} $Z$-errors of weight larger than 1 on $\ket{0}_{\mathrm{T}}$ must be prevented from propagating to the Steane code during the course of code switching although they could be correctable in the Tetrahedral code. It is clear from Figs.~\ref{fig:scheme} and \ref{fig:logerrprop}b that propagation of such an error to the Steane code via the first logical CNOT gate, may cause a logically-flipped measurement result and subsequent erroneous application of a logical $Z$-operator. We find that, in our state preparation subroutine, three flags are necessary to catch all high-weight $Z$-errors. These flags can be chosen to measure the stabilizers $X_8X_9X_{11}X_{12}$, $X_2X_9X_{12}X_{14}$ and $X_0X_3X_{6}X_{7}$. 

\textbf{3) Mixed-type $X_iZ_j$-errors with $i \neq j$:} Although in CSS codes a single $X$-error on qubit $i$ and another single $Z$-error on qubit $j$ can in principle be corrected independently, our FT $T$-gate circuit may not respect the separation of errors into distinct $X$- and $Z$-sectors. Since we apply physical $T$-gates to the 15-qubit code state, the logical $T$-gate is not a Clifford circuit. A single $Z$-error on qubit $j$ is mapped by a single $T$-gate\footnote{Remember that our physical $T$-gates are $X$-rotations.} to the operator $(Z_j-Y_j)/\sqrt{2}$. Combined with another single $X$-operator on qubit $i$, there is a finite probability that the error superposition in conjunction with such $X$-error has the effect of a weight-2 uncorrectable error $X_iX_j$ ($i \neq j$), which breaks fault tolerance. If not sorted out, such an error would erroneously flip the logical $Z$-measurement after the second logical CNOT in Fig.~\ref{fig:scheme} and a logical $X$-operator would wrongly be applied when switching back to the Steane code. An additional flag qubit that measures the complementary $Z$-operator $Z_0Z_2Z_4Z_6Z_{8}Z_{10}Z_{11}Z_{14}$ achieves this goal. As mentioned above, some \mbox{weight-1} $Z$-errors propagate through the first logical CNOT of Fig.~\ref{fig:scheme} and could be corrected in the teleportation step by decoding the measurement result of the Steane code logical qubit. Since the logical CNOT is only comprised of seven physical CNOT gates, unfortunately not \emph{all} weight-1 $Z$-errors can be eliminated this way to prevent the occurrence of mixed-type $X_iZ_j$-errors.

In App.~\ref{sec:circs}, we give the non-FT state preparation circuit (Fig.~\ref{fig:init15}) and the total flag verification circuit (Fig.~\ref{fig:flags15}), which together render the logical auxiliary state preparation subroutine FT for transversal code switching. Note that measurements of the three $X$-flags with bare auxiliary qubits could again cause propagation of $X$-faults back to high-weight $X$-errors on the data qubits. Therefore two additional CNOT gates per flag measurement are required that again act as flag qubits in these subcircuits. However, no new qubits are needed but the measurement qubits employed previously to flag high-weight $X$-errors can be re-used here without reset. These three $X$-flags are also tailored to detect high-weight $Z$-errors that may have been propagated from the measurement qubits for the two $Z$-operators described above. 

For practical purposes, we suggest to use these circuits for non-deterministic logical state preparation because the acceptance rate of approximately $95\%$\footnote{This value is estimated from a circuit-level depolarizing noise model with parameters $p_1 = p_i = p_m = 10^{-4}$ and $p_2 = 10^{-3}$. More detail will be provided in Sec.~\ref{sec:noise}.} is reasonably close to unity for state-of-the-art quantum processors where the entangling gate error rate of order $p_2 \approx 10^{-3}$ is the limiting factor \cite{heussen2023strategies}. No mid-circuit measurements or branching logic are required to correctly flag all dangerous faults. Deterministic variants of logical state preparation might as well be employed but we suspect this approach to require a larger number of entangling gates in general for low physical error rates. 

At this point, we already notice that a large portion of the overall noise will stem from the logical state preparation of the triorthogonal, i.e.~the Tetrahedral, code state because the largest fraction of entangling gates is required in this step. In the distance-3 implementation, 69 CNOT gates out of 83 CNOT gates for the whole scheme are taken up by state preparation of both the Tetrahedral code state and the Steane code state, which we teleport onto in the second code switching step. In spite of this gate overhead, performing logical state teleportation onto a (relatively) fresh set of auxiliary qubits could yield lower failure rates than keeping a logical state on the same physical qubits throughout the whole scheme, as, e.g., in FT magic state preparation \cite{goto2016minimizing, chamberland2019fault, postler2022demonstration}, because intermediate errors may be erased this way; similar to teleportation-based Knill-type QEC compared to Steane-EC \cite{ryananderson2024highfidelity}. 

\subsection{Parallelized implementation} 

Transversality arguably constitutes the simplest method to render logical gates fault-tolerant. Not only are transversal gates conceptually elegant but they also do not require any additional qubit or repetition overhead. A small number of physical gates is typically sufficient to implement a transversal gate that appears hard to beat with alternative FT gate constructions. Within physical hardware architectures that carry out quantum error correction routines relying on transversal CNOT gates anyway, it is advisable to also use our transversal code switching scheme to implement a universal set of gates. It has already been shown that using transversal gates in general and especially a transversal entangling gate can have advantages in hardware platforms that potentially allow for highly parallelized gate operations such as neutral atom \cite{bluvstein2022quantum, evered2023high, bluvstein2024logical} or ion trap devices \cite{figgatt2019parallel, pino2021demonstration, zhu2023pairwise, valentini2024demonstration}. 

Given logical auxiliary qubits and parallelized physical-level gates, the transversal code switching scheme can be implemented in constant circuit depth w.r.t.~the code size. The first logical CNOT requires one time step, the transversal measurements and corrections require one time step each, the physical $T$-gates are all applied in parallel and the second code switching step then is done again in three time steps; the full logical $T$-gate circuitry is thus performed in seven steps regardless of the code distances, assuming a supply of logical auxiliary qubits is available. Unitary encoding circuits for the preparation of $n$-qubit logical auxiliary states with depth $\mathcal{O}(n)$ can be found for 2D topological codes with nearest-neighbor connectivity \cite{higgott2021optimal} and might be improved to $\mathcal{O}(\log n)$ by allowing for non-local qubit connectivity, exploiting symmetries \cite{moore1998parallel} or with the help of physical auxiliary qubits \cite{zhang2022quantum}. 

We emphasize that the logical qubit states have an auxiliary function only. This means that the physical qubits that hold these states can in principle be reused an arbitrary number of times. Also, if the physical hardware architecture is capable of such operations, many logical auxiliary qubit states for both codes could be prepared and stored in separate regions of the quantum computer in parallel and be supplied to the processing unit on demand. While conceptually similar to the idea of magic state factories, only Clifford circuits are necessary to prepare such logical auxiliary qubit states. Since our auxiliary states are stabilizer states, no-go theorems related to the non-reusability of magic states do not apply \cite{anderson2012power}.

In a practical implementation, the time $t_M$ to perform a mid-circuit measurement might be much longer than the time $t_G$ it takes to apply a single two-qubit gate. Therefore, the relatively large gate overhead induced by the necessity to prepare logical auxiliary qubits compared to the one-way transversal CNOT gate may not slow down the logical $T$-gate at all as long as the state preparation time $t_G \times L$ is smaller than the measurement time $t_M$, where $L$ is the number of CNOT layers in the logical auxiliary state preparation subroutine. Especially for quantum processors based on neutral atoms, where entangling gates only take time on the order of nanoseconds but measurements require several hundred microseconds, execution of the transversal code switching protocol will be limited by the mid-circuit measurement for low-distance codes \cite{evered2023high, bluvstein2024logical}.  In modern ion trap quantum processors on the other hand, single-qubit gates are generally fast but entangling gates and measurements can both take several hundred microseconds such that, here, offline preparation of logical auxiliary qubit states may provide an advantage.

\subsection{Limited connectivity}\label{sec:conn}

Today's quantum computers are to varying degrees limited in their capabilities to perform entangling gates between arbitrary pairs of the individual qubits. We now briefly discuss some implications from limited connectivity between physical qubits in the most prevalent types of quantum computing hardware. 

\subsubsection{Superconducting architecture}
State-of-the-art superconducting architectures are essentially restricted to nearest-neighbor connectivity and long-range coupling is a work in progress \cite{kjaergaard2020superconducting, bravyi2022future}. One option to practically implement transversal entangling gates nonetheless is to stack two 2D layers of superconducting transmon qubits and drive the nearest-neighbor entangling gate in the third, inter-layer, spatial dimension. If the spatial dimensionality shall be retained, one can instead make use of a lattice surgery protocol to perform logical CNOT gates at the price of an increased FT overhead \cite{horsman2012surface}. 

The long-range connectivity required for unitary encoding circuits could in theory be circumvented with the help of circuit knitting \cite{piveteau2023circuit}. One may resort to a stabilizer-measurement-based state preparation routine altogether if non-destructive operator measurements can be executed fast and with high fidelity. These can be executed largely in parallel on non-overlapping stabilizers. Parallelization of flag-FT stabilizer measurements has been recently shown for distance-5 codes \cite{liou2024reducingquantumerrorcorrection}. Also on stabilizers, which overlap on a number of data qubits, measurement circuits may be executed in parallel with appropriate gate schedules. Note that in a stabilizer-measurement-based state preparation routine, entangling gates are always only applied between data qubits and auxiliary qubits but never between two data qubits. As a consequence, the occurrence of dangerous mixed-type $X_iZ_j$ errors, as described in Sec.~\ref{sec:ftstateprep}, is naturally prevented for $d=3$ without additional flag qubits (apart from those that might be needed for FT stabilizer measurements, depending on the specific code).

\subsubsection{Neutral atom platform}
Atomic platforms natively implement long-range interactions due to their ability to shuttle large registers of atoms in parallel combined with entangling operations via Rydberg states \cite{henriet2020quantum, wintersperger2023neutral}. We suggest that our scheme, and especially the transversal CNOT gate, would ideally be implemented in a neutral atom quantum processor \cite{bluvstein2022quantum}. Their often-claimed ``all-to-all'' connectivity might be limited in practice by the necessity to physically move qubits into close proximity because the physical distance to cover by such operations grows with the code distance.

While a relatively small code state could be prepared with flag qubits, we alternatively envision a Steane-type verification of logical auxiliary qubit states, which leverages the parallelized transversal CNOT gate, as demonstrated recently \cite{bluvstein2024logical}. Here, several $\ket{0}_{\mathrm{T}}$ states are prepared non-fault-tolerantly and are then connected via transversal CNOT gates. Transversal measurements can be used to verify the absence of uncorrectable errors. Also, a combination of, say, flag qubits that catch $X$-errors and a half-cycle of Steane-type EC that removes all correctable $Z$-errors could be advantageous in practice. 

\begin{figure}\centering
	\includegraphics[width=0.99\linewidth]{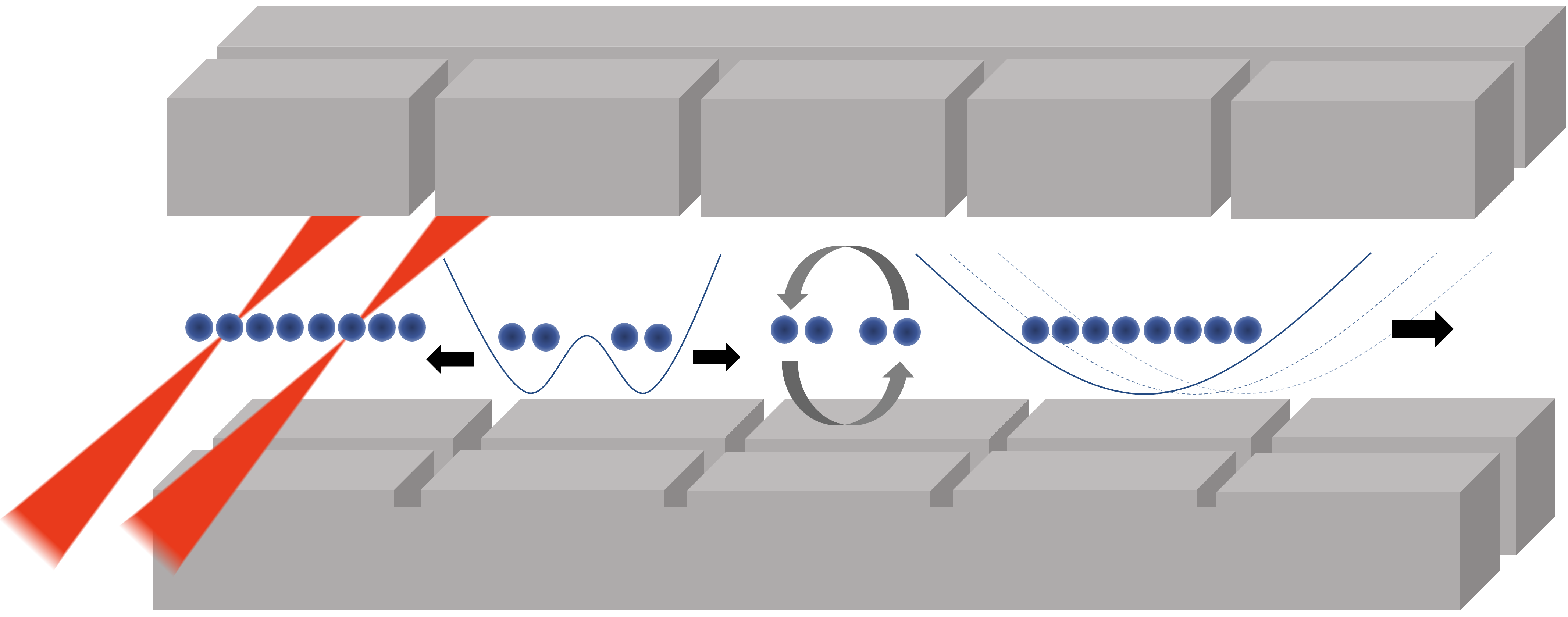}
	\caption{Illustration of a segmented ion trap with five zones (grey bars), each holding a linear ion crystal (blue dots). Focused laser beams can be pointed at any pair of ions to perform an entangling gate (red). Any ion chain can be reconfigured within a single crystal (grey arrows) and can be shuttled from one zone to another (black arrow) by modifying the trapping potential (blue lines).}
	\label{fig:yellow_trap}
\end{figure}

\subsubsection{Ion trap processor}
Entangling gates in trapped-ion quantum processors, such as the one sketched in Fig.~\ref{fig:yellow_trap}, are conducted by leveraging the Coulomb interaction between the electrically charged particles \cite{molmer1999multiparticle, sorensen2000entanglement, ballance2017high}. This way one achieves the ability to entangle arbitrary pairs of ion qubits in a single ion crystal. Technical problems may arise when aiming to execute such geometric phase gates with high fidelity once more than 10-20 ions are trapped in the same potential well. 

Our transversal code switching scheme can be conveniently partitioned into a segmented ion trap architecture where several ion crystals are connected via junctions that allow one to physically move ions between laser-interaction zones. We provide a detailed example of partitioning the $d=3$ instance of transversal code switching into a segmented ion trap in App.~\ref{sec:ion3d}.

The transversal CNOT gate can be performed with any number of trap segments as long as one control qubit and its corresponding target qubit can be brought into same zone. In the extremal case of two-ion crystals, all physical entangling gates could be performed in parallel in just a single time step. For an implementation in just two size-$n$ ion crystals, $\lceil n/2 \rceil$ sequential physical entangling gates must be performed (see Fig.~\ref{fig:tv}).

Regarding logical auxiliary state preparation, let us consider unitary encoding circuits created with the Latin rectangle method or related techniques \cite{steane2004fastfaulttolerantfilteringquantum, cross2009comparativecodestudyquantum}. To construct these densely connected circuits, for each $X$-type stabilizer a single physical qubit is initialized in the $\ket{+}$ state and serves as control qubit for several physical CNOT gates connected to the rest of the qubits that form the stabilizer. In general, we can assign a trap segment to each control qubit/stabilizer\footnote{For this reason, the original Tetrahedral code defined by four $X$- and ten $Z$-stabilizers \cite{butt2024fault} might be preferable to reduce the number of CNOT gates and ion reconfigurations if this state preparation method is employed.}. Then, by devising a shuttling schedule that allows each target qubit to get entangled with the control qubit that represents the stabilizer it belongs to, the state preparation circuit can be implemented with the help of ion reconfigurations. Additional shuttling moves that enable measurement of logical operators, as for flag verification, might be required. An embedding to initialize the logical zero state of the Tetrahedral code is given in App.~\ref{sec:ion3d}.

On a more general note, when performing state preparation via stabilizer measurements, it is recommended to assign qubits of disjoint stabilizer operators to ions in different gate zones in order to be able to measure them in parallel. This works as long as the highest-weight stabilizer to be measured fits into a single trap segment.

Wrapping up, we point out that all numerical tools are publicly available to construct the code switching protocol on the physical qubit level. The python package \emph{StabGraph} can be used to find unitary encoding circuits \cite{amaro2019stabgraph, amaro2020scalable}. Fault-tolerant versions of these circuits may be found in an automated way with the help of Refs.~\cite{zen2024quantum, zen2024rlftqc, peham2024automatedsynthesisfaulttolerantstate}. A shuttling compiler assists in devising a concrete schedule of moving ions or atoms with short shuttling paths \cite{kreppel2023quantum, schoenberger2024using}. 

\section{Numerical estimation of logical failure rates}\label{sec:noise}

In this section we investigate the logical failure rate of the fault-tolerant $T$-gate implementation for the Steane code enabled via transversal code switching. The relatively low number of physical CNOT gates and its transversality properties play out in favor of achieving lower logical failure rates than the state-of-the-art flag-FT code switching scheme \cite{butt2024fault, pogorelov2024experimental} and even magic state injection \cite{postler2022demonstration, heussen2023strategies}. We do not include the intermediate error correction, described in Sec.~\ref{sec:steanetetraswitch}, in our numerical analysis.

In what follows we first employ a generic, architecture-agnostic depolarizing circuit-level noise model with a single parameter $p$, which represents the probability of applying faults to circuit operations. In particular, we apply one of the three possible Pauli operators $X, Y$ or $Z$ after any physical single-qubit gate, state initialization or before a measurement according to the channel $\curlE(\rho) = (1-p) \rho + \frac{p}{3} \left( X \rho X + Y \rho Y + Z \rho Z \right)$. In addition, physical two-qubit gates are followed by a two-qubit fault operator that is randomly drawn from the set $\Lambda = \{I, X, Y, Z\} \otimes \{I, X, Y, Z\} \,\backslash\, \{I \otimes I\}$ as prescribed by the depolarizing channel $\curlE(\rho) = (1-p) \rho + \frac{p}{15} \sum_{P \in \Lambda} P \rho P$.

Subsequently, we employ a more detailed incoherent multi-parameter Pauli noise model. It has been shown before that such twirled noise models are capable of approximating logical failure rates in state-of-the-art ion trap quantum processors well enough \cite{postler2022demonstration} to estimate the scaling behavior of the noisy circuit's failure rate in the low-error-rate-limit and allow to conduct a break-even analysis \cite{heussen2023strategies}, while granting numerically efficient classical stabilizer simulations \cite{aaronson2004improved}. In our noise model we include depolarizing noise with four different parameters on single-qubit gates ($p_1$), two-qubit gates ($p_2$), initializations ($p_i$) and measurements ($p_m$) as well as different error rates on idling locations that belong to different types of operations ($p_{\text{idle},1}, p_{\text{idle},2}, p_{\text{idle},m}$ respectively), during which an idling qubit experiences a dephasing channel $\curlE(\rho) = (1-p_{\text{idle}}) \rho + p_{\text{idle}} Z \rho Z$. In an ion trap, this could be caused by magnetic field fluctuations that are assumed to be uncorrelated between individual ion positions.

\begin{table}\begin{center}
\begin{tabular}{ c || c | c}
    Rate set & \emph{high} ($T_2 = \SI{100}{ms}$) & \emph{low} ($T_2 = \SI{2}{s}$) \\ \hhline{=|=|=}
    \rule{0pt}{2.5ex} $p_1$ & $5 \times 10^{-3}$ & $10^{-4}$ \\ 
    $p_2$ & $2.5 \times 10^{-2}$ & $10^{-3}$ \\ 
    $p_i$ & $4.5 \times 10^{-3}$ & $10^{-4}$ \\ 
    $p_m$ & $4.5 \times 10^{-3}$ & $10^{-4}$ \\ 
    $p_{\text{idle}, 1}$ & $7.5 \times 10^{-5} $ & $3.75 \times 10^{-6} $ \\ 
    $p_{\text{idle}, 2}$ & $10^{-3}$ & $10^{-4}$ \\ 
    $p_{\text{idle}, m}$ & $1.5 \times 10^{-3}$ & $10^{-4}$
\end{tabular}
\end{center}
\caption{We perform numerical simulations of incoherent Pauli noise with two sets of parameters that are representative for state-of-the-art or near-term ion trap quantum processors. The operation times $t$ translate to physical idling error rates $p_\text{idle}$ via the coherence time $T_2$ as $p_\text{idle} = \left[1-\exp{\left(-t/T_2\right)}\right]/2$ for a dephasing channel applied on idling qubits. We assume operation times $t_1 = \SI{15}{\micro s}, t_2 = \SI{200}{\micro s}$ and $ t_m = \SI{300}{\micro s}$ for the \textit{high} rates and $t_1 = \SI{15}{\micro s}$ and $t_2 = t_m = \SI{400}{\micro s}$ for the \textit{low} rates.}
\label{tab:errorrates}
\end{table}

Two sets of error rates, called \emph{high} and \emph{low}, that we use for our analysis are given in Tab.~\ref{tab:errorrates}. Statevector simulations in \texttt{PECOS} \cite{ryan2018quantum, pecos} of the logical $T$-gate from Fig.~\ref{fig:scheme} applied to the logical zero state of the Steane code yield failure rates of $(16.6 \pm 1.2) \times 10^{-2}$ and $(8.2 \pm 1.6) \times 10^{-4}$ for the two parameter sets respectively. We find that a stabilizer simulation of the Clifford circuit where we replace the physical $T$-gates with noisy identity gates and apply the circuit to the logical $Y$-eigenstate \mbox{$\ket{+\ii}_{\mathrm{S}} = \left( \ket{0}_{\mathrm{S}} + \ii \ket{1}_{\mathrm{S}}\right)/\sqrt{2}$} yields logical failure rates $(15.3 \pm 1.1) \times 10^{-2}$ and $(6.2 \pm 0.8) \times 10^{-4}$. These are in good agreement with the statevector simulation results. The state $\ket{+\ii}_{\mathrm{S}}$ is susceptible to both $X$- and $Z$-errors, just as the intermediate non-stabilizer state generated in the code switching scheme. We therefore use the stabilizer simulation as a proxy to full statevector simulations in order to allow for faster numerical simulations. The following simulation results are obtained with the python package \texttt{qsample} \cite{heussen2024dynamical, qsample}. All errorbars in this work are $68\%$ confidence intervals. 

\subsection{Depolarizing noise model}

\begin{figure}\centering
	\includegraphics[width=0.99\linewidth]{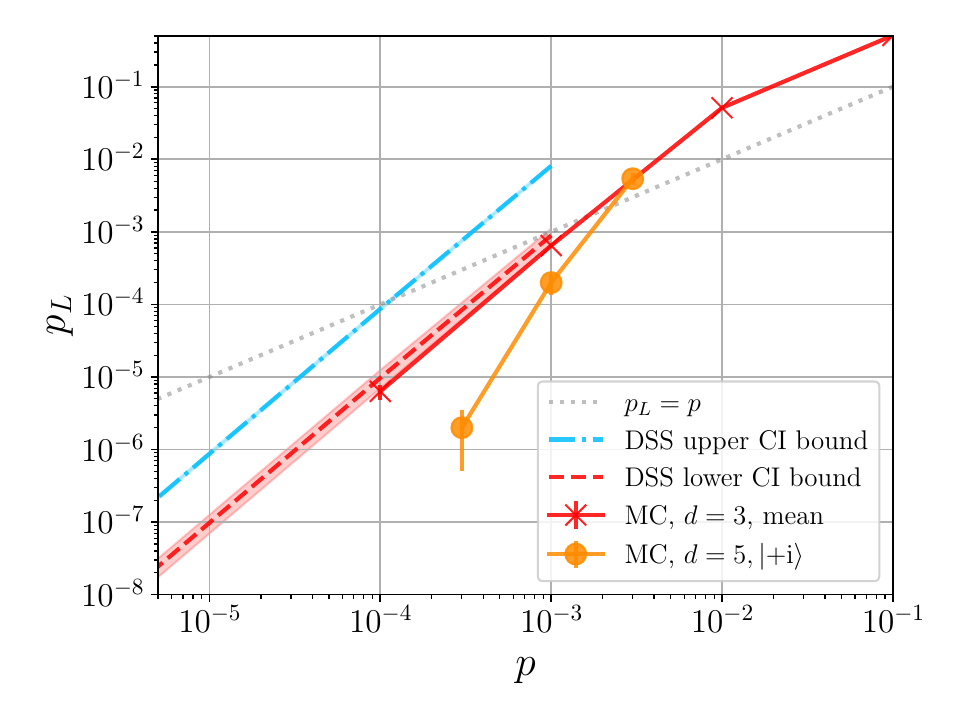}
	\caption{Logical failure rate estimation of the logical $T$-gate for a single-parameter depolarizing noise model. Direct Monte Carlo (MC) simulation with up to $10^6$ shots suggests that the break-even point $p_L = p$ lies at $p \approx 2 \times 10^{-3}$ when averaging failure rates (red line with crosses) for the three logical input states $\ket{+}_\mathrm{S}, \ket{+\ii}_\mathrm{S}$ and $\ket{0}_\mathrm{S}$. Individual failure rates are $(3.1 \pm 0.2) \times 10^{-4}, (9.3\pm 0.5) \times 10^{-4}$ and $(7.0\pm 0.3) \times 10^{-4}$ respectively at $p = 10^{-3}$. For the $[[17,1,5]]$ code, discussed in Sec.~\ref{sec:scaleup}, we observe that a better performance can be expected for the $\ket{+\ii}_\mathrm{S}$ state, which is the most noise-susceptible of the three Pauli eigenstates, at $p = 10^{-3}$ (orange line with circles). With $10^3$ shots of dynamical subset sampling (DSS), we extract the low-$p$ scaling behavior $p_L = \mathcal{O}(p^2)$ for the FT protocol of distance $d=3$. In this regime, the MC logical failure rate estimation coincides with the lower bound of the DSS confidence interval (CI, dashed line). The relatively large gap to the DSS upper CI bound (dash-dotted line) is an artifact of the DSS implementation in \texttt{qsample}.}
	\label{fig:scaling_singleP}
\end{figure}

We show numerical estimations of logical failure rates for the single-parameter depolarizing noise model in Fig.~\ref{fig:scaling_singleP}. Both a direct Monte Carlo (MC) sampling approach and dynamical subset sampling (DSS) are employed and yield consistent results. The former is used to accurately sample data points at relatively large $p$ where the sampling error can be made small with at most $10^6$ shots. The latter enables extracting the quadratic low-$p$ scaling behavior analytically from only a small number of $10^3$ shots (see Ref.~\cite{heussen2024dynamical} for details). 

The break-even point where $p_L = p$ can be identified at $p \approx 2 \times 10^{-3}$. This point is a meaningful comparison of an FT vs.~physical-qubit implementation in the following sense: In a current or near-term quantum processor, a quantum algorithm performed with physical qubits will be dominated by the entangling gate error $p \sim p_2$. For an implementation with 2D color codes, a quantum algorithm performed with logical qubits will be dominated by the logical $T$-gate error since this is by far the most complicated logical gate to execute for this class of QEC codes. Note that this break-even point is very close to the threshold of QEC cycles with the 2D color code under circuit-level depolarizing noise, $p \approx 2-3 \times 10^{-3}$, reported in Refs.~\cite{kubica2018abcs, chamberland2020triangular, takada2024improving}, which reflects the low overhead of the code switching scheme and its advantageous design for operational noise.

Simulation of a distance-5 version of the protocol yields a lower logical failure rate than the distance-3 implementation for a physical error rate below $p \approx 2 \times 10^{-3}$. A more detailed discussion of the distance-5 simulation is postponed to Sec.~\ref{sec:scaleup}.

\subsection{Multi-parameter noise model}

Let us now employ a more elaborate noise model described by operational noise as well as dephasing noise on idling qubits. We restrict ourselves to a single idling noise rate $p_\text{idle}$ in the following since the dominant idling noise stems from waiting during relatively slow two-qubit gates and measurements.

\begin{figure}\centering
	\includegraphics[width=0.99\linewidth]{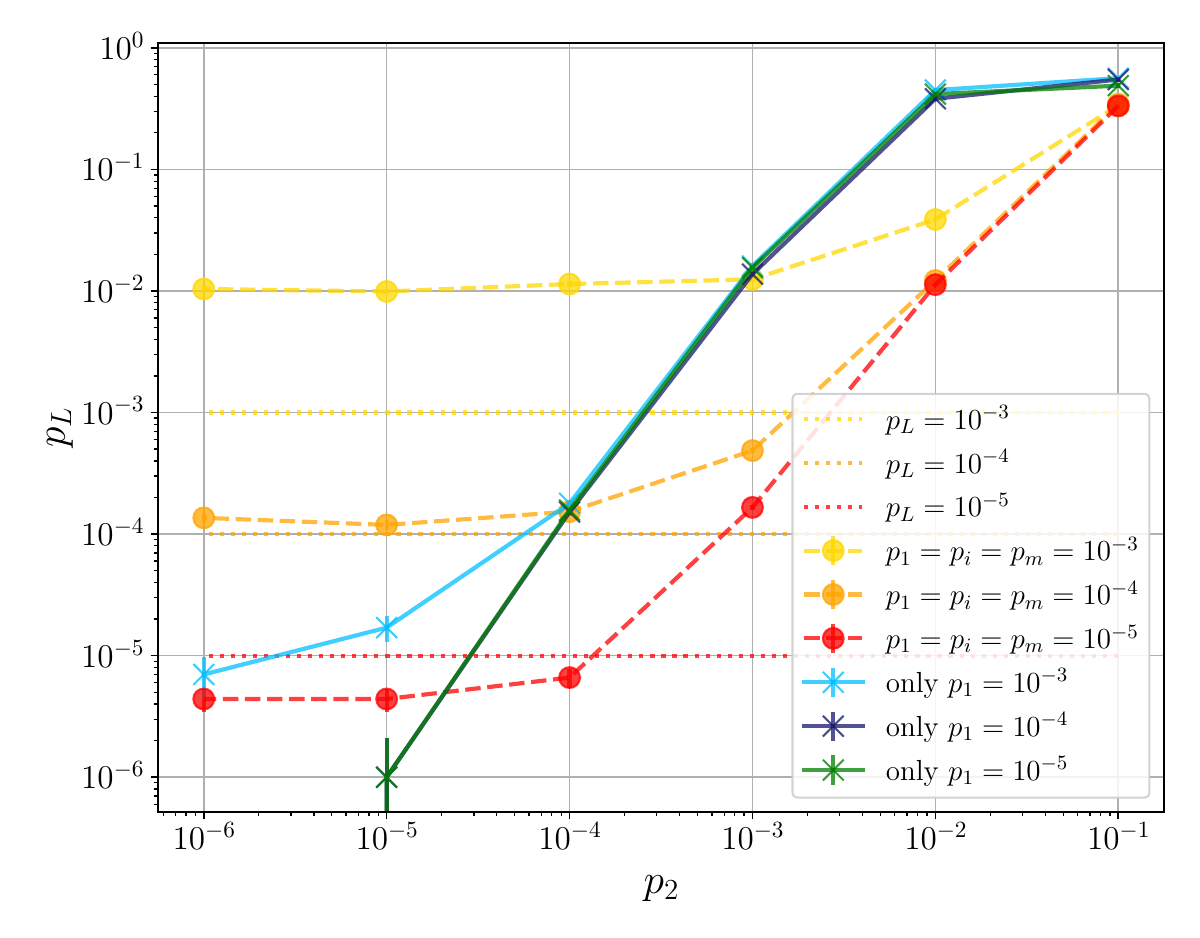}
	\caption{Logical failure rate estimation of the logical $T$-gate for various ion trap parameters. Dashed lines with circles represent a scenario where all but the entangling gate error rate $p_2$ are fixed and $p_2$ is varied. For all these curves we set $p_\text{idle} = 10^{-4}$. At low $p_2$, we observe saturation since noisy single-qubit operations do not permit a further decrease of the logical failure rate. When we also scale the other parameters $p_i = p_m = p_\text{idle} = p_2$ and \textit{only} $p_1$ is kept fixed, as shown by the solid lines with crosses, logical failure rates can be suppressed below the error rate of a physical $T$-gate. For instance, \mbox{$p_2 \leq 2 \times 10^{-4}\,(7 \times 10^{-5})$} is sufficient to yield $p_L < p_1 = 10^{-3}\,(10^{-4})$.}
	\label{fig:scaling_ionparams}
\end{figure}

Figure \ref{fig:scaling_ionparams} depicts a collection of scaling different parameter subsets. When only the entangling gate error rate $p_2$ is scaled and all other parameters are held constant, the failure rate of the logical $T$-gate exhibits a noise floor that cannot be undercut. This noise floor is determined by the number of uncorrectable errors that stem from two single-qubit operation/idling locations. Especially, for our example rates $p_1 = p_i = p_m = 10^{-3}$ and $10^{-4}$, we see that the logical failure rate does not drop below the physical $T$-gate error rate $p_1$. Only $p_1 = p_i = p_m = 10^{-5}$ is small enough so that the total scheme can actually yield a lower logical failure rate. We keep $p_\text{idle} = 10^{-4}$ fixed for all these three cases. 

In a different scenario, we only keep $p_1$ fixed and scale all other error rates $p_2 = p_i = p_m = p_\text{idle}$ uniformly. A clear advantage $p_L < p_1$ can be seen in this case in Fig.~\ref{fig:scaling_ionparams} as soon as $p_2$ drops below $2 \times 10^{-4}\,(7 \times 10^{-5})$ for $p_1 = 10^{-3}\,(10^{-4})$ although here, also, a saturation of logical failure rates seems to set in eventually as $p_2 \rightarrow 0$. The logical $T$-gate performs advantageously even when comparing to physical $T$-gates that only fail with a rate of $p_1 = 10^{-5}$.

We acknowledge that reaching entangling gate error rates this low presents considerable experimental challenges. Note, however, that it is not strictly deemed necessary to run a logical $T$-gate with lower failure rate than the physical $T$-gate for a fault tolerance advantage. It is sufficient that the failure rate of a logical state after applying the logical $T$-gate is low enough such that it can be further reduced by a subsequent round of QEC that will likely be required in any practical algorithm performed with logical qubits anyway.

\section{Comparison to other universality strategies}\label{sec:comp}

\begin{table*}\begin{center}
\begin{tabular}{ c || c | c | c | c}
    \multirow{2}{3.5cm}{~~~~$(d = 3)~T$-gate via...} & \multicolumn{2}{c|}{Failure rate $p_L$} & \multirow{2}{1.5cm}{\#\!\,CNOTs} & \multirow{2}{1.5cm}{\#\!\,qubits} \\
    & \emph{high} & \emph{low} & & \\ \hhline{=|=|=|=|=}
    \rule{0pt}{2.5ex} Transversal CS & $(16 \pm 1) \times 10^{-2}$ & $(8 \pm 2) \times 10^{-4}$ & 83 & 24 \\ \hline
    \rule{0pt}{2.5ex} Flag-based CS & $\gtrsim 3 \times 10^{-2}$ & $\gtrsim 2 \times 10^{-3}$ & 108 (70) & 17 (12) \\
    Magic state injection& $(4.9 \pm 0.4) \times 10^{-2}$ & $(3.1 \pm 0.1) \times 10^{-3}$ & 119 (55) & 18 (16) \\
    Concatenation & - & - & $\gtrsim$350 (154)& 105 (70)
\end{tabular}
\end{center}
\caption{Comparison of performance and required resources for different small-scale universality strategies. For \emph{low} physical error rates, transversal code switching (CS) yields the lowest logical failure rate, obtained via statevector simulations, but can be improved upon by deterministic flag-based CS or non-deterministic magic state injection for the \emph{high} physical error rate set. Nonetheless, transversal CS is implemented with the lowest number of CNOT gates in case deterministic protocols are used. We assume the typical protocol case where no errors occur for this figure of merit. The values in brackets refer to non-deterministic versions of the indicated schemes, which may save CNOT gates at the price of introducing post-selection. Transversal CS is only mildly non-deterministic in the sense that logical state preparations may need to be repeated in case flags are triggered. The slightly higher number of physical qubits seems acceptable in the light of recent hardware developments.}
\label{tab:comp}
\end{table*}

\begin{figure*}\centering
	\includegraphics[width=0.99\linewidth]{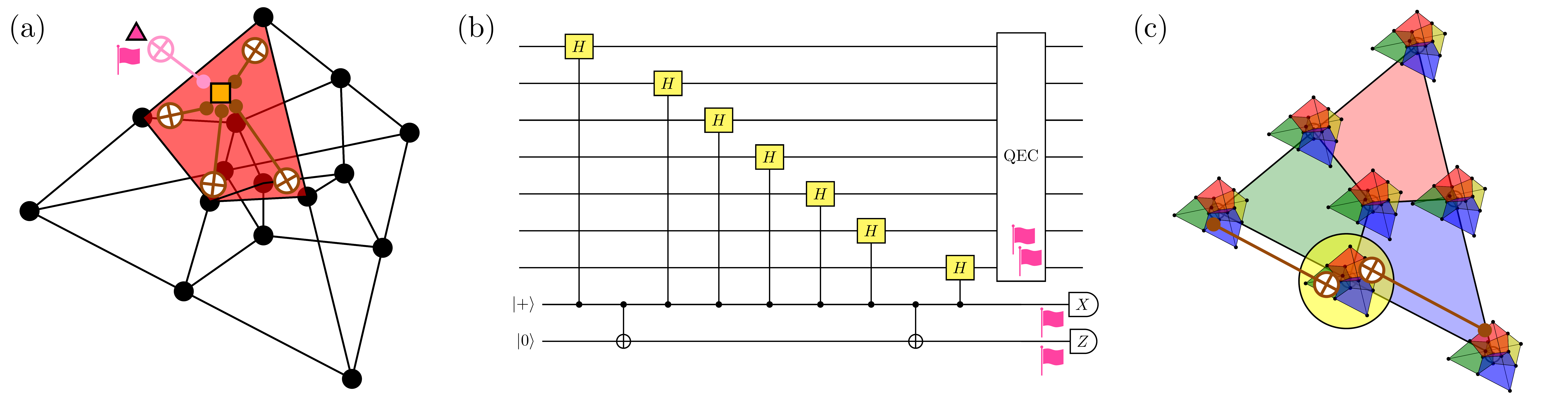}
	\caption{Illustrations of common FT universality strategies. (a) For flag-based code switching, individual stabilizers are measured with the help of single auxiliary qubits (orange square) that are also connected to flag qubits (pink triangle). (b) FT magic state preparation relies on fault-tolerantly measuring an operator whose eigenstate is a magic state. One example is the Hadamard operator, which has eigenbasis $\rho_{\pm H} = ( I \pm (X + Z)/\sqrt{2})/2$. (c) Replacing each physical qubit of the Steane code with a logical qubit encoded in the Tetrahedral code, provides a recipe to perform the combined set of transversal gates of both codes fault-tolerantly. The authors of Ref.~\cite{chamberland2017overhead} propose a simplification that uses only three Tetrahedral codes. Fault tolerance must be ensured on the circuit level, for instance by performing a round of QEC on the outer Steane code.}
	\label{fig:techcomp}
\end{figure*}

In this section we provide a quantitative comparison of our transversal code switching scheme with other state-of-the-art techniques to achieve an FT universal gate set: Flag-based code switching in Sec.~\ref{sec:flagcs}, magic state injection in Sec.~\ref{sec:msi} and concatenation in Sec.~\ref{sec:concat}. All three techniques allow one to establish FT circuits on the physical level without employing complicated magic state distillation schemes, which are only FT when applied to logical qubits. A comparison of logical failure rates and resource count for specific small instances of these schemes is given in Tab.~\ref{tab:comp}. Transversal code switching from the Steane code to the Tetrahedral code and back requires at least 24 qubits\footnote{It is assumed that flag qubits are re-used after state preparation of either logical auxiliary qubit, otherwise the qubit count might increase to 27. One may use a completely fresh set of qubits to prepare the Steane code auxiliary qubit in the second switching step if no fast reset operation is available. This would add another 7 data qubits and 1 flag qubit.} and 83 CNOT gates. 

For comparison with flag-based code switching, we consult the numerical simulation results for logical failure rates that are given in Ref.~\cite{butt2024fault}. The authors use a similar noise model but different error rates so that we only extract a rough lower bound estimation for our noise parameters based on their results. They conclude that flag-based code switching does not outperform FT magic state preparation and injection in near-term devices. 

To compare our scheme to FT magic state preparation and injection \cite{postler2022demonstration}, which implements a logical $\pi/4$-rotation about the $Y$-axis $T_Y = R^Y(\pi/4)$, we run our own simulations with the circuitry and noise model of Ref.~\cite{postler2022demonstration} and report the mean logical infidelity of the states $T_Y\!\ket{0}_{\mathrm{S}}$ and $T_Y\!\ket{1}_{\mathrm{S}}$ in Tab.~\ref{tab:comp}. 

The comparison to a concatenation-enabled universal set of gates indicates that a prohibitive number of CNOT gates and physical qubits would be required to implement such a scheme with state-of-the-art quantum processors. Therefore, we refrain from explicitly investigating their logical failures rates numerically in this work.

\subsection{Flag-style code switching}\label{sec:flagcs}

The first fully FT code switching scheme was given in Ref.~\cite{butt2024fault} and shortly after demonstrated experimentally \cite{pogorelov2024experimental}. Here, the authors provide deterministic protocols to switch between the Steane code and the Tetrahedral code that rely on stabilizer measurements with flag circuits, sketched in Fig.~\ref{fig:techcomp}a, and the application of gauge operators in the subsystem code that contains both stabilizer codes, each as a particular gauge. A large reduction of resource overhead is achieved by allowing for post-selection and using a morphed 10-qubit code that has distance $d = 2$ and therefore only serves to detect errors but can not correct them \cite{vasmer2022morphing}. 

Using the Tetrahedral code, the given scheme requires more CNOT gates than our transversal code switching protocol: In case that no faults occur, 108 CNOT gates are needed for this deterministic variant. However, in the pathological case where flags are triggered early in the protocol, a large portion of measurements can subsequently be omitted and in these special cases the protocol might terminate faster and with fewer CNOT gates. This can lead to a better performance at \emph{high} physical error rates. With entangling gates typically being the most noisy operations, we anticipate a worse overall performance of this scheme in the \emph{low} physical error rate regime. However, a major advantage of the protocol for the implementation in small-scale devices is that two auxiliary qubits are sufficient for the flagged stabilizer measurements. Additionally, decoding of higher weight errors is relatively simple; only the flag error set needs to be determined. Since all circuits only contain Clifford gates, this step can be assisted by efficient numerical stabilizer simulations when scaling up to larger codes.

We point out that the number of stabilizers $n - k$ that might need to be measured quickly grows for topological codes with $k=1$ since the number of qubits $n$ in a 2D code scales as $\mathcal{O}(d^2)$. With a Shor-type measurement routine, each stabilizer needs to be measured $\mathcal{O}(d)$ times to maintain fault tolerance so the total number of CNOT gates is at least $\mathcal{O}(d^3)$, assuming the number of CNOTs per stabilizer measurement is constant as for qLDPC codes. It might also be challenging in practice to measure high-weight operators with flag circuits.

\subsection{Magic state injection}\label{sec:msi}

Another commonly used technique to implement the $T$-gate using only Clifford gates is known as magic state preparation and injection \cite{reichardt2004improved, bravyi2005universal, chamberland2019fault, postler2022demonstration, gupta2024encoding}. Fault-tolerant preparation of a logical magic state is followed by a teleportation circuit, through which the $T$-gate is injected and the magic state is consumed\footnote{Referring back to our transversal code switching scheme, we note that one may prepare a magic state directly in the Tetrahedral code and teleport it one-way to the Steane code to obtain the logical magic state.}. For its first FT experimental demonstration, a non-deterministic FT magic state preparation scheme for the Steane code that uses a total of eight flags was employed \cite{postler2022demonstration}. Fault-tolerant measurement of the logical Hadamard operator is a centerpiece subroutine of this protocol and depicted in Fig.~\ref{fig:techcomp}b. A large fraction of runs is discarded due to post-selection but a high-fidelity magic state is achieved this way using only 10 qubits. Subsequently, an additional logical CNOT is required between the Steane code magic state and the target state $\ket{\psi}_\mathrm{S}$, which is encoded in another Steane code. Simulations suggest that the logical state fidelity worsens dramatically if the FT magic state preparation scheme is made deterministic by repeatedly measuring the logical Hadamard operator \cite{heussen2023strategies}.

Even more flags would be required to go to larger distances \cite{chamberland2019fault, chamberland2018flag}. Unfortunately, measurement of the logical Hadamard operator and full rounds of FT QEC need to be repeated $(d-1)/2$ times \cite{chamberland2020very}. Since the controlled-$H$ gate is not a Clifford gate, decoding under circuit-level noise may become more complicated \cite{heussen2023strategies}.

We note that one may directly opt to use a triorthogonal code as the ``default'' code so that the universal gate set, except the Hadamard gate, is transversal. The Hadamard gate might be performed fault-tolerantly via magic state injection \cite{jones2024hadamard}, lattice surgery \cite{cohen2022low, cowtan2024css} or using logical auxiliary qubits \cite{paetznick2013universal}.

\subsection{Concatenation}\label{sec:concat}

By concatenating two different QEC codes that each have a set of transversal gates, one obtains a larger code on which logical gates -- previously non-FT on the separated codes -- may naturally become FT \cite{oconnor2014using, yoder2016universal}. A dangerous error on the inner code, resulting in logical failure, can be corrected by the outer code. Faults on the outer code only spread between code blocks of the inner code and can be separately corrected by each instance of the inner code. Decoding of concatenated codes is straight-forward by decoding the individual QEC codes on each level.  

If both codes can be implemented in, say, a 1D hardware architecture, then the concatenated code has a natural embedding in 2D. A very simple example for this could be a 1D bitflip code in the horizontal direction and a 1D phaseflip code in the vertical direction of a 2D square lattice, which, concatenated together, can correct an arbitrary Pauli error. 

However, a potentially large gate overhead may be induced by the necessity to perform QEC on each level of concatenation. Figure \ref{fig:techcomp}c illustrates a well-known implementation of a logical $T$-gate on the Steane code, which only becomes FT when applied to logical qubits encoded in the Tetrahedral code \cite{oconnor2014using}\footnote{Concatenation of a Tetrahedral code state with the Steane code can be done via the circuit given in Ref.~\cite{buchbinder2013encoding}. We are not aware of a state-agnostic encoding circuit for the Tetrahedral code or the morphed 10-qubit code \cite{vasmer2022morphing} but use the CNOT count of their logical zero state preparation circuits as a lower bound estimation in Tab.~\ref{tab:comp}.}. Note that intermediate rounds of QEC in each of the used codes may further increase the CNOT gate overhead. Crucially, any logical operator at the end of a Tetrahedral code's encoding procedure must be corrected by performing QEC on the Steane code. Otherwise, for instance, a logical $Y$ error on a Tetrahedral code may spread maliciously through the subsequent logical CNOT gates and lead to failure of the Steane code. 

We emphasize that, by concatenating two codes to achieve a universal gate set, the minimal distance of the resulting code might not change. Only further concatenation may yield the famous double-exponential suppression of noise \cite{aliferis2006quantum}. Here lies the central practical problem with concatenation: While, already for distance $d = 3$, the concatenation of the Steane code and the Tetrahedral code (see Fig.~\ref{fig:techcomp}c) leads to a 105-qubit code -- or 70-qubit code when the $[[10,1,2]]$ code is used instead of the Tetrahedral code --, the number of qubits keeps scaling exponentially with $d$ \cite{chamberland2017overhead}. Recent developments suggest, however, that an advantage of concatenated codes might not only be expected asymptotically but also in finite size realizations that could become available in real hardware in the near future \cite{yoshida2024concatenate, goto2024manyhypercube, yamasaki2024time}.

\section{Scaling up}\label{sec:scaleup}
The challenge to build a scalable quantum computer requires the development of practical schemes that can be adjusted to a given number of qubits at will. Physical hardware should be designed in a modular fashion so that existing small-scale architectures can be upgraded to larger systems in the future. The QEC primitives that can be run on such devices should scale in conjunction with hardware improvements. 

Today's hardware typically suffers from deterioration of fidelities of physical operations when adding more qubits to the system \cite{katabarwa2023early}. It is already a noticeable experimental achievement to keep physical error rates constant while adding more qubits to the setup \cite{decross2024computational}.
Once physical error rates $p$ fall below the fault tolerance threshold, it is useful to employ codes with larger distances $d$ in order to achieve stronger suppression of noise in their logical failure rates $p_L = \mathcal{O}(p^{\lfloor (d + 1) / 2\rfloor})$ as $p \rightarrow 0$.

In this section we discuss the scalability of transversal code switching and aspects of embedding it into a 2D hardware layout.

\subsection{Larger-distance codes}

The number of physical CNOT gates per logical CNOT gate is $n = \mathcal{O}(d^2)$ for the transversal CNOT gate and preparation of an $n$-qubit logical code state is known to be possible with at most $\mathcal{O}(n^2/ \log n)$ physical CNOT gates\footnote{The logical auxiliary qubit state for the used 3D code requires \mbox{$n' = \mathcal{O}(d^3) = \mathcal{O}(n^{3/2})$} physical qubits. A potential gate overhead for fault tolerance must be added to the above given CNOT count.} \cite{aaronson2004improved}. Therefore, one can expect the fraction of physical CNOT gates taken up by logical state preparation in transversal code switching to approach 1 when scaling up our scheme if no specialized state preparation subroutine is used. 

A stricter definition of fault tolerance is required in order to patch subroutines of FT circuits together into a larger FT algorithm, as is the case for our transversal code switching scheme: Here we combine the FT preparation of logical auxiliary qubits with transversal CNOT gates. The latter are FT in the strictest sense \cite{gottesman2009introductionquantumerrorcorrection, chamberland2018flag, zhou2024algorithmicfaulttolerancefast}: Any number $s \leq t$ of faults that happen within the transversal CNOT gate leads to at most $s$ errors on any output code state. This need not be true for a flagged encoding circuit. Here, one typically only requires that for all integer number of faults $0 < s \leq t$ within the circuit, the correct logical state is prepared with at most $t$ errors that are correctable by the QEC code. It is obvious that in this scenario the combined execution of state preparation and transversal CNOT may break fault tolerance. 

Consider, as an example, a $d = 5$ QEC code that corrects $t = 2$ errors. Say that there exists a single fault in the encoding circuit that causes two errors. For the individual encoding circuit, this is sufficient to call the circuit FT since these two errors will be correctable by the distance-5 code. Now assume that a second fault may happen during the subsequent transversal CNOT such that the total number of errors after both circuits is 3 and can thus lead to failure of the code. In this stricter sense of ``sequential fault tolerance'', the single fault in the encoding circuit should at most propagate to a single error, despite $t = 2$. This could be achieved by using flag circuits \cite{chamberland2018flag} or Steane-type state preparation, as mentioned in Sec.~\ref{sec:conn}.

Transversal code switching with distance $d=5$ can, for instance, be performed with the 2D self-dual $[[17,1,5]]$ color code \cite{bombin2006distillation} and the triorthogonal $[[49,1,5]]$ code \cite{bravyi2012magic}. The latter, despite being already established in the literature, can be obtained via code doubling from the $[[15,1,3]]$ code. We recommend to consult Ref.~\cite{sullivan2024code} for more detail and another example for distance $d=7$. Additionally, we also consider the triangular $[[19,1,5]]$ color code \cite{kubica2015universal, bombin2015gauge} in two spatial dimensions, for which the tetrahedral $[[65,1,5]]$ three-dimensional color code \cite{bombin2018transversalgateserrorpropagation} can be used to perform the transversal $T$-gate. 

\begin{figure}\centering
	\includegraphics[width=0.99\linewidth]{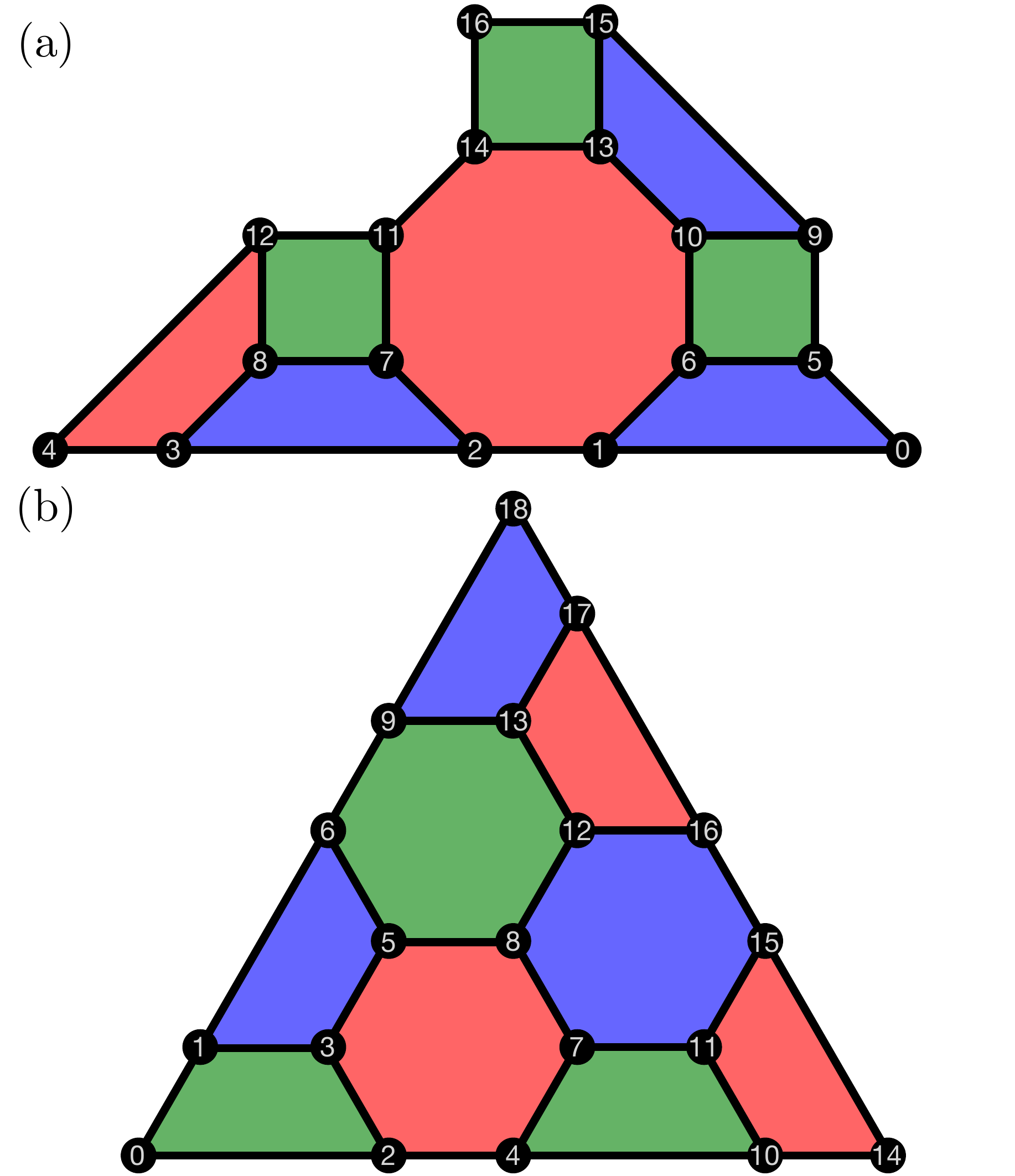}
	\caption{Two common distance-5 2D color codes with qubit labels according to the parity check matrices stated in App.~\ref{sec:circs}. (a) A 4.8.8 color code instance with code parameters $[[17,1,5]]$. A unitary encoding circuit is given in Fig.~\ref{fig:init17}. Its complementing doubled code, which has a transversal $T$-gate, is the $[[49,1,5]]$ code. (b) The $[[19,1,5]]$ color code, also referred to as the 6.6.6 color code, corresponds to a hexagonal tiling of the 2D plane. A unitary encoding circuit is given in Fig.~\ref{fig:init19}. In 3D, a $[[65, 1, 5]]$ triorthogonal code is used to perform the logical $T$-gate.}
	\label{fig:d5codes}
\end{figure}

We numerically verify that transversal code switching in both directions indeed works for $d=5$ with the aforementioned codes and the one-way transversal CNOT gates consisting of 17 (19) physical CNOT gates respectively. To prepare both code switching auxiliary qubits fault-tolerantly, we employ non-FT unitary state preparation followed by Steane-type verification (see Fig.~\ref{fig:steane_verification}). In total, the protocols operate with 164 (217) physical qubits and 2317 (3797) CNOT gates\footnote{Our non-FT encoding circuits of the $\ket{0}_\mathrm{S}$ state of the $[[17,1,5]]$ ($[[19,1,5]]$) and the $\ket{+}_\mathrm{T}$ state of the $[[49,1,5]]$ ($[[65,1,5]]$) codes have 36 (45) and 159 (298) CNOT gates respectively. Physical qubits from auxiliary state verification are reset and reused where possible.} in the typical case where logical auxiliary qubits are verified at the first trial; only 34 (38) CNOTs are used for the actual code switching steps. 

We show the scaling behavior $p_L = \mathcal{O}(p^3)$ for the $[[17,1,5]]$ code in Fig.~\ref{fig:scaling_singleP}. The code switching protocol is only applied to the logical input state, which we expect to be least protected by the triorthogonal code, i.e.~$\ket{+\ii}_\mathrm{S}$ (see App.~\ref{sec:circs}). It is numerically challenging to collect enough Monte Carlo samples in the low-$p$ regime \cite{heussen2024dynamical}. We stress that, at a realistically attainable entangling gate error rate of $p_2 = 10^{-3}$, our stabilizer simulations suggest lower logical failure rates for the distance-5 codes, $p_L = (2.0 \pm 0.6) \times 10^{-4}$ and $p_L = (5.8 \pm 1.1) \times 10^{-4}$ respectively, as compared to the distance-3 logical $T$-gate, which achieves $p_L = (9.3 \pm 0.5) \times 10^{-4}$ for the least protected logical input state\footnote{For the three input logical Pauli states on the $[[17,1,5]]$ code, we find failure rates $(7.0 \pm 2.6) \times 10^{-6}, (2.0 \pm 0.6) \times 10^{-4}$ and $(1.5 \pm 0.4) \times 10^{-4}$ respectively. We also point out that our scheme seems competitive with Ref.~\cite{chamberland2020very} for $d=5$, where only non-deterministic FT magic state preparation was considered but not the full $T$-gate implementation.}. It is expected that the scheme performs slightly worse for the $[[19,1,5]]$ code than for the $[[17,1,5]]$ code due to the larger CNOT gate and qubit overhead. For larger physical error rates $p$, the acceptance rate of logical auxiliary states vanishes quickly for Steane-type state preparation (see Tab.~\ref{tab:acc_reps}). So, it is imperative to devise FT state preparation circuits for larger-distance codes and methods to synthesize such circuits that are practical for the relatively large numbers of physical qubits required by the appropriate 3D codes \cite{peham2024automatedsynthesisfaulttolerantstate}.

\subsection{Two-dimensional hardware layout}

As a concrete scenario for operating our proposed scheme on a future scalable architecture, we consider a trapped-ion quantum processor based on one-dimensional segmented ion traps, which are arranged in a two-dimensional square lattice, similar to the architectures described in  \cite{hensinger2006tjunction, blakestadt2009highfidelity, Wright_2013, shu2014heating, bautista2019multilayer, Kaushal2020shuttlingbasedtiqc, malinowski2023howtowire, cai2023looped} (an alternative could be, e.g., the Quantum Spring Array \cite{valentini2024demonstration}). Ion shuttling operations enable dynamic reconfiguration of the qubits, 
establishing effective all-to-all connectivity. 
On such a layout, we envisage at least two different use cases of the transversal code switching scheme with an arbitrary number of distance-3 logical qubits.

\begin{figure}\centering
	\includegraphics[width=0.99\linewidth]{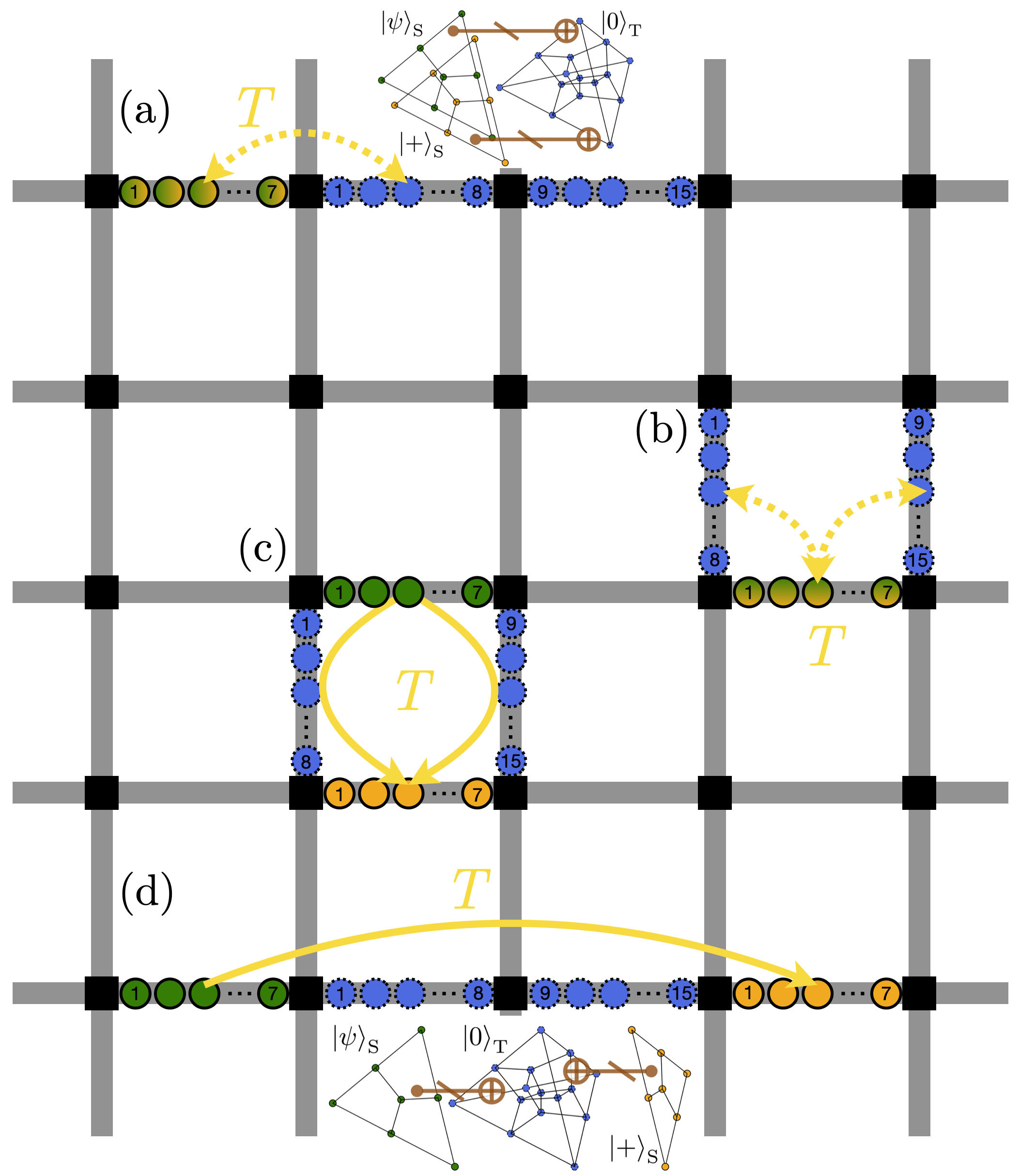}
	\caption{Possible arrangements of 1D ion crystals (colored circles) in a 2D lattice structure where trap segments (grey, cf.~Fig.~\ref{fig:yellow_trap}) are connected via junctions (black). (a) A linear arrangement of three zones can be used to teleport (arrows) a Steane code state (solid-line circles) to the 15-qubit state (blue dashed-line circles) and back while executing the logical $T$-gate. We split the 15-qubit state into two ion crystal due to limited control capabilities when crystals are too large (see App.~\ref{sec:ion3d}). These ion chains can also be shuttled close to the Steane code state from two sides as depicted by the configuration in (b). A circular four-zone arrangement (c) can be used to move the initial Steane code state (green, $\ket{\psi}_{\mathrm{S}}$) down by one row (orange, $T\!\ket{\psi}_{\mathrm{S}}$) while applying the logical $T$-gate via switching to the 15-qubit state. (d) Arranging the four zones linearly allows one to quickly move the state $\ket{\psi}_{\mathrm{S}}$ three columns to the right while transforming it to $T\!\ket{\psi}_{\mathrm{S}}$.}
	\label{fig:embed}
\end{figure}

One option would be to keep the ions of the logical qubit $\ket{\psi}_{\mathrm{S}}$ static and move the ions that constitute the Tetrahedral code auxiliary qubit $\ket{0}_{\mathrm{T}}$ into close proximity when a logical $T$-gate should be executed. The transversal CNOT gate is enabled by local ion crystal reconfigurations that establish the connectivity between the Steane code ions and a suitable subset of the Tetrahedral code ions: One may move some ions of the 15-qubit state into the Steane code zone or move subsets of ions from both crystals into adjacent empty zones in order to apply the appropriate physical entangling gates. After the first code switching step, the Steane-qubit is (re-)initialized to $\ket{+}_{\mathrm{S}}$ on the same physical qubits that held the Steane code state $\ket{\psi}_{\mathrm{S}}$ previously. This is illustrated in Fig.~\ref{fig:embed}a+b. The Tetrahedral code ions may be moved away again after having performed the second code switching step. 

Apart from implementing such a stationary logical $T$-gate fault-tolerantly and enabling universal FT quantum computation via color codes, the teleportation subroutines could be utilized to move the logical Steane code state $\ket{\psi}_{\mathrm{S}}$ through the lattice quickly while applying the FT $T$-gate, as shown in Fig.~\ref{fig:embed}c+d. Different subsets of physical qubits of the 15-qubit code, corresponding to different sides of the Tetrahedral code, can be used to perform the transversal CNOT gates in the two code switching steps. 

The above described routines could be used in parallel in distinct regions of the lattice, allowing one to scale the system up to an arbitrary number of logical qubits that can each be controlled via an FT universal gate set. Since our Tetrahedral code has $d_Z = 7$, the logical auxiliary qubit offers a better protection against dephasing noise. Up to 3 $Z$-errors could, in principle, be removed from the logical auxiliary qubit \emph{before} it is coupled to the Steane code state. 

\section{Conclusions and outlook}\label{sec:outlook}

In this work, we have investigated a new FT code switching scheme \cite{sullivan2024code} based on transversal CNOT gates, which are applied in one orientation, between the two quantum error correcting codes. While it has been known before that teleportation-based error correction can be combined with logical gate operations \cite{knill2004scalable, knill2005quantum, gottesman2014faulttolerant}, the one-way transversal CNOT gate between the Steane code and the Tetrahedral code has been missing to construct a protocol that is practically useful \cite{sullivan2024code}. 

We focus on using the scheme as an enabler to perform the logical $T$-gate fault-tolerantly on a low-distance 2D color code state with smaller gate overhead than previously known deterministic schemes such as flag-FT code switching and FT magic state preparation and injection. The FT code switching protocol has been shown suitable for parallelized implementation in hardware architectures with limited qubit connectivity. Microscopically motivated noise simulations suggest that our logical $T$-gate could be applied successfully in near-term quantum processors with improved entangling gate error rates and fast measurements on the order of the entangling gate time. Only a moderate increase of qubits is required compared to other state-of-the-art universality strategies when using QEC codes of distance $d=3$. Since the actual code switching procedure entirely consists of transversal operations, it naturally allows for scaling up to larger code distances. We identify the repetition overhead of ``repeat until success'' logical auxiliary state preparation as the relevant bottleneck for the practical implementation of distance-5 protocols on near-term hardware.

We have not explored the effect of coherent errors in the transversal code switching circuit since we expect the mid-circuit measurements to destroy such unwanted coherences so that no build-up of noise takes place.

In all shown code instances, we made use of standard look-up-table decoders, which are known to grow exponentially in size when increasing the code distance. Exploring the use of efficient decoders and the potential of correlated decoding in the context of logical teleportation could be a promising direction for further improving scalability by, for instance, simplifying the logical auxiliary state verification \cite{sullivan2024code, cain2024correlateddecodinglogicalalgorithms, wan2024iterativetransversalcnotdecoder, sahay2024errorcorrectiontransversalcontrollednot}. Additionally, further attention is needed to optimally leverage the potential of using the logical measurements of the teleportation subroutines for intermediate error correction. Possibilities of re-using the logical auxiliary qubits or at least restore their state after the logical teleportation step with potentially fewer physical operations could be another line of future research.

A measurement-free code switching scheme would facilitate the practical implementation of the FT $T$-gate in near-term quantum computers by avoiding potentially long measurement times, measurement-induced errors or the need for feed-forward logical operations. We spare some additional thoughts in App.~\ref{sec:mf}.

With the help of systematic circuit synthesis tools, such as a recently presented approach based on reinforcement learning \cite{zen2024quantum}, a more economical and/or deterministic FT state preparation circuit, for instance using single-shot flags \cite{veroni2024optimized}, could be discovered. Efficient state preparation circuits would be needed to further decrease logical failure rates and increase acceptance rates of logical auxiliary qubit states in practical implementations beyond distance-3 \cite{peham2024automatedsynthesisfaulttolerantstate}. 

Our perspective on concatenated codes might be overly pessimistic as there could exist decoding strategies that allow for the correction of higher-weight errors even if only small-distance QEC codes, such as $C_4, C_6$ and/or the Steane code as suggested in Ref.~\cite{yoshida2024concatenate}, are concatenated to obtain an FT universal gate set.

An interesting avenue of future work would be to shed more light on a systematic categorization of appropriate codes for transversal code switching, potentially without the need of explicitly constructing a new doubled code \cite{jain2024highdistancecodestransversalclifford}. This should include codes with more than $k=1$ logical qubit. We speculate that this could be possible for surface-code-like QEC codes, which may even include high-rate qLDPC codes \cite{zhu2023noncliffordparallelizablefaulttolerantlogical}, such as hypergraph product codes or lift-connected surface codes \cite{Old_2024}. The homomorphic CNOT gate, a generalization of the transversal CNOT gate recently employed to perform generalized lattice surgery with encoded auxiliary qubits, could offer a path forward \cite{huang2022homomorphiclogicalmeasurements, cross2024linearsizeancillasystemslogical, xu2024fastparallelizablelogicalcomputation}. 

Despite not being LDPC, an emphasis is put on the distance-7 Golay code in Ref.~\cite{sullivan2024code}, which could be a good candidate to take the first steps towards larger-distance FT quantum computation once enough physical qubits with (almost) all-to-all connectivity are available in practice \cite{paetznick2013faulttolerant}. Exploration of systematic scale-up of all transversal operations in the 2D color code, including the transversal code-switching-based $T$-gate, in an appropriate Clifford+$T$ quantum algorithm that solves a practical problem, should be a future goal.

\section*{Code availability}
All software code used in this project is available from the corresponding author upon reasonable request.

\section*{Author contributions}
SH conceptualized the project, analyzed the scheme, designed the circuits and performed all numerical simulations. SH wrote the manuscript with input and feedback from JH.

\section*{Acknowledgements}
We thank Thomas Scruby, Arthur Pesah, Tom Peham and Janis Wagner for fruitful discussions. We appreciate useful feedback on the manuscript by Ulrich Poschinger and Markus Müller. The python code to create Fig.~\ref{fig:codes} has initially been drafted with the help of ChatGPT and was then manually adjusted and extended.

JH gratefully acknowledges support by the European Union’s Horizon Europe research and innovation program under Grant Agreement Number 101114305 (“MILLENION-SGA1” EU Project) and the US Army Research Office through Grant Number W911NF-21-1-0007. The views and conclusions contained in this document are those of the authors and should not be interpreted as representing the official policies, either expressed or implied, of IARPA, the Army Research Office, or the U.S. Government. The U.S. Government is authorized to reproduce and distribute reprints for Government purposes notwithstanding any copyright notation herein.

\bibliographystyle{bibstyle}

\begin{thebibliography}{136}\makeatletter
	\providecommand \@ifxundefined [1]{\@ifx{#1\undefined}
	}\providecommand \@ifnum [1]{\ifnum #1\expandafter \@firstoftwo
		\else \expandafter \@secondoftwo
		\fi
	}\providecommand \@ifx [1]{\ifx #1\expandafter \@firstoftwo
		\else \expandafter \@secondoftwo
		\fi
	}\providecommand \natexlab [1]{#1}\providecommand \enquote  [1]{``#1''}\providecommand \bibnamefont  [1]{#1}\providecommand \bibfnamefont [1]{#1}\providecommand \citenamefont [1]{#1}\providecommand \href@noop [0]{\@secondoftwo}\providecommand \href [0]{\begingroup \@sanitize@url \@href}\providecommand \@href[1]{\@@startlink{#1}\@@href}\providecommand \@@href[1]{\endgroup#1\@@endlink}\providecommand \@sanitize@url [0]{\catcode `\\12\catcode `\$12\catcode `\&12\catcode `\#12\catcode `\^12\catcode `\_12\catcode `\%12\relax}\providecommand \@@startlink[1]{}\providecommand \@@endlink[0]{}\providecommand \url  [0]{\begingroup\@sanitize@url \@url }\providecommand \@url [1]{\endgroup\@href {#1}{\urlprefix }}\providecommand \urlprefix  [0]{URL }\providecommand \Eprint [0]{\href }\providecommand \doibase [0]{https://doi.org/}\providecommand \selectlanguage [0]{\@gobble}\providecommand \bibinfo  [0]{\@secondoftwo}\providecommand \bibfield  [0]{\@secondoftwo}\providecommand \translation [1]{[#1]}\providecommand \BibitemOpen [0]{}\providecommand \bibitemStop [0]{}\providecommand \bibitemNoStop [0]{.\EOS\space}\providecommand \EOS [0]{\spacefactor3000\relax}\providecommand \BibitemShut  [1]{\csname bibitem#1\endcsname}\let\auto@bib@innerbib\@empty
	\bibitem [{\citenamefont {Campbell}\ \emph {et~al.}(2017)\citenamefont {Campbell}, \citenamefont {Terhal},\ and\ \citenamefont {Vuillot}}]{campbell2017roads}\BibitemOpen
	\bibfield  {author} {\bibinfo {author} {E.~T. Campbell}, \bibinfo {author} {B.~M. Terhal},\ and\ \bibinfo {author} {C.~Vuillot},\ }\emph {{Roads towards fault-tolerant universal quantum computation}},\ \href {https://doi.org/10.1038/nature23460} {\bibfield  {journal} {\bibinfo  {journal} {Nature}\ }\textbf {\bibinfo {volume} {549}},\ \bibinfo {pages} {172} (\bibinfo {year} {2017})}\BibitemShut {NoStop}\bibitem [{\citenamefont {Terhal}(2015)}]{terhal2015quantum}\BibitemOpen
	\bibfield  {author} {\bibinfo {author} {B.~M. Terhal},\ }\emph {Quantum error correction for quantum memories},\ \href {https://doi.org/10.1103/RevModPhys.87.307} {\bibfield  {journal} {\bibinfo  {journal} {Reviews of Modern Physics}\ }\textbf {\bibinfo {volume} {87}},\ \bibinfo {pages} {307} (\bibinfo {year} {2015})}\BibitemShut {NoStop}\bibitem [{\citenamefont {Gottesman}(2014)}]{gottesman2014faulttolerant}\BibitemOpen
	\bibfield  {author} {\bibinfo {author} {D.~Gottesman},\ }\href@noop {} {\emph {Fault-tolerant quantum computation with constant overhead}} (\bibinfo {year} {2014}),\ \Eprint {https://arxiv.org/abs/1310.2984} {arXiv:1310.2984 [quant-ph]} \BibitemShut {NoStop}\bibitem [{\citenamefont {Ryan-Anderson}\ \emph {et~al.}(2021)\citenamefont {Ryan-Anderson}, \citenamefont {Bohnet}, \citenamefont {Lee}, \citenamefont {Gresh}, \citenamefont {Hankin}, \citenamefont {Gaebler}, \citenamefont {Francois}, \citenamefont {Chernoguzov}, \citenamefont {Lucchetti}, \citenamefont {Brown} \emph {et~al.}}]{ryan2021realization}\BibitemOpen
	\bibfield  {author} {\bibinfo {author} {C.~Ryan-Anderson}, et~al.,\ }\emph {Realization of real-time fault-tolerant quantum error correction},\ \href {https://doi.org/10.1103/PhysRevX.11.041058} {\bibfield  {journal} {\bibinfo  {journal} {Physical Review X}\ }\textbf {\bibinfo {volume} {11}},\ \bibinfo {pages} {041058} (\bibinfo {year} {2021})}\BibitemShut {NoStop}\bibitem [{\citenamefont {Ryan-Anderson}\ \emph {et~al.}(2022)\citenamefont {Ryan-Anderson}, \citenamefont {Brown}, \citenamefont {Allman}, \citenamefont {Arkin}, \citenamefont {Asa-Attuah}, \citenamefont {Baldwin}, \citenamefont {Berg}, \citenamefont {Bohnet}, \citenamefont {Braxton}, \citenamefont {Burdick}, \citenamefont {Campora}, \citenamefont {Chernoguzov}, \citenamefont {Esposito}, \citenamefont {Evans}, \citenamefont {Francois}, \citenamefont {Gaebler}, \citenamefont {Gatterman}, \citenamefont {Gerber}, \citenamefont {Gilmore}, \citenamefont {Gresh}, \citenamefont {Hall}, \citenamefont {Hankin}, \citenamefont {Hostetter}, \citenamefont {Lucchetti}, \citenamefont {Mayer}, \citenamefont {Myers}, \citenamefont {Neyenhuis}, \citenamefont {Santiago}, \citenamefont {Sedlacek}, \citenamefont {Skripka}, \citenamefont {Slattery}, \citenamefont {Stutz}, \citenamefont {Tait}, \citenamefont {Tobey}, \citenamefont {Vittorini}, \citenamefont {Walker},\ and\ \citenamefont {Hayes}}]{ryan2022implementing}\BibitemOpen
	\bibfield  {author} {\bibinfo {author} {C.~Ryan-Anderson}, et~al.,\ }\href@noop {} {\emph {Implementing fault-tolerant entangling gates on the five-qubit code and the color code}} (\bibinfo {year} {2022}),\ \Eprint {https://arxiv.org/abs/2208.01863} {arXiv:2208.01863 [quant-ph]} \BibitemShut {NoStop}\bibitem [{\citenamefont {Postler}\ \emph {et~al.}(2022)\citenamefont {Postler}, \citenamefont {Heu{\ss}en}, \citenamefont {Pogorelov}, \citenamefont {Rispler}, \citenamefont {Feldker}, \citenamefont {Meth}, \citenamefont {Marciniak}, \citenamefont {Stricker}, \citenamefont {Ringbauer}, \citenamefont {Blatt}, \citenamefont {Schindler}, \citenamefont {Müller},\ and\ \citenamefont {Monz}}]{postler2022demonstration}\BibitemOpen
	\bibfield  {author} {\bibinfo {author} {L.~Postler}, et~al.,\ }\emph {Demonstration of fault-tolerant universal quantum gate operations},\ \href {https://doi.org/10.1038/s41586-022-04721-1} {\bibfield  {journal} {\bibinfo  {journal} {Nature}\ }\textbf {\bibinfo {volume} {605}},\ \bibinfo {pages} {675} (\bibinfo {year} {2022})}\BibitemShut {NoStop}\bibitem [{\citenamefont {{{Google Quantum AI}}}(2023)}]{google2023suppressing}\BibitemOpen
	\bibfield  {author} {\bibinfo {author} {{{Google Quantum AI}}},\ }\emph {Suppressing quantum errors by scaling a surface code logical qubit},\ \href {https://doi.org/10.1038/s41586-022-05434-1} {\bibfield  {journal} {\bibinfo  {journal} {Nature}\ }\textbf {\bibinfo {volume} {614}},\ \bibinfo {pages} {676} (\bibinfo {year} {2023})}\BibitemShut {NoStop}\bibitem [{\citenamefont {da~Silva}\ \emph {et~al.}(2024)\citenamefont {da~Silva}, \citenamefont {Ryan-Anderson}, \citenamefont {Bello-Rivas}, \citenamefont {Chernoguzov}, \citenamefont {Dreiling}, \citenamefont {Foltz}, \citenamefont {Frachon}, \citenamefont {Gaebler}, \citenamefont {Gatterman}, \citenamefont {Grans-Samuelsson}, \citenamefont {Hayes}, \citenamefont {Hewitt}, \citenamefont {Johansen}, \citenamefont {Lucchetti}, \citenamefont {Mills}, \citenamefont {Moses}, \citenamefont {Neyenhuis}, \citenamefont {Paz}, \citenamefont {Pino}, \citenamefont {Siegfried}, \citenamefont {Strabley}, \citenamefont {Sundaram}, \citenamefont {Tom}, \citenamefont {Wernli}, \citenamefont {Zanner}, \citenamefont {Stutz},\ and\ \citenamefont {Svore}}]{dasilva2024demonstrationlogicalqubitsrepeated}\BibitemOpen
	\bibfield  {author} {\bibinfo {author} {M.~P. da~Silva}, et~al.,\ }\href@noop {} {\emph {Demonstration of logical qubits and repeated error correction with better-than-physical error rates}} (\bibinfo {year} {2024}),\ \Eprint {https://arxiv.org/abs/2404.02280} {arXiv:2404.02280 [quant-ph]} \BibitemShut {NoStop}\bibitem [{\citenamefont {Mayer}\ \emph {et~al.}(2024)\citenamefont {Mayer}, \citenamefont {Ryan-Anderson}, \citenamefont {Brown}, \citenamefont {Durso-Sabina}, \citenamefont {Baldwin}, \citenamefont {Hayes}, \citenamefont {Dreiling}, \citenamefont {Foltz}, \citenamefont {Gaebler}, \citenamefont {Gatterman}, \citenamefont {Gerber}, \citenamefont {Gilmore}, \citenamefont {Gresh}, \citenamefont {Hewitt}, \citenamefont {Horst}, \citenamefont {Johansen}, \citenamefont {Mengle}, \citenamefont {Mills}, \citenamefont {Moses}, \citenamefont {Siegfried}, \citenamefont {Neyenhuis}, \citenamefont {Pino},\ and\ \citenamefont {Stutz}}]{mayer2024benchmarkinglogicalthreequbitquantum}\BibitemOpen
	\bibfield  {author} {\bibinfo {author} {K.~Mayer}, et~al.,\ }\href@noop {} {\emph {Benchmarking logical three-qubit quantum Fourier transform encoded in the Steane code on a trapped-ion quantum computer}} (\bibinfo {year} {2024}),\ \Eprint {https://arxiv.org/abs/2404.08616} {arXiv:2404.08616 [quant-ph]} \BibitemShut {NoStop}\bibitem [{\citenamefont {Ryan-Anderson}\ \emph {et~al.}(2024)\citenamefont {Ryan-Anderson}, \citenamefont {Brown}, \citenamefont {Baldwin}, \citenamefont {Dreiling}, \citenamefont {Foltz}, \citenamefont {Gaebler}, \citenamefont {Gatterman}, \citenamefont {Hewitt}, \citenamefont {Holliman}, \citenamefont {Horst}, \citenamefont {Johansen}, \citenamefont {Lucchetti}, \citenamefont {Mengle}, \citenamefont {Matheny}, \citenamefont {Matsuoka}, \citenamefont {Mayer}, \citenamefont {Mills}, \citenamefont {Moses}, \citenamefont {Neyenhuis}, \citenamefont {Pino}, \citenamefont {Siegfried}, \citenamefont {Stutz}, \citenamefont {Walker},\ and\ \citenamefont {Hayes}}]{ryananderson2024highfidelity}\BibitemOpen
	\bibfield  {author} {\bibinfo {author} {C.~Ryan-Anderson}, et~al.,\ }\href@noop {} {\emph {High-fidelity and fault-tolerant teleportation of a logical qubit using transversal gates and lattice surgery on a trapped-ion quantum computer}} (\bibinfo {year} {2024}),\ \Eprint {https://arxiv.org/abs/2404.16728} {arXiv:2404.16728 [quant-ph]} \BibitemShut {NoStop}\bibitem [{ach(2025)}]{acharya2024quantumerrorcorrectionsurface}\BibitemOpen
	\emph {Quantum error correction below the surface code threshold},\ \href {https://doi.org/10.1038/s41586-024-08449-y} {\bibfield  {journal} {\bibinfo  {journal} {Nature}\ }\textbf {\bibinfo {volume} {638}},\ \bibinfo {pages} {920} (\bibinfo {year} {2025})}\BibitemShut {NoStop}\bibitem [{\citenamefont {Goto}(2016)}]{goto2016minimizing}\BibitemOpen
	\bibfield  {author} {\bibinfo {author} {H.~Goto},\ }\emph {Minimizing resource overheads for fault-tolerant preparation of encoded states of the Steane code},\ \href {https://doi.org/10.1038/srep19578} {\bibfield  {journal} {\bibinfo  {journal} {Scientific Reports}\ }\textbf {\bibinfo {volume} {6}},\ \bibinfo {pages} {1} (\bibinfo {year} {2016})}\BibitemShut {NoStop}\bibitem [{\citenamefont {Yoder}\ and\ \citenamefont {Kim}(2017)}]{yoder2017surface}\BibitemOpen
	\bibfield  {author} {\bibinfo {author} {T.~J. Yoder}\ and\ \bibinfo {author} {I.~H. Kim},\ }\emph {The surface code with a twist},\ \href {https://doi.org/10.22331/q-2017-04-25-2} {\bibfield  {journal} {\bibinfo  {journal} {Quantum}\ }\textbf {\bibinfo {volume} {1}},\ \bibinfo {pages} {2} (\bibinfo {year} {2017})}\BibitemShut {NoStop}\bibitem [{\citenamefont {Chao}\ and\ \citenamefont {Reichardt}(2018)}]{chao2018fault}\BibitemOpen
	\bibfield  {author} {\bibinfo {author} {R.~Chao}\ and\ \bibinfo {author} {B.~W. Reichardt},\ }\emph {Fault-tolerant quantum computation with few qubits},\ \href {https://doi.org/10.1038/s41534-018-0085-z} {\bibfield  {journal} {\bibinfo  {journal} {npj Quantum Information}\ }\textbf {\bibinfo {volume} {4}},\ \bibinfo {pages} {42} (\bibinfo {year} {2018})}\BibitemShut {NoStop}\bibitem [{\citenamefont {Chamberland}\ and\ \citenamefont {Beverland}(2018)}]{chamberland2018flag}\BibitemOpen
	\bibfield  {author} {\bibinfo {author} {C.~Chamberland}\ and\ \bibinfo {author} {M.~E. Beverland},\ }\emph {Flag fault-tolerant error correction with arbitrary distance codes},\ \href {https://doi.org/https://doi.org/10.22331/q-2018-02-08-53} {\bibfield  {journal} {\bibinfo  {journal} {Quantum}\ }\textbf {\bibinfo {volume} {2}},\ \bibinfo {pages} {53} (\bibinfo {year} {2018})}\BibitemShut {NoStop}\bibitem [{\citenamefont {Reichardt}(2020)}]{reichardt2020fault}\BibitemOpen
	\bibfield  {author} {\bibinfo {author} {B.~W. Reichardt},\ }\emph {{Fault-tolerant quantum error correction for Steane{'}s seven-qubit color code with few or no extra qubits}},\ \href {https://doi.org/10.1088/2058-9565/abc6f4} {\bibfield  {journal} {\bibinfo  {journal} {Quantum Science and Technology}\ }\textbf {\bibinfo {volume} {6}},\ \bibinfo {pages} {015007} (\bibinfo {year} {2020})}\BibitemShut {NoStop}\bibitem [{\citenamefont {Prabhu}\ and\ \citenamefont {Reichardt}(2021)}]{prabhu2021fault}\BibitemOpen
	\bibfield  {author} {\bibinfo {author} {P.~Prabhu}\ and\ \bibinfo {author} {B.~W. Reichardt},\ }\emph {{Fault-tolerant syndrome extraction and cat state preparation with fewer qubits}},\ \href {https://doi.org/10.4230/LIPIcs.TQC.2021.5} {\bibfield  {journal} {\bibinfo  {journal} {16th Conference on the Theory of Quantum Computation, Communication and Cryptography\!\!}\ ,\ \bibinfo {pages} {5:1}} (\bibinfo {year} {2021})}\BibitemShut {NoStop}\bibitem [{\citenamefont {Bhatnagar}\ \emph {et~al.}(2023)\citenamefont {Bhatnagar}, \citenamefont {Steinberg}, \citenamefont {Elkouss}, \citenamefont {Almudever},\ and\ \citenamefont {Feld}}]{Bhatnagar_2023}\BibitemOpen
	\bibfield  {author} {\bibinfo {author} {D.~Bhatnagar}, \bibinfo {author} {M.~Steinberg}, \bibinfo {author} {D.~Elkouss}, \bibinfo {author} {C.~G. Almudever},\ and\ \bibinfo {author} {S.~Feld},\ }\emph {Low-depth flag-style syndrome extraction for small quantum error-correction cdes},\ \href {https://doi.org/10.1109/qce57702.2023.00016} {\bibfield  {journal} {\bibinfo  {journal} {2023 IEEE International Conference on Quantum Computing and Engineering\!}\ ,\ \bibinfo {pages} {63}} (\bibinfo {year} {2023})}\BibitemShut {NoStop}\bibitem [{\citenamefont {Du}\ \emph {et~al.}(2024)\citenamefont {Du}, \citenamefont {Ma}, \citenamefont {Liu}, \citenamefont {Duan},\ and\ \citenamefont {Fei}}]{du2024parallel}\BibitemOpen
	\bibfield  {author} {\bibinfo {author} {C.~Du}, \bibinfo {author} {Z.~Ma}, \bibinfo {author} {Y.~Liu}, \bibinfo {author} {Q.~Duan},\ and\ \bibinfo {author} {Y.~Fei},\ }\emph {Parallel flagged fault-tolerant error correction for stabilizer codes of distance 5},\ \href {https://doi.org/https://doi.org/10.1002/qute.202300291} {\bibfield  {journal} {\bibinfo  {journal} {Advanced Quantum Technologies\!}\ ,\ \bibinfo {pages} {2300291}} (\bibinfo {year} {2024})}\BibitemShut {NoStop}\bibitem [{\citenamefont {Liou}\ and\ \citenamefont {Lai}(2024)}]{liou2024reducingquantumerrorcorrection}\BibitemOpen
	\bibfield  {author} {\bibinfo {author} {P.-H. Liou}\ and\ \bibinfo {author} {C.-Y. Lai},\ }\href@noop {} {\emph {Reducing quantum error correction overhead with versatile flag-sharing syndrome extraction circuits}} (\bibinfo {year} {2024}),\ \Eprint {https://arxiv.org/abs/2407.00607} {arXiv:2407.00607 [quant-ph]} \BibitemShut {NoStop}\bibitem [{\citenamefont {Bravyi}\ \emph {et~al.}(2022)\citenamefont {Bravyi}, \citenamefont {Dial}, \citenamefont {Gambetta}, \citenamefont {Gil},\ and\ \citenamefont {Nazario}}]{bravyi2022future}\BibitemOpen
	\bibfield  {author} {\bibinfo {author} {S.~Bravyi}, \bibinfo {author} {O.~Dial}, \bibinfo {author} {J.~M. Gambetta}, \bibinfo {author} {D.~Gil},\ and\ \bibinfo {author} {Z.~Nazario},\ }\emph {The future of quantum computing with superconducting qubits},\ \href@noop {} {\bibfield  {journal} {\bibinfo  {journal} {\href{http://dx.doi.org/10.1063/5.0082975}{Journal of Applied Physics}}\ }\textbf {\bibinfo {volume} {132}} (\bibinfo {year} {2022})}\BibitemShut {NoStop}\bibitem [{\citenamefont {Bluvstein}\ \emph {et~al.}(2022)\citenamefont {Bluvstein}, \citenamefont {Levine}, \citenamefont {Semeghini}, \citenamefont {Wang}, \citenamefont {Ebadi}, \citenamefont {Kalinowski}, \citenamefont {Keesling}, \citenamefont {Maskara}, \citenamefont {Pichler}, \citenamefont {Greiner} \emph {et~al.}}]{bluvstein2022quantum}\BibitemOpen
	\bibfield  {author} {\bibinfo {author} {D.~Bluvstein}, et~al.,\ }\emph {A quantum processor based on coherent transport of entangled atom arrays},\ \href {https://doi.org/https://doi.org/10.1038/s41586-022-04592-6} {\bibfield  {journal} {\bibinfo  {journal} {Nature}\ }\textbf {\bibinfo {volume} {604}},\ \bibinfo {pages} {451} (\bibinfo {year} {2022})}\BibitemShut {NoStop}\bibitem [{\citenamefont {Pause}\ \emph {et~al.}(2024)\citenamefont {Pause}, \citenamefont {Sturm}, \citenamefont {Mittenbühler}, \citenamefont {Amann}, \citenamefont {Preuschoff}, \citenamefont {Schäffner}, \citenamefont {Schlosser},\ and\ \citenamefont {Birkl}}]{Pause_2024}\BibitemOpen
	\bibfield  {author} {\bibinfo {author} {L.~Pause}, et~al.,\ }\emph {Supercharged two-dimensional tweezer array with more than 1000 atomic qubits},\ \href {https://doi.org/10.1364/optica.513551} {\bibfield  {journal} {\bibinfo  {journal} {Optica}\ }\textbf {\bibinfo {volume} {11}},\ \bibinfo {pages} {222} (\bibinfo {year} {2024})}\BibitemShut {NoStop}\bibitem [{\citenamefont {Bluvstein}\ \emph {et~al.}(2024)\citenamefont {Bluvstein}, \citenamefont {Evered}, \citenamefont {Geim}, \citenamefont {Li}, \citenamefont {Zhou}, \citenamefont {Manovitz}, \citenamefont {Ebadi}, \citenamefont {Cain}, \citenamefont {Kalinowski}, \citenamefont {Hangleiter} \emph {et~al.}}]{bluvstein2024logical}\BibitemOpen
	\bibfield  {author} {\bibinfo {author} {D.~Bluvstein}, et~al.,\ }\emph {Logical quantum processor based on reconfigurable atom arrays},\ \href {https://doi.org/https://doi.org/10.1038/s41586-023-06927-3} {\bibfield  {journal} {\bibinfo  {journal} {Nature}\ }\textbf {\bibinfo {volume} {626}},\ \bibinfo {pages} {58} (\bibinfo {year} {2024})}\BibitemShut {NoStop}\bibitem [{\citenamefont {Manetsch}\ \emph {et~al.}(2024)\citenamefont {Manetsch}, \citenamefont {Nomura}, \citenamefont {Bataille}, \citenamefont {Leung}, \citenamefont {Lv},\ and\ \citenamefont {Endres}}]{manetsch2024tweezerarray6100highly}\BibitemOpen
	\bibfield  {author} {\bibinfo {author} {H.~J. Manetsch}, et~al.,\ }\href@noop {} {\emph {A tweezer array with 6100 highly coherent atomic qubits}} (\bibinfo {year} {2024}),\ \Eprint {https://arxiv.org/abs/2403.12021} {arXiv:2403.12021 [quant-ph]} \BibitemShut {NoStop}\bibitem [{\citenamefont {Steane}(1997)}]{steane1997active}\BibitemOpen
	\bibfield  {author} {\bibinfo {author} {A.~M. Steane},\ }\emph {Active stabilization, quantum computation, and quantum state synthesis},\ \href {https://doi.org/10.1103/PhysRevLett.78.2252} {\bibfield  {journal} {\bibinfo  {journal} {Physical Review Letters}\ }\textbf {\bibinfo {volume} {78}},\ \bibinfo {pages} {2252} (\bibinfo {year} {1997})}\BibitemShut {NoStop}\bibitem [{\citenamefont {Postler}\ \emph {et~al.}(2024)\citenamefont {Postler}, \citenamefont {Butt}, \citenamefont {Pogorelov}, \citenamefont {Marciniak}, \citenamefont {Heu\ss{}en}, \citenamefont {Blatt}, \citenamefont {Schindler}, \citenamefont {Rispler}, \citenamefont {M\"uller},\ and\ \citenamefont {Monz}}]{postler2024demonstration}\BibitemOpen
	\bibfield  {author} {\bibinfo {author} {L.~Postler}, et~al.,\ }\emph {Demonstration of fault-tolerant Steane quantum error correction},\ \href {https://doi.org/10.1103/PRXQuantum.5.030326} {\bibfield  {journal} {\bibinfo  {journal} {PRX Quantum}\ }\textbf {\bibinfo {volume} {5}},\ \bibinfo {pages} {030326} (\bibinfo {year} {2024})}\BibitemShut {NoStop}\bibitem [{\citenamefont {Knill}(2004)}]{knill2004scalable}\BibitemOpen
	\bibfield  {author} {\bibinfo {author} {E.~Knill},\ }\href@noop {} {\emph {Scalable quantum computation in the presence of large detected-error rates}} (\bibinfo {year} {2004}),\ \Eprint {https://arxiv.org/abs/quant-ph/0312190} {arXiv:quant-ph/0312190 [quant-ph]} \BibitemShut {NoStop}\bibitem [{\citenamefont {Knill}(2005)}]{knill2005quantum}\BibitemOpen
	\bibfield  {author} {\bibinfo {author} {E.~Knill},\ }\emph {Quantum computing with realistically noisy devices},\ \href {https://doi.org/https://doi.org/10.1038/nature03350} {\bibfield  {journal} {\bibinfo  {journal} {Nature}\ }\textbf {\bibinfo {volume} {434}},\ \bibinfo {pages} {39} (\bibinfo {year} {2005})}\BibitemShut {NoStop}\bibitem [{\citenamefont {Horsman}\ \emph {et~al.}(2012)\citenamefont {Horsman}, \citenamefont {Fowler}, \citenamefont {Devitt},\ and\ \citenamefont {Van~Meter}}]{horsman2012surface}\BibitemOpen
	\bibfield  {author} {\bibinfo {author} {C.~Horsman}, \bibinfo {author} {A.~G. Fowler}, \bibinfo {author} {S.~Devitt},\ and\ \bibinfo {author} {R.~Van~Meter},\ }\emph {Surface code quantum computing by lattice surgery},\ \href {https://doi.org/10.1088/1367-2630/14/12/123011} {\bibfield  {journal} {\bibinfo  {journal} {New Journal of Physics}\ }\textbf {\bibinfo {volume} {14}},\ \bibinfo {pages} {123011} (\bibinfo {year} {2012})}\BibitemShut {NoStop}\bibitem [{\citenamefont {Cohen}\ \emph {et~al.}(2022)\citenamefont {Cohen}, \citenamefont {Kim}, \citenamefont {Bartlett},\ and\ \citenamefont {Brown}}]{cohen2022low}\BibitemOpen
	\bibfield  {author} {\bibinfo {author} {L.~Z. Cohen}, \bibinfo {author} {I.~H. Kim}, \bibinfo {author} {S.~D. Bartlett},\ and\ \bibinfo {author} {B.~J. Brown},\ }\emph {Low-overhead fault-tolerant quantum computing using long-range connectivity},\ \href {https://doi.org/https://doi.org/10.1126/sciadv.abn1717} {\bibfield  {journal} {\bibinfo  {journal} {Science Advances}\ }\textbf {\bibinfo {volume} {8}},\ \bibinfo {pages} {eabn1717} (\bibinfo {year} {2022})}\BibitemShut {NoStop}\bibitem [{\citenamefont {Cowtan}\ and\ \citenamefont {Burton}(2024)}]{cowtan2024css}\BibitemOpen
	\bibfield  {author} {\bibinfo {author} {A.~Cowtan}\ and\ \bibinfo {author} {S.~Burton},\ }\emph {CSS code surgery as a universal construction},\ \href {https://doi.org/https://doi.org/10.22331/q-2024-05-14-1344} {\bibfield  {journal} {\bibinfo  {journal} {Quantum}\ }\textbf {\bibinfo {volume} {8}},\ \bibinfo {pages} {1344} (\bibinfo {year} {2024})}\BibitemShut {NoStop}\bibitem [{\citenamefont {Eastin}\ and\ \citenamefont {Knill}(2009)}]{eastin2009restrictions}\BibitemOpen
	\bibfield  {author} {\bibinfo {author} {B.~Eastin}\ and\ \bibinfo {author} {E.~Knill},\ }\emph {Restrictions on transversal encoded quantum gate sets},\ \href {https://doi.org/10.1103/PhysRevLett.102.110502} {\bibfield  {journal} {\bibinfo  {journal} {Physical Review Letters}\ }\textbf {\bibinfo {volume} {102}},\ \bibinfo {pages} {110502} (\bibinfo {year} {2009})}\BibitemShut {NoStop}\bibitem [{\citenamefont {Gottesman}\ and\ \citenamefont {Chuang}(1999)}]{gottesman1999demonstrating}\BibitemOpen
	\bibfield  {author} {\bibinfo {author} {D.~Gottesman}\ and\ \bibinfo {author} {I.~L. Chuang},\ }\emph {Demonstrating the viability of universal quantum computation using teleportation and single-qubit operations},\ \href {https://doi.org/10.1038/46503} {\bibfield  {journal} {\bibinfo  {journal} {Nature}\ }\textbf {\bibinfo {volume} {402}},\ \bibinfo {pages} {390} (\bibinfo {year} {1999})}\BibitemShut {NoStop}\bibitem [{\citenamefont {Fowler}\ \emph {et~al.}(2012)\citenamefont {Fowler}, \citenamefont {Mariantoni}, \citenamefont {Martinis},\ and\ \citenamefont {Cleland}}]{fowler2012surface}\BibitemOpen
	\bibfield  {author} {\bibinfo {author} {A.~G. Fowler}, \bibinfo {author} {M.~Mariantoni}, \bibinfo {author} {J.~M. Martinis},\ and\ \bibinfo {author} {A.~N. Cleland},\ }\emph {Surface codes: Towards practical large-scale quantum computation},\ \href {https://doi.org/10.1103/PhysRevA.86.032324} {\bibfield  {journal} {\bibinfo  {journal} {Physical Review A}\ }\textbf {\bibinfo {volume} {86}},\ \bibinfo {pages} {032324} (\bibinfo {year} {2012})}\BibitemShut {NoStop}\bibitem [{\citenamefont {Anderson}\ \emph {et~al.}(2014)\citenamefont {Anderson}, \citenamefont {Duclos-Cianci},\ and\ \citenamefont {Poulin}}]{anderson2014fault}\BibitemOpen
	\bibfield  {author} {\bibinfo {author} {J.~T. Anderson}, \bibinfo {author} {G.~Duclos-Cianci},\ and\ \bibinfo {author} {D.~Poulin},\ }\emph {Fault-tolerant conversion between the Steane and Reed-Muller quantum codes},\ \href {https://doi.org/10.1103/PhysRevLett.113.080501} {\bibfield  {journal} {\bibinfo  {journal} {Physical Review Letters}\ }\textbf {\bibinfo {volume} {113}},\ \bibinfo {pages} {080501} (\bibinfo {year} {2014})}\BibitemShut {NoStop}\bibitem [{\citenamefont {Bomb{\'\i}n}(2015)}]{bombin2015gauge}\BibitemOpen
	\bibfield  {author} {\bibinfo {author} {H.~Bomb{\'\i}n},\ }\emph {Gauge color codes: Optimal transversal gates and gauge fixing in topological stabilizer codes},\ \href {https://doi.org/10.1088/1367-2630/17/8/083002} {\bibfield  {journal} {\bibinfo  {journal} {New Journal of Physics}\ }\textbf {\bibinfo {volume} {17}},\ \bibinfo {pages} {083002} (\bibinfo {year} {2015})}\BibitemShut {NoStop}\bibitem [{\citenamefont {Kubica}\ and\ \citenamefont {Beverland}(2015)}]{kubica2015universal}\BibitemOpen
	\bibfield  {author} {\bibinfo {author} {A.~Kubica}\ and\ \bibinfo {author} {M.~E. Beverland},\ }\emph {Universal transversal gates with color codes: A simplified approach},\ \href {https://doi.org/10.1103/PhysRevA.91.032330} {\bibfield  {journal} {\bibinfo  {journal} {Physical Review A}\ }\textbf {\bibinfo {volume} {91}},\ \bibinfo {pages} {032330} (\bibinfo {year} {2015})}\BibitemShut {NoStop}\bibitem [{\citenamefont {Bomb\'{\i}n}\ and\ \citenamefont {Mart\'{\i}n-Delgado}(2006)}]{bombin2006distillation}\BibitemOpen
	\bibfield  {author} {\bibinfo {author} {H.~Bomb\'{\i}n}\ and\ \bibinfo {author} {M.~A. Mart\'{\i}n-Delgado},\ }\emph {Topological quantum distillation},\ \href {https://doi.org/10.1103/PhysRevLett.97.180501} {\bibfield  {journal} {\bibinfo  {journal} {Physical Review Letters}\ }\textbf {\bibinfo {volume} {97}},\ \bibinfo {pages} {180501} (\bibinfo {year} {2006})}\BibitemShut {NoStop}\bibitem [{\citenamefont {Bombin}\ and\ \citenamefont {Martin-Delgado}(2007)}]{bombin2007exact}\BibitemOpen
	\bibfield  {author} {\bibinfo {author} {H.~Bombin}\ and\ \bibinfo {author} {M.~A. Martin-Delgado},\ }\emph {Exact topological quantum order in $D=3$ and beyond: Branyons and brane-net condensates},\ \href {https://doi.org/10.1103/PhysRevB.75.075103} {\bibfield  {journal} {\bibinfo  {journal} {Phys. Rev. B}\ }\textbf {\bibinfo {volume} {75}},\ \bibinfo {pages} {075103} (\bibinfo {year} {2007})}\BibitemShut {NoStop}\bibitem [{\citenamefont {Butt}\ \emph {et~al.}(2024)\citenamefont {Butt}, \citenamefont {Heu\ss{}en}, \citenamefont {Rispler},\ and\ \citenamefont {M\"uller}}]{butt2024fault}\BibitemOpen
	\bibfield  {author} {\bibinfo {author} {F.~Butt}, \bibinfo {author} {S.~Heu\ss{}en}, \bibinfo {author} {M.~Rispler},\ and\ \bibinfo {author} {M.~M\"uller},\ }\emph {Fault-tolerant code-switching protocols for near-term quantum processors},\ \href {https://doi.org/10.1103/PRXQuantum.5.020345} {\bibfield  {journal} {\bibinfo  {journal} {PRX Quantum}\ }\textbf {\bibinfo {volume} {5}},\ \bibinfo {pages} {020345} (\bibinfo {year} {2024})}\BibitemShut {NoStop}\bibitem [{\citenamefont {Pogorelov}\ \emph {et~al.}(2025)\citenamefont {Pogorelov}, \citenamefont {Butt}, \citenamefont {Postler}, \citenamefont {Marciniak}, \citenamefont {Schindler}, \citenamefont {M{\"u}ller},\ and\ \citenamefont {Monz}}]{pogorelov2024experimental}\BibitemOpen
	\bibfield  {author} {\bibinfo {author} {I.~Pogorelov}, et~al.,\ }\emph {Experimental fault-tolerant code switching},\ \href {https://doi.org/10.1038/s41567-024-02727-2} {\bibfield  {journal} {\bibinfo  {journal} {Nature Physics}\ }\textbf {\bibinfo {volume} {21}},\ \bibinfo {pages} {298} (\bibinfo {year} {2025})}\BibitemShut {NoStop}\bibitem [{\citenamefont {Betsumiya}\ and\ \citenamefont {Munemasa}(2012)}]{betsumiya2012triply}\BibitemOpen
	\bibfield  {author} {\bibinfo {author} {K.~Betsumiya}\ and\ \bibinfo {author} {A.~Munemasa},\ }\emph {On triply even binary codes},\ \href {https://doi.org/https://doi.org/10.1112/jlms/jdr054} {\bibfield  {journal} {\bibinfo  {journal} {Journal of the London Mathematical Society}\ }\textbf {\bibinfo {volume} {86}},\ \bibinfo {pages} {1} (\bibinfo {year} {2012})}\BibitemShut {NoStop}\bibitem [{\citenamefont {Bravyi}\ and\ \citenamefont {Cross}(2015)}]{bravyi2015doubledcolorcodes}\BibitemOpen
	\bibfield  {author} {\bibinfo {author} {S.~Bravyi}\ and\ \bibinfo {author} {A.~Cross},\ }\href@noop {} {\emph {Doubled color codes}} (\bibinfo {year} {2015}),\ \Eprint {https://arxiv.org/abs/1509.03239} {arXiv:1509.03239 [quant-ph]} \BibitemShut {NoStop}\bibitem [{\citenamefont {Shi}\ \emph {et~al.}(2024)\citenamefont {Shi}, \citenamefont {Lu}, \citenamefont {Kim},\ and\ \citenamefont {Sol{\'e}}}]{shi2024triorthogonal}\BibitemOpen
	\bibfield  {author} {\bibinfo {author} {M.~Shi}, \bibinfo {author} {H.~Lu}, \bibinfo {author} {J.-L. Kim},\ and\ \bibinfo {author} {P.~Sol{\'e}},\ }\emph {Triorthogonal codes and self-dual codes},\ \href {https://doi.org/https://doi.org/10.1007/s11128-024-04485-9} {\bibfield  {journal} {\bibinfo  {journal} {Quantum Information Processing}\ }\textbf {\bibinfo {volume} {23}},\ \bibinfo {pages} {1} (\bibinfo {year} {2024})}\BibitemShut {NoStop}\bibitem [{\citenamefont {Sullivan}(2024)}]{sullivan2024code}\BibitemOpen
	\bibfield  {author} {\bibinfo {author} {M.~Sullivan},\ }\emph {Code conversion with the quantum Golay code for a universal transversal gate set},\ \href {https://doi.org/10.1103/PhysRevA.109.042416} {\bibfield  {journal} {\bibinfo  {journal} {Physical Review A}\ }\textbf {\bibinfo {volume} {109}},\ \bibinfo {pages} {042416} (\bibinfo {year} {2024})}\BibitemShut {NoStop}\bibitem [{\citenamefont {Paetznick}\ and\ \citenamefont {Reichardt}(2013{\natexlab{a}})}]{paetznick2013universal}\BibitemOpen
	\bibfield  {author} {\bibinfo {author} {A.~Paetznick}\ and\ \bibinfo {author} {B.~W. Reichardt},\ }\emph {Universal fault-tolerant quantum computation with only transversal gates and error correction},\ \href {https://doi.org/10.1103/PhysRevLett.111.090505} {\bibfield  {journal} {\bibinfo  {journal} {Physical Review Letters}\ }\textbf {\bibinfo {volume} {111}},\ \bibinfo {pages} {090505} (\bibinfo {year} {2013}{\natexlab{a}})}\BibitemShut {NoStop}\bibitem [{\citenamefont {Steane}(1996)}]{steane1996multiple}\BibitemOpen
	\bibfield  {author} {\bibinfo {author} {A.~Steane},\ }\emph {{Multiple-particle interference and quantum error correction}},\ \href {https://doi.org/10.1098/rspa.1996.0136} {\bibfield  {journal} {\bibinfo  {journal} {Proceedings of the Royal Society A}\ }\textbf {\bibinfo {volume} {452}},\ \bibinfo {pages} {2551} (\bibinfo {year} {1996})}\BibitemShut {NoStop}\bibitem [{\citenamefont {Steane}(1999)}]{steane1999quantum}\BibitemOpen
	\bibfield  {author} {\bibinfo {author} {A.~M. Steane},\ }\emph {Quantum Reed-Muller codes},\ \href {https://doi.org/doi.org/10.1109/18.771249} {\bibfield  {journal} {\bibinfo  {journal} {IEEE Transactions on Information Theory}\ }\textbf {\bibinfo {volume} {45}},\ \bibinfo {pages} {1701} (\bibinfo {year} {1999})}\BibitemShut {NoStop}\bibitem [{\citenamefont {Gottesman}(2009)}]{gottesman2009introductionquantumerrorcorrection}\BibitemOpen
	\bibfield  {author} {\bibinfo {author} {D.~Gottesman},\ }\href@noop {} {\emph {An introduction to quantum error correction and fault-tolerant quantum computation}} (\bibinfo {year} {2009}),\ \Eprint {https://arxiv.org/abs/0904.2557} {arXiv:0904.2557 [quant-ph]} \BibitemShut {NoStop}\bibitem [{\citenamefont {Zhou}\ \emph {et~al.}(2024)\citenamefont {Zhou}, \citenamefont {Zhao}, \citenamefont {Cain}, \citenamefont {Bluvstein}, \citenamefont {Duckering}, \citenamefont {Hu}, \citenamefont {Wang}, \citenamefont {Kubica},\ and\ \citenamefont {Lukin}}]{zhou2024algorithmicfaulttolerancefast}\BibitemOpen
	\bibfield  {author} {\bibinfo {author} {H.~Zhou}, et~al.,\ }\href@noop {} {\emph {Algorithmic fault tolerance for fast quantum computing}} (\bibinfo {year} {2024}),\ \Eprint {https://arxiv.org/abs/2406.17653} {arXiv:2406.17653 [quant-ph]} \BibitemShut {NoStop}\bibitem [{\citenamefont {Heu\ss{}en}\ \emph {et~al.}(2023)\citenamefont {Heu\ss{}en}, \citenamefont {Postler}, \citenamefont {Rispler}, \citenamefont {Pogorelov}, \citenamefont {Marciniak}, \citenamefont {Monz}, \citenamefont {Schindler},\ and\ \citenamefont {M\"uller}}]{heussen2023strategies}\BibitemOpen
	\bibfield  {author} {\bibinfo {author} {S.~Heu\ss{}en}, et~al.,\ }\emph {Strategies for a practical advantage of fault-tolerant circuit design in noisy trapped-ion quantum computers},\ \href {https://doi.org/10.1103/PhysRevA.107.042422} {\bibfield  {journal} {\bibinfo  {journal} {Physical Review A}\ }\textbf {\bibinfo {volume} {107}},\ \bibinfo {pages} {042422} (\bibinfo {year} {2023})}\BibitemShut {NoStop}\bibitem [{\citenamefont {Chamberland}\ and\ \citenamefont {Cross}(2019)}]{chamberland2019fault}\BibitemOpen
	\bibfield  {author} {\bibinfo {author} {C.~Chamberland}\ and\ \bibinfo {author} {A.~W. Cross},\ }\emph {Fault-tolerant magic state preparation with flag qubits},\ \href {https://doi.org/https://doi.org/10.22331/q-2019-05-20-143} {\bibfield  {journal} {\bibinfo  {journal} {Quantum}\ }\textbf {\bibinfo {volume} {3}},\ \bibinfo {pages} {143} (\bibinfo {year} {2019})}\BibitemShut {NoStop}\bibitem [{\citenamefont {Evered}\ \emph {et~al.}(2023)\citenamefont {Evered}, \citenamefont {Bluvstein}, \citenamefont {Kalinowski}, \citenamefont {Ebadi}, \citenamefont {Manovitz}, \citenamefont {Zhou}, \citenamefont {Li}, \citenamefont {Geim}, \citenamefont {Wang}, \citenamefont {Maskara} \emph {et~al.}}]{evered2023high}\BibitemOpen
	\bibfield  {author} {\bibinfo {author} {S.~J. Evered}, et~al.,\ }\emph {High-fidelity parallel entangling gates on a neutral-atom quantum computer},\ \href {https://doi.org/https://doi.org/10.1038/s41586-023-06481-y} {\bibfield  {journal} {\bibinfo  {journal} {Nature}\ }\textbf {\bibinfo {volume} {622}},\ \bibinfo {pages} {268} (\bibinfo {year} {2023})}\BibitemShut {NoStop}\bibitem [{\citenamefont {Figgatt}\ \emph {et~al.}(2019)\citenamefont {Figgatt}, \citenamefont {Ostrander}, \citenamefont {Linke}, \citenamefont {Landsman}, \citenamefont {Zhu}, \citenamefont {Maslov},\ and\ \citenamefont {Monroe}}]{figgatt2019parallel}\BibitemOpen
	\bibfield  {author} {\bibinfo {author} {C.~Figgatt}, et~al.,\ }\emph {Parallel entangling operations on a universal ion-trap quantum computer},\ \href {https://doi.org/https://doi.org/10.1038/s41586-019-1427-5} {\bibfield  {journal} {\bibinfo  {journal} {Nature}\ }\textbf {\bibinfo {volume} {572}},\ \bibinfo {pages} {368} (\bibinfo {year} {2019})}\BibitemShut {NoStop}\bibitem [{\citenamefont {Pino}\ \emph {et~al.}(2021)\citenamefont {Pino}, \citenamefont {Dreiling}, \citenamefont {Figgatt}, \citenamefont {Gaebler}, \citenamefont {Moses}, \citenamefont {Allman}, \citenamefont {Baldwin}, \citenamefont {Foss-Feig}, \citenamefont {Hayes}, \citenamefont {Mayer} \emph {et~al.}}]{pino2021demonstration}\BibitemOpen
	\bibfield  {author} {\bibinfo {author} {J.~M. Pino}, et~al.,\ }\emph {Demonstration of the trapped-ion quantum CCD computer architecture},\ \href {https://doi.org/https://doi.org/10.1038/s41586-021-03318-4} {\bibfield  {journal} {\bibinfo  {journal} {Nature}\ }\textbf {\bibinfo {volume} {592}},\ \bibinfo {pages} {209} (\bibinfo {year} {2021})}\BibitemShut {NoStop}\bibitem [{\citenamefont {Zhu}\ \emph {et~al.}(2023{\natexlab{a}})\citenamefont {Zhu}, \citenamefont {Green}, \citenamefont {Nguyen}, \citenamefont {Huerta~Alderete}, \citenamefont {Mossman},\ and\ \citenamefont {Linke}}]{zhu2023pairwise}\BibitemOpen
	\bibfield  {author} {\bibinfo {author} {Y.~Zhu}, et~al.,\ }\emph {Pairwise-parallel entangling gates on orthogonal modes in a trapped-ion chain},\ \href {https://doi.org/https://doi.org/10.1002/qute.202300056} {\bibfield  {journal} {\bibinfo  {journal} {Advanced Quantum Technologies}\ }\textbf {\bibinfo {volume} {6}},\ \bibinfo {pages} {2300056} (\bibinfo {year} {2023}{\natexlab{a}})}\BibitemShut {NoStop}\bibitem [{\citenamefont {Valentini}\ \emph {et~al.}(2024)\citenamefont {Valentini}, \citenamefont {van Mourik}, \citenamefont {Butt}, \citenamefont {Wahl}, \citenamefont {Dietl}, \citenamefont {Pfeifer}, \citenamefont {Anmasser}, \citenamefont {Colombe}, \citenamefont {Rössler}, \citenamefont {Holz}, \citenamefont {Blatt}, \citenamefont {Müller}, \citenamefont {Monz},\ and\ \citenamefont {Schindler}}]{valentini2024demonstration}\BibitemOpen
	\bibfield  {author} {\bibinfo {author} {M.~Valentini}, et~al.,\ }\href@noop {} {\emph {Demonstration of two-dimensional connectivity for a scalable error-corrected ion-trap quantum processor architecture}} (\bibinfo {year} {2024}),\ \Eprint {https://arxiv.org/abs/2406.02406} {arXiv:2406.02406 [quant-ph]} \BibitemShut {NoStop}\bibitem [{\citenamefont {Higgott}\ \emph {et~al.}(2021)\citenamefont {Higgott}, \citenamefont {Wilson}, \citenamefont {Hefford}, \citenamefont {Dborin}, \citenamefont {Hanif}, \citenamefont {Burton},\ and\ \citenamefont {Browne}}]{higgott2021optimal}\BibitemOpen
	\bibfield  {author} {\bibinfo {author} {O.~Higgott}, et~al.,\ }\emph {Optimal local unitary encoding circuits for the surface code},\ \href {https://doi.org/https://doi.org/10.22331/q-2021-08-05-517} {\bibfield  {journal} {\bibinfo  {journal} {Quantum}\ }\textbf {\bibinfo {volume} {5}},\ \bibinfo {pages} {517} (\bibinfo {year} {2021})}\BibitemShut {NoStop}\bibitem [{\citenamefont {Moore}\ and\ \citenamefont {Nilsson}(1998)}]{moore1998parallel}\BibitemOpen
	\bibfield  {author} {\bibinfo {author} {C.~Moore}\ and\ \bibinfo {author} {M.~Nilsson},\ }\href@noop {} {\emph {Parallel quantum computation and quantum codes}} (\bibinfo {year} {1998}),\ \Eprint {https://arxiv.org/abs/quant-ph/9808027} {arXiv:quant-ph/9808027 [quant-ph]} \BibitemShut {NoStop}\bibitem [{\citenamefont {Zhang}\ \emph {et~al.}(2022)\citenamefont {Zhang}, \citenamefont {Li},\ and\ \citenamefont {Yuan}}]{zhang2022quantum}\BibitemOpen
	\bibfield  {author} {\bibinfo {author} {X.-M. Zhang}, \bibinfo {author} {T.~Li},\ and\ \bibinfo {author} {X.~Yuan},\ }\emph {Quantum state preparation with optimal circuit depth: Implementations and applications},\ \href {https://doi.org/10.1103/PhysRevLett.129.230504} {\bibfield  {journal} {\bibinfo  {journal} {Physical Review Letters}\ }\textbf {\bibinfo {volume} {129}},\ \bibinfo {pages} {230504} (\bibinfo {year} {2022})}\BibitemShut {NoStop}\bibitem [{\citenamefont {Anderson}(2012)}]{anderson2012power}\BibitemOpen
	\bibfield  {author} {\bibinfo {author} {J.~T. Anderson},\ }\href@noop {} {\emph {On the power of reusable magic states}} (\bibinfo {year} {2012}),\ \Eprint {https://arxiv.org/abs/1205.0289} {arXiv:1205.0289 [quant-ph]} \BibitemShut {NoStop}\bibitem [{\citenamefont {Kjaergaard}\ \emph {et~al.}(2020)\citenamefont {Kjaergaard}, \citenamefont {Schwartz}, \citenamefont {Braum{\"u}ller}, \citenamefont {Krantz}, \citenamefont {Wang}, \citenamefont {Gustavsson},\ and\ \citenamefont {Oliver}}]{kjaergaard2020superconducting}\BibitemOpen
	\bibfield  {author} {\bibinfo {author} {M.~Kjaergaard}, et~al.,\ }\emph {Superconducting qubits: Current state of play},\ \href {https://doi.org/10.1146/annurev-conmatphys-031119-050605} {\bibfield  {journal} {\bibinfo  {journal} {Annual Review of Condensed Matter Physics}\ }\textbf {\bibinfo {volume} {11}},\ \bibinfo {pages} {369} (\bibinfo {year} {2020})}\BibitemShut {NoStop}\bibitem [{\citenamefont {Piveteau}\ and\ \citenamefont {Sutter}(2024)}]{piveteau2023circuit}\BibitemOpen
	\bibfield  {author} {\bibinfo {author} {C.~Piveteau}\ and\ \bibinfo {author} {D.~Sutter},\ }\emph {Circuit knitting with classical communication},\ \href {https://doi.org/10.1109/TIT.2023.3310797} {\bibfield  {journal} {\bibinfo  {journal} {IEEE Transactions on Information Theory}\ }\textbf {\bibinfo {volume} {70}},\ \bibinfo {pages} {2734} (\bibinfo {year} {2024})}\BibitemShut {NoStop}\bibitem [{\citenamefont {Henriet}\ \emph {et~al.}(2020)\citenamefont {Henriet}, \citenamefont {Beguin}, \citenamefont {Signoles}, \citenamefont {Lahaye}, \citenamefont {Browaeys}, \citenamefont {Reymond},\ and\ \citenamefont {Jurczak}}]{henriet2020quantum}\BibitemOpen
	\bibfield  {author} {\bibinfo {author} {L.~Henriet}, et~al.,\ }\emph {Quantum computing with neutral atoms},\ \href {https://doi.org/10.22331/q-2020-09-21-327} {\bibfield  {journal} {\bibinfo  {journal} {Quantum}\ }\textbf {\bibinfo {volume} {4}},\ \bibinfo {pages} {327} (\bibinfo {year} {2020})}\BibitemShut {NoStop}\bibitem [{\citenamefont {Wintersperger}\ \emph {et~al.}(2023)\citenamefont {Wintersperger}, \citenamefont {Dommert}, \citenamefont {Ehmer}, \citenamefont {Hoursanov}, \citenamefont {Klepsch}, \citenamefont {Mauerer}, \citenamefont {Reuber}, \citenamefont {Strohm}, \citenamefont {Yin},\ and\ \citenamefont {Luber}}]{wintersperger2023neutral}\BibitemOpen
	\bibfield  {author} {\bibinfo {author} {K.~Wintersperger}, et~al.,\ }\emph {Neutral atom quantum computing hardware: Performance and end-user perspective},\ \href {https://doi.org/10.1140/epjqt/s40507-023-00190-1} {\bibfield  {journal} {\bibinfo  {journal} {EPJ Quantum Technology}\ }\textbf {\bibinfo {volume} {10}},\ \bibinfo {pages} {1} (\bibinfo {year} {2023})}\BibitemShut {NoStop}\bibitem [{\citenamefont {M{\o}lmer}\ and\ \citenamefont {S{\o}rensen}(1999)}]{molmer1999multiparticle}\BibitemOpen
	\bibfield  {author} {\bibinfo {author} {K.~M{\o}lmer}\ and\ \bibinfo {author} {A.~S{\o}rensen},\ }\emph {Multiparticle entanglement of hot trapped ions},\ \href {https://doi.org/10.1103/PhysRevLett.82.1835} {\bibfield  {journal} {\bibinfo  {journal} {Physical Review Letters}\ }\textbf {\bibinfo {volume} {82}},\ \bibinfo {pages} {1835} (\bibinfo {year} {1999})}\BibitemShut {NoStop}\bibitem [{\citenamefont {S\o{}rensen}\ and\ \citenamefont {M\o{}lmer}(2000)}]{sorensen2000entanglement}\BibitemOpen
	\bibfield  {author} {\bibinfo {author} {A.~S\o{}rensen}\ and\ \bibinfo {author} {K.~M\o{}lmer},\ }\emph {Entanglement and quantum computation with ions in thermal motion},\ \href {https://doi.org/10.1103/PhysRevA.62.022311} {\bibfield  {journal} {\bibinfo  {journal} {Physical Review A}\ }\textbf {\bibinfo {volume} {62}},\ \bibinfo {pages} {022311} (\bibinfo {year} {2000})}\BibitemShut {NoStop}\bibitem [{\citenamefont {Ballance}(2017)}]{ballance2017high}\BibitemOpen
	\bibfield  {author} {\bibinfo {author} {C.~J. Ballance},\ }\href {https://doi.org/10.1007/978-3-319-68216-7} {\emph {\bibinfo {title} {High-fidelity quantum logic in Ca$^+$}}}\ (\bibinfo  {publisher} {Springer},\ \bibinfo {year} {2017})\BibitemShut {NoStop}\bibitem [{\citenamefont {Steane}(2004)}]{steane2004fastfaulttolerantfilteringquantum}\BibitemOpen
	\bibfield  {author} {\bibinfo {author} {A.~M. Steane},\ }\href@noop {} {\emph {Fast fault-tolerant filtering of quantum codewords}} (\bibinfo {year} {2004}),\ \Eprint {https://arxiv.org/abs/quant-ph/0202036} {arXiv:quant-ph/0202036 [quant-ph]} \BibitemShut {NoStop}\bibitem [{\citenamefont {Cross}\ \emph {et~al.}(2009)\citenamefont {Cross}, \citenamefont {DiVincenzo},\ and\ \citenamefont {Terhal}}]{cross2009comparativecodestudyquantum}\BibitemOpen
	\bibfield  {author} {\bibinfo {author} {A.~W. Cross}, \bibinfo {author} {D.~P. DiVincenzo},\ and\ \bibinfo {author} {B.~M. Terhal},\ }\href@noop {} {\emph {A comparative code study for quantum fault-tolerance}} (\bibinfo {year} {2009}),\ \Eprint {https://arxiv.org/abs/0711.1556} {arXiv:0711.1556 [quant-ph]} \BibitemShut {NoStop}\bibitem [{\citenamefont {Amaro}(2019)}]{amaro2019stabgraph}\BibitemOpen
	\bibfield  {author} {\bibinfo {author} {D.~Amaro},\ }\href@noop {} {\emph {StabGraph}},\ \bibinfo {howpublished} {\url{https://github.com/davamaro/stabgraph}} (\bibinfo {year} {2019})\BibitemShut {NoStop}\bibitem [{\citenamefont {Amaro}\ \emph {et~al.}(2020)\citenamefont {Amaro}, \citenamefont {M{\"u}ller},\ and\ \citenamefont {Pal}}]{amaro2020scalable}\BibitemOpen
	\bibfield  {author} {\bibinfo {author} {D.~Amaro}, \bibinfo {author} {M.~M{\"u}ller},\ and\ \bibinfo {author} {A.~K. Pal},\ }\emph {Scalable characterization of localizable entanglement in noisy topological quantum codes},\ \href {https://doi.org/doi.org/10.1088/1367-2630/ab84b3} {\bibfield  {journal} {\bibinfo  {journal} {New Journal of Physics}\ }\textbf {\bibinfo {volume} {22}},\ \bibinfo {pages} {053038} (\bibinfo {year} {2020})}\BibitemShut {NoStop}\bibitem [{\citenamefont {Zen}\ \emph {et~al.}(2024)\citenamefont {Zen}, \citenamefont {Olle}, \citenamefont {Colmenarez}, \citenamefont {Puviani}, \citenamefont {Müller},\ and\ \citenamefont {Marquardt}}]{zen2024quantum}\BibitemOpen
	\bibfield  {author} {\bibinfo {author} {R.~Zen}, et~al.,\ }\href@noop {} {\emph {Quantum circuit discovery for fault-tolerant logical state preparation with reinforcement learning}} (\bibinfo {year} {2024}),\ \Eprint {https://arxiv.org/abs/2402.17761} {arXiv:2402.17761 [quant-ph]} \BibitemShut {NoStop}\bibitem [{\citenamefont {Zen}(2024)}]{zen2024rlftqc}\BibitemOpen
	\bibfield  {author} {\bibinfo {author} {R.~Zen},\ }\href@noop {} {\emph {Reinforcement learning for fault-tolerant quantum circuit discovery}},\ \bibinfo {howpublished} {\url{https://github.com/remmyzen/rlftqc}} (\bibinfo {year} {2024})\BibitemShut {NoStop}\bibitem [{\citenamefont {Peham}\ \emph {et~al.}(2025)\citenamefont {Peham}, \citenamefont {Schmid}, \citenamefont {Berent}, \citenamefont {M{\"u}ller},\ and\ \citenamefont {Wille}}]{peham2024automatedsynthesisfaulttolerantstate}\BibitemOpen
	\bibfield  {author} {\bibinfo {author} {T.~Peham}, \bibinfo {author} {L.~Schmid}, \bibinfo {author} {L.~Berent}, \bibinfo {author} {M.~M{\"u}ller},\ and\ \bibinfo {author} {R.~Wille},\ }\emph {Automated synthesis of fault-tolerant state preparation circuits for quantum error-correction codes},\ \href {https://doi.org/10.1103/PRXQuantum.6.020330} {\bibfield  {journal} {\bibinfo  {journal} {PRX Quantum}\ }\textbf {\bibinfo {volume} {6}},\ \bibinfo {pages} {020330} (\bibinfo {year} {2025})}\BibitemShut {NoStop}\bibitem [{\citenamefont {Kreppel}\ \emph {et~al.}(2023)\citenamefont {Kreppel}, \citenamefont {Melzer}, \citenamefont {Mill{\'a}n}, \citenamefont {Wagner}, \citenamefont {Hilder}, \citenamefont {Poschinger}, \citenamefont {Schmidt-Kaler},\ and\ \citenamefont {Brinkmann}}]{kreppel2023quantum}\BibitemOpen
	\bibfield  {author} {\bibinfo {author} {F.~Kreppel}, et~al.,\ }\emph {Quantum circuit compiler for a shuttling-based trapped-ion quantum computer},\ \href {https://doi.org/https://doi.org/10.22331/q-2023-11-08-1176} {\bibfield  {journal} {\bibinfo  {journal} {Quantum}\ }\textbf {\bibinfo {volume} {7}},\ \bibinfo {pages} {1176} (\bibinfo {year} {2023})}\BibitemShut {NoStop}\bibitem [{\citenamefont {Schoenberger}\ \emph {et~al.}(2024{\natexlab{a}})\citenamefont {Schoenberger}, \citenamefont {Hillmich}, \citenamefont {Brandl},\ and\ \citenamefont {Wille}}]{schoenberger2024using}\BibitemOpen
	\bibfield  {author} {\bibinfo {author} {D.~Schoenberger}, \bibinfo {author} {S.~Hillmich}, \bibinfo {author} {M.~Brandl},\ and\ \bibinfo {author} {R.~Wille},\ }\emph {Using Boolean satisfiability for exact shuttling in trapped-ion quantum computers},\ \href {https://doi.org/https://doi.org/10.1109/ASP-DAC58780.2024.10473902} {\bibfield  {journal} {\bibinfo  {journal} {29th Asia and South Pacific Design Automation Conference}\ ,\ \bibinfo {pages} {127}} (\bibinfo {year} {2024}{\natexlab{a}})}\BibitemShut {NoStop}\bibitem [{\citenamefont {Aaronson}\ and\ \citenamefont {Gottesman}(2004)}]{aaronson2004improved}\BibitemOpen
	\bibfield  {author} {\bibinfo {author} {S.~Aaronson}\ and\ \bibinfo {author} {D.~Gottesman},\ }\emph {Improved simulation of stabilizer circuits},\ \href {https://doi.org/10.1103/PhysRevA.70.052328} {\bibfield  {journal} {\bibinfo  {journal} {Physical Review A}\ }\textbf {\bibinfo {volume} {70}},\ \bibinfo {pages} {052328} (\bibinfo {year} {2004})}\BibitemShut {NoStop}\bibitem [{\citenamefont {Ryan-Anderson}(2018)}]{ryan2018quantum}\BibitemOpen
	\bibfield  {author} {\bibinfo {author} {C.~Ryan-Anderson},\ }\emph {\bibinfo {title} {Quantum algorithms, architecture, and error correction}},\ \href {https://digitalrepository.unm.edu/phyc_etds/203/} {Ph.D. thesis},\ \bibinfo  {school} {The University of New Mexico} (\bibinfo {year} {2018})\BibitemShut {NoStop}\bibitem [{\citenamefont {Ryan-Anderson}(2019)}]{pecos}\BibitemOpen
	\bibfield  {author} {\bibinfo {author} {C.~Ryan-Anderson},\ }\href@noop {} {\emph {PECOS: Performance estimator of codes on surfaces}},\ \bibinfo {howpublished} {\url{https://github.com/PECOS-packages/PECOS}} (\bibinfo {year} {2019})\BibitemShut {NoStop}\bibitem [{\citenamefont {Heu{\ss}en}\ \emph {et~al.}(2024)\citenamefont {Heu{\ss}en}, \citenamefont {Winter}, \citenamefont {Rispler},\ and\ \citenamefont {M\"uller}}]{heussen2024dynamical}\BibitemOpen
	\bibfield  {author} {\bibinfo {author} {S.~Heu{\ss}en}, \bibinfo {author} {D.~Winter}, \bibinfo {author} {M.~Rispler},\ and\ \bibinfo {author} {M.~M\"uller},\ }\emph {Dynamical subset sampling of quantum error-correcting protocols},\ \href {https://doi.org/10.1103/PhysRevResearch.6.013177} {\bibfield  {journal} {\bibinfo  {journal} {Physical Review Research}\ }\textbf {\bibinfo {volume} {6}},\ \bibinfo {pages} {013177} (\bibinfo {year} {2024})}\BibitemShut {NoStop}\bibitem [{\citenamefont {Winter}\ and\ \citenamefont {Heu{\ss}en}(2023)}]{qsample}\BibitemOpen
	\bibfield  {author} {\bibinfo {author} {D.~Winter}\ and\ \bibinfo {author} {S.~Heu{\ss}en},\ }\href@noop {} {\emph {qsample}},\ \bibinfo {howpublished} {\url{https://github.com/dpwinter/qsample}} (\bibinfo {year} {2023})\BibitemShut {NoStop}\bibitem [{\citenamefont {Kubica}(2018)}]{kubica2018abcs}\BibitemOpen
	\bibfield  {author} {\bibinfo {author} {A.~Kubica},\ }\emph {\bibinfo {title} {The ABCs of the color code: A study of topological quantum codes as toy models for fault-tolerant quantum computation and quantum phases of matter}},\ \href {https://thesis.library.caltech.edu/10955/} {Ph.D. thesis},\ \bibinfo  {school} {California Institute of Technology} (\bibinfo {year} {2018})\BibitemShut {NoStop}\bibitem [{\citenamefont {Chamberland}\ \emph {et~al.}(2020)\citenamefont {Chamberland}, \citenamefont {Kubica}, \citenamefont {Yoder},\ and\ \citenamefont {Zhu}}]{chamberland2020triangular}\BibitemOpen
	\bibfield  {author} {\bibinfo {author} {C.~Chamberland}, \bibinfo {author} {A.~Kubica}, \bibinfo {author} {T.~J. Yoder},\ and\ \bibinfo {author} {G.~Zhu},\ }\emph {Triangular color codes on trivalent graphs with flag qubits},\ \href {https://doi.org/10.1088/1367-2630/ab68fd} {\bibfield  {journal} {\bibinfo  {journal} {New Journal of Physics}\ }\textbf {\bibinfo {volume} {22}},\ \bibinfo {pages} {023019} (\bibinfo {year} {2020})}\BibitemShut {NoStop}\bibitem [{\citenamefont {Takada}\ and\ \citenamefont {Fujii}(2024)}]{takada2024improving}\BibitemOpen
	\bibfield  {author} {\bibinfo {author} {Y.~Takada}\ and\ \bibinfo {author} {K.~Fujii},\ }\emph {Improving threshold for fault-tolerant color-code quantum computing by flagged weight optimization},\ \href {https://doi.org/10.1103/PRXQuantum.5.030352} {\bibfield  {journal} {\bibinfo  {journal} {PRX Quantum}\ }\textbf {\bibinfo {volume} {5}},\ \bibinfo {pages} {030352} (\bibinfo {year} {2024})}\BibitemShut {NoStop}\bibitem [{\citenamefont {Chamberland}\ \emph {et~al.}(2017)\citenamefont {Chamberland}, \citenamefont {Jochym-O'Connor},\ and\ \citenamefont {Laflamme}}]{chamberland2017overhead}\BibitemOpen
	\bibfield  {author} {\bibinfo {author} {C.~Chamberland}, \bibinfo {author} {T.~Jochym-O'Connor},\ and\ \bibinfo {author} {R.~Laflamme},\ }\emph {Overhead analysis of universal concatenated quantum codes},\ \href {https://doi.org/10.1103/PhysRevA.95.022313} {\bibfield  {journal} {\bibinfo  {journal} {Physical Review A}\ }\textbf {\bibinfo {volume} {95}},\ \bibinfo {pages} {022313} (\bibinfo {year} {2017})}\BibitemShut {NoStop}\bibitem [{\citenamefont {Vasmer}\ and\ \citenamefont {Kubica}(2022)}]{vasmer2022morphing}\BibitemOpen
	\bibfield  {author} {\bibinfo {author} {M.~Vasmer}\ and\ \bibinfo {author} {A.~Kubica},\ }\emph {Morphing quantum codes},\ \href {https://doi.org/10.1103/PRXQuantum.3.030319} {\bibfield  {journal} {\bibinfo  {journal} {PRX Quantum}\ }\textbf {\bibinfo {volume} {3}},\ \bibinfo {pages} {030319} (\bibinfo {year} {2022})}\BibitemShut {NoStop}\bibitem [{\citenamefont {Reichardt}(2005)}]{reichardt2004improved}\BibitemOpen
	\bibfield  {author} {\bibinfo {author} {B.~W. Reichardt},\ }\emph {Quantum universality from magic states distillation applied to {CSS} codes},\ \href {https://doi.org/10.1007/s11128-005-7654-8} {\bibfield  {journal} {\bibinfo  {journal} {Quantum Information Processing}\ }\textbf {\bibinfo {volume} {4}},\ \bibinfo {pages} {251} (\bibinfo {year} {2005})}\BibitemShut {NoStop}\bibitem [{\citenamefont {Bravyi}\ and\ \citenamefont {Kitaev}(2005)}]{bravyi2005universal}\BibitemOpen
	\bibfield  {author} {\bibinfo {author} {S.~Bravyi}\ and\ \bibinfo {author} {A.~Kitaev},\ }\emph {{Universal quantum computation with ideal Clifford gates and noisy ancillas}},\ \href {https://doi.org/10.1103/PhysRevA.71.022316} {\bibfield  {journal} {\bibinfo  {journal} {Physical Review A}\ }\textbf {\bibinfo {volume} {71}},\ \bibinfo {pages} {022316} (\bibinfo {year} {2005})}\BibitemShut {NoStop}\bibitem [{\citenamefont {Gupta}\ \emph {et~al.}(2024)\citenamefont {Gupta}, \citenamefont {Sundaresan}, \citenamefont {Alexander}, \citenamefont {Wood}, \citenamefont {Merkel}, \citenamefont {Healy}, \citenamefont {Hillenbrand}, \citenamefont {Jochym-O’Connor}, \citenamefont {Wootton}, \citenamefont {Yoder} \emph {et~al.}}]{gupta2024encoding}\BibitemOpen
	\bibfield  {author} {\bibinfo {author} {R.~S. Gupta}, et~al.,\ }\emph {Encoding a magic state with beyond break-even fidelity},\ \href {https://doi.org/10.1038/s41586-023-06846-3} {\bibfield  {journal} {\bibinfo  {journal} {Nature}\ }\textbf {\bibinfo {volume} {625}},\ \bibinfo {pages} {259--263} (\bibinfo {year} {2024})}\BibitemShut {NoStop}\bibitem [{\citenamefont {Chamberland}\ and\ \citenamefont {Noh}(2020)}]{chamberland2020very}\BibitemOpen
	\bibfield  {author} {\bibinfo {author} {C.~Chamberland}\ and\ \bibinfo {author} {K.~Noh},\ }\emph {Very low overhead fault-tolerant magic state preparation using redundant ancilla encoding and flag qubits},\ \href {https://doi.org/https://doi.org/10.1038/s41534-020-00319-5} {\bibfield  {journal} {\bibinfo  {journal} {npj Quantum Information}\ }\textbf {\bibinfo {volume} {6}},\ \bibinfo {pages} {91} (\bibinfo {year} {2020})}\BibitemShut {NoStop}\bibitem [{\citenamefont {Jones}\ \emph {et~al.}(2024)\citenamefont {Jones}, \citenamefont {Linden},\ and\ \citenamefont {Skrzypczyk}}]{jones2024hadamard}\BibitemOpen
	\bibfield  {author} {\bibinfo {author} {B.~D. Jones}, \bibinfo {author} {N.~Linden},\ and\ \bibinfo {author} {P.~Skrzypczyk},\ }\emph {The Hadamard gate cannot be replaced by a resource state in universal quantum computation},\ \href {https://doi.org/10.22331/q-2024-09-11-1470} {\bibfield  {journal} {\bibinfo  {journal} {Quantum}\ }\textbf {\bibinfo {volume} {8}},\ \bibinfo {pages} {1470} (\bibinfo {year} {2024})}\BibitemShut {NoStop}\bibitem [{\citenamefont {Jochym-O'Connor}\ and\ \citenamefont {Laflamme}(2014)}]{oconnor2014using}\BibitemOpen
	\bibfield  {author} {\bibinfo {author} {T.~Jochym-O'Connor}\ and\ \bibinfo {author} {R.~Laflamme},\ }\emph {Using concatenated quantum codes for universal fault-tolerant quantum gates},\ \href {https://doi.org/10.1103/PhysRevLett.112.010505} {\bibfield  {journal} {\bibinfo  {journal} {Physical Review Letters}\ }\textbf {\bibinfo {volume} {112}},\ \bibinfo {pages} {010505} (\bibinfo {year} {2014})}\BibitemShut {NoStop}\bibitem [{\citenamefont {Yoder}\ \emph {et~al.}(2016)\citenamefont {Yoder}, \citenamefont {Takagi},\ and\ \citenamefont {Chuang}}]{yoder2016universal}\BibitemOpen
	\bibfield  {author} {\bibinfo {author} {T.~J. Yoder}, \bibinfo {author} {R.~Takagi},\ and\ \bibinfo {author} {I.~L. Chuang},\ }\emph {Universal fault-tolerant gates on concatenated stabilizer codes},\ \href {https://doi.org/10.1103/PhysRevX.6.031039} {\bibfield  {journal} {\bibinfo  {journal} {Physical Review X}\ }\textbf {\bibinfo {volume} {6}},\ \bibinfo {pages} {031039} (\bibinfo {year} {2016})}\BibitemShut {NoStop}\bibitem [{\citenamefont {Buchbinder}\ \emph {et~al.}(2013)\citenamefont {Buchbinder}, \citenamefont {Huang},\ and\ \citenamefont {Weinstein}}]{buchbinder2013encoding}\BibitemOpen
	\bibfield  {author} {\bibinfo {author} {S.~D. Buchbinder}, \bibinfo {author} {C.~L. Huang},\ and\ \bibinfo {author} {Y.~S. Weinstein},\ }\emph {Encoding an arbitrary state in a [7, 1, 3] quantum error correction code},\ \href {https://doi.org/https://doi.org/10.1007/s11128-012-0414-7} {\bibfield  {journal} {\bibinfo  {journal} {Quantum information processing}\ }\textbf {\bibinfo {volume} {12}},\ \bibinfo {pages} {699} (\bibinfo {year} {2013})}\BibitemShut {NoStop}\bibitem [{\citenamefont {Aliferis}\ \emph {et~al.}(2006)\citenamefont {Aliferis}, \citenamefont {Gottesman},\ and\ \citenamefont {Preskill}}]{aliferis2006quantum}\BibitemOpen
	\bibfield  {author} {\bibinfo {author} {P.~Aliferis}, \bibinfo {author} {D.~Gottesman},\ and\ \bibinfo {author} {J.~Preskill},\ }\emph {{Quantum accuracy threshold for concatenated distance-3 codes}},\ \href {https://doi.org/10.26421/QIC6.2-1} {\bibfield  {journal} {\bibinfo  {journal} {Quantum Information and Computation}\ }\textbf {\bibinfo {volume} {6}},\ \bibinfo {pages} {97} (\bibinfo {year} {2006})}\BibitemShut {NoStop}\bibitem [{\citenamefont {Yoshida}\ \emph {et~al.}(2025)\citenamefont {Yoshida}, \citenamefont {Tamiya},\ and\ \citenamefont {Yamasaki}}]{yoshida2024concatenate}\BibitemOpen
	\bibfield  {author} {\bibinfo {author} {S.~Yoshida}, \bibinfo {author} {S.~Tamiya},\ and\ \bibinfo {author} {H.~Yamasaki},\ }\emph {Concatenate codes, save qubits},\ \href {https://doi.org/10.1038/s41534-025-01035-8} {\bibfield  {journal} {\bibinfo  {journal} {npj Quantum Information}\ }\textbf {\bibinfo {volume} {11}},\ \bibinfo {pages} {88} (\bibinfo {year} {2025})}\BibitemShut {NoStop}\bibitem [{\citenamefont {Goto}(2024)}]{goto2024manyhypercube}\BibitemOpen
	\bibfield  {author} {\bibinfo {author} {H.~Goto},\ }\emph {High-performance fault-tolerant quantum computing with many-hypercube codes},\ \href {https://doi.org/10.1126/sciadv.adp6388} {\bibfield  {journal} {\bibinfo  {journal} {Science Advances}\ }\textbf {\bibinfo {volume} {10}},\ \bibinfo {pages} {eadp6388} (\bibinfo {year} {2024})}\BibitemShut {NoStop}\bibitem [{\citenamefont {Yamasaki}\ and\ \citenamefont {Koashi}(2024)}]{yamasaki2024time}\BibitemOpen
	\bibfield  {author} {\bibinfo {author} {H.~Yamasaki}\ and\ \bibinfo {author} {M.~Koashi},\ }\emph {Time-efficient constant-space-overhead fault-tolerant quantum computation},\ \href {https://doi.org/https://doi.org/10.1038/s41567-023-02325-8} {\bibfield  {journal} {\bibinfo  {journal} {Nature Physics\!}\ ,\ \bibinfo {pages} {1}} (\bibinfo {year} {2024})}\BibitemShut {NoStop}\bibitem [{\citenamefont {Katabarwa}\ \emph {et~al.}(2024)\citenamefont {Katabarwa}, \citenamefont {Gratsea}, \citenamefont {Caesura},\ and\ \citenamefont {Johnson}}]{katabarwa2023early}\BibitemOpen
	\bibfield  {author} {\bibinfo {author} {A.~Katabarwa}, \bibinfo {author} {K.~Gratsea}, \bibinfo {author} {A.~Caesura},\ and\ \bibinfo {author} {P.~D. Johnson},\ }\emph {Early fault-tolerant quantum computing},\ \href {https://doi.org/10.1103/PRXQuantum.5.020101} {\bibfield  {journal} {\bibinfo  {journal} {PRX Quantum}\ }\textbf {\bibinfo {volume} {5}},\ \bibinfo {pages} {020101} (\bibinfo {year} {2024})}\BibitemShut {NoStop}\bibitem [{\citenamefont {DeCross}\ \emph {et~al.}(2025)\citenamefont {DeCross}, \citenamefont {Haghshenas}, \citenamefont {Liu}, \citenamefont {Rinaldi}, \citenamefont {Gray}, \citenamefont {Alexeev}, \citenamefont {Baldwin}, \citenamefont {Bartolotta}, \citenamefont {Bohn}, \citenamefont {Chertkov} \emph {et~al.}}]{decross2024computational}\BibitemOpen
	\bibfield  {author} {\bibinfo {author} {M.~DeCross}, et~al.,\ }\emph {Computational power of random quantum circuits in abitrary geometries},\ \href {https://doi.org/10.1103/PhysRevX.15.021052} {\bibfield  {journal} {\bibinfo  {journal} {Physical Review X}\ }\textbf {\bibinfo {volume} {15}},\ \bibinfo {pages} {021052} (\bibinfo {year} {2025})}\BibitemShut {NoStop}\bibitem [{\citenamefont {Bravyi}\ and\ \citenamefont {Haah}(2012)}]{bravyi2012magic}\BibitemOpen
	\bibfield  {author} {\bibinfo {author} {S.~Bravyi}\ and\ \bibinfo {author} {J.~Haah},\ }\emph {Magic-state distillation with low overhead},\ \href {https://doi.org/10.1103/PhysRevA.86.052329} {\bibfield  {journal} {\bibinfo  {journal} {Physical Review A}\ }\textbf {\bibinfo {volume} {86}},\ \bibinfo {pages} {052329} (\bibinfo {year} {2012})}\BibitemShut {NoStop}\bibitem [{\citenamefont {Bombin}(2018)}]{bombin2018transversalgateserrorpropagation}\BibitemOpen
	\bibfield  {author} {\bibinfo {author} {H.~Bombin},\ }\href@noop {} {\emph {Transversal gates and error propagation in 3D topological codes}} (\bibinfo {year} {2018}),\ \Eprint {https://arxiv.org/abs/1810.09575} {arXiv:1810.09575 [quant-ph]} \BibitemShut {NoStop}\bibitem [{\citenamefont {Hensinger}\ \emph {et~al.}(2006)\citenamefont {Hensinger}, \citenamefont {Olmschenk}, \citenamefont {Stick}, \citenamefont {Hucul}, \citenamefont {Yeo}, \citenamefont {Acton}, \citenamefont {Deslauriers}, \citenamefont {Monroe},\ and\ \citenamefont {Rabchuk}}]{hensinger2006tjunction}\BibitemOpen
	\bibfield  {author} {\bibinfo {author} {W.~Hensinger}, et~al.,\ }\emph {T-junction ion trap array for two-dimensional ion shuttling, storage, and manipulation},\ \href {https://doi.org/https://doi.org/10.1063/1.2164910} {\bibfield  {journal} {\bibinfo  {journal} {Applied Physics Letters}\ }\textbf {\bibinfo {volume} {88}},\ \bibinfo {pages} {034101} (\bibinfo {year} {2006})}\BibitemShut {NoStop}\bibitem [{\citenamefont {Blakestad}\ \emph {et~al.}(2009)\citenamefont {Blakestad}, \citenamefont {Ospelkaus}, \citenamefont {VanDevender}, \citenamefont {Amini}, \citenamefont {Britton}, \citenamefont {Leibfried},\ and\ \citenamefont {Wineland}}]{blakestadt2009highfidelity}\BibitemOpen
	\bibfield  {author} {\bibinfo {author} {R.~B. Blakestad}, et~al.,\ }\emph {High-fidelity transport of trapped-ion qubits through an $X$-junction trap array},\ \href {https://doi.org/10.1103/PhysRevLett.102.153002} {\bibfield  {journal} {\bibinfo  {journal} {Physical Review Letters}\ }\textbf {\bibinfo {volume} {102}},\ \bibinfo {pages} {153002} (\bibinfo {year} {2009})}\BibitemShut {NoStop}\bibitem [{\citenamefont {Wright}\ \emph {et~al.}(2013)\citenamefont {Wright}, \citenamefont {Amini}, \citenamefont {Faircloth}, \citenamefont {Volin}, \citenamefont {Doret}, \citenamefont {Hayden}, \citenamefont {Pai}, \citenamefont {Landgren}, \citenamefont {Denison}, \citenamefont {Killian}, \citenamefont {Slusher},\ and\ \citenamefont {Harter}}]{Wright_2013}\BibitemOpen
	\bibfield  {author} {\bibinfo {author} {K.~Wright}, et~al.,\ }\emph {Reliable transport through a microfabricated X-junction surface-electrode ion trap},\ \href {https://doi.org/10.1088/1367-2630/15/3/033004} {\bibfield  {journal} {\bibinfo  {journal} {New Journal of Physics}\ }\textbf {\bibinfo {volume} {15}},\ \bibinfo {pages} {033004} (\bibinfo {year} {2013})}\BibitemShut {NoStop}\bibitem [{\citenamefont {Shu}\ \emph {et~al.}(2014)\citenamefont {Shu}, \citenamefont {Vittorini}, \citenamefont {Buikema}, \citenamefont {Nichols}, \citenamefont {Volin}, \citenamefont {Stick},\ and\ \citenamefont {Brown}}]{shu2014heating}\BibitemOpen
	\bibfield  {author} {\bibinfo {author} {G.~Shu}, et~al.,\ }\emph {Heating rates and ion-motion control in a $Y$-junction surface-electrode trap},\ \href {https://doi.org/10.1103/PhysRevA.89.062308} {\bibfield  {journal} {\bibinfo  {journal} {Physical Review A}\ }\textbf {\bibinfo {volume} {89}},\ \bibinfo {pages} {062308} (\bibinfo {year} {2014})}\BibitemShut {NoStop}\bibitem [{\citenamefont {Bautista-Salvador}\ \emph {et~al.}(2019)\citenamefont {Bautista-Salvador}, \citenamefont {Zarantonello}, \citenamefont {Hahn}, \citenamefont {Preciado-Grijalva}, \citenamefont {Morgner}, \citenamefont {Wahnschaffe},\ and\ \citenamefont {Ospelkaus}}]{bautista2019multilayer}\BibitemOpen
	\bibfield  {author} {\bibinfo {author} {A.~Bautista-Salvador}, et~al.,\ }\emph {Multilayer ion trap technology for scalable quantum computing and quantum simulation},\ \href {https://doi.org/10.1088/1367-2630/ab0e46} {\bibfield  {journal} {\bibinfo  {journal} {New Journal of Physics}\ }\textbf {\bibinfo {volume} {21}},\ \bibinfo {pages} {043011} (\bibinfo {year} {2019})}\BibitemShut {NoStop}\bibitem [{\citenamefont {Kaushal}\ \emph {et~al.}(2020)\citenamefont {Kaushal}, \citenamefont {Lekitsch}, \citenamefont {Stahl}, \citenamefont {Hilder}, \citenamefont {Pijn}, \citenamefont {Schmiegelow}, \citenamefont {Bermudez}, \citenamefont {Müller}, \citenamefont {Schmidt-Kaler},\ and\ \citenamefont {Poschinger}}]{Kaushal2020shuttlingbasedtiqc}\BibitemOpen
	\bibfield  {author} {\bibinfo {author} {V.~Kaushal}, et~al.,\ }\emph {{Shuttling-based trapped-ion quantum information processing}},\ \href {https://doi.org/10.1116/1.5126186} {\bibfield  {journal} {\bibinfo  {journal} {AVS Quantum Science}\ }\textbf {\bibinfo {volume} {2}},\ \bibinfo {pages} {014101} (\bibinfo {year} {2020})}\BibitemShut {NoStop}\bibitem [{\citenamefont {Malinowski}\ \emph {et~al.}(2023)\citenamefont {Malinowski}, \citenamefont {Allcock},\ and\ \citenamefont {Ballance}}]{malinowski2023howtowire}\BibitemOpen
	\bibfield  {author} {\bibinfo {author} {M.~Malinowski}, \bibinfo {author} {D.~Allcock},\ and\ \bibinfo {author} {C.~Ballance},\ }\emph {How to wire a $1000$-qubit trapped-ion quantum computer},\ \href {https://doi.org/10.1103/PRXQuantum.4.040313} {\bibfield  {journal} {\bibinfo  {journal} {PRX Quantum}\ }\textbf {\bibinfo {volume} {4}},\ \bibinfo {pages} {040313} (\bibinfo {year} {2023})}\BibitemShut {NoStop}\bibitem [{\citenamefont {Cai}\ \emph {et~al.}(2023)\citenamefont {Cai}, \citenamefont {Siegel},\ and\ \citenamefont {Benjamin}}]{cai2023looped}\BibitemOpen
	\bibfield  {author} {\bibinfo {author} {Z.~Cai}, \bibinfo {author} {A.~Siegel},\ and\ \bibinfo {author} {S.~Benjamin},\ }\emph {Looped pipelines enabling effective 3D qubit lattices in a strictly 2D device},\ \href {https://doi.org/10.1103/PRXQuantum.4.020345} {\bibfield  {journal} {\bibinfo  {journal} {PRX Quantum}\ }\textbf {\bibinfo {volume} {4}},\ \bibinfo {pages} {020345} (\bibinfo {year} {2023})}\BibitemShut {NoStop}\bibitem [{\citenamefont {Cain}\ \emph {et~al.}(2024)\citenamefont {Cain}, \citenamefont {Zhao}, \citenamefont {Zhou}, \citenamefont {Meister}, \citenamefont {Ataides}, \citenamefont {Jaffe}, \citenamefont {Bluvstein},\ and\ \citenamefont {Lukin}}]{cain2024correlateddecodinglogicalalgorithms}\BibitemOpen
	\bibfield  {author} {\bibinfo {author} {M.~Cain}, et~al.,\ }\href@noop {} {\emph {Correlated decoding of logical algorithms with transversal gates}} (\bibinfo {year} {2024}),\ \Eprint {https://arxiv.org/abs/2403.03272} {arXiv:2403.03272 [quant-ph]} \BibitemShut {NoStop}\bibitem [{\citenamefont {Wan}\ \emph {et~al.}(2024)\citenamefont {Wan}, \citenamefont {Webber}, \citenamefont {Fowler},\ and\ \citenamefont {Hensinger}}]{wan2024iterativetransversalcnotdecoder}\BibitemOpen
	\bibfield  {author} {\bibinfo {author} {K.~H. Wan}, \bibinfo {author} {M.~Webber}, \bibinfo {author} {A.~G. Fowler},\ and\ \bibinfo {author} {W.~K. Hensinger},\ }\href@noop {} {\emph {An iterative transversal CNOT decoder}} (\bibinfo {year} {2024}),\ \Eprint {https://arxiv.org/abs/2407.20976} {arXiv:2407.20976 [quant-ph]} \BibitemShut {NoStop}\bibitem [{\citenamefont {Sahay}\ \emph {et~al.}(2024)\citenamefont {Sahay}, \citenamefont {Lin}, \citenamefont {Huang}, \citenamefont {Brown},\ and\ \citenamefont {Puri}}]{sahay2024errorcorrectiontransversalcontrollednot}\BibitemOpen
	\bibfield  {author} {\bibinfo {author} {K.~Sahay}, \bibinfo {author} {Y.~Lin}, \bibinfo {author} {S.~Huang}, \bibinfo {author} {K.~R. Brown},\ and\ \bibinfo {author} {S.~Puri},\ }\href@noop {} {\emph {Error correction of transversal controlled-NOT gates for scalable surface code computation}} (\bibinfo {year} {2024}),\ \Eprint {https://arxiv.org/abs/2408.01393} {arXiv:2408.01393 [quant-ph]} \BibitemShut {NoStop}\bibitem [{\citenamefont {Veroni}\ \emph {et~al.}(2024)\citenamefont {Veroni}, \citenamefont {M{\"u}ller},\ and\ \citenamefont {Giudice}}]{veroni2024optimized}\BibitemOpen
	\bibfield  {author} {\bibinfo {author} {S.~Veroni}, \bibinfo {author} {M.~M{\"u}ller},\ and\ \bibinfo {author} {G.~Giudice},\ }\emph {Optimized measurement-free and fault-tolerant quantum error correction for neutral atoms},\ \href {https://doi.org/10.1103/PhysRevResearch.6.043253} {\bibfield  {journal} {\bibinfo  {journal} {Physical Review Research}\ }\textbf {\bibinfo {volume} {6}},\ \bibinfo {pages} {043253} (\bibinfo {year} {2024})}\BibitemShut {NoStop}\bibitem [{\citenamefont {Jain}\ and\ \citenamefont {Albert}(2025)}]{jain2024highdistancecodestransversalclifford}\BibitemOpen
	\bibfield  {author} {\bibinfo {author} {S.~P. Jain}\ and\ \bibinfo {author} {V.~V. Albert},\ }\emph {Transversal Clifford and T-Gate codes of short length and high distance},\ \href {https://doi.org/10.1109/jsait.2025.3570832} {\bibfield  {journal} {\bibinfo  {journal} {IEEE Journal on Selected Areas in Information Theory}\ }\textbf {\bibinfo {volume} {6}},\ \bibinfo {pages} {127} (\bibinfo {year} {2025})}\BibitemShut {NoStop}\bibitem [{\citenamefont {Zhu}\ \emph {et~al.}(2023{\natexlab{b}})\citenamefont {Zhu}, \citenamefont {Sikander}, \citenamefont {Portnoy}, \citenamefont {Cross},\ and\ \citenamefont {Brown}}]{zhu2023noncliffordparallelizablefaulttolerantlogical}\BibitemOpen
	\bibfield  {author} {\bibinfo {author} {G.~Zhu}, \bibinfo {author} {S.~Sikander}, \bibinfo {author} {E.~Portnoy}, \bibinfo {author} {A.~W. Cross},\ and\ \bibinfo {author} {B.~J. Brown},\ }\href@noop {} {\emph {Non-Clifford and parallelizable fault-tolerant logical gates on constant and almost-constant rate homological quantum LDPC codes via higher symmetries}} (\bibinfo {year} {2023}{\natexlab{b}}),\ \Eprint {https://arxiv.org/abs/2310.16982} {arXiv:2310.16982 [quant-ph]} \BibitemShut {NoStop}\bibitem [{\citenamefont {Old}\ \emph {et~al.}(2024)\citenamefont {Old}, \citenamefont {Rispler},\ and\ \citenamefont {Müller}}]{Old_2024}\BibitemOpen
	\bibfield  {author} {\bibinfo {author} {J.~Old}, \bibinfo {author} {M.~Rispler},\ and\ \bibinfo {author} {M.~Müller},\ }\emph {Lift-connected surface codes},\ \href {https://doi.org/10.1088/2058-9565/ad5eb6} {\bibfield  {journal} {\bibinfo  {journal} {Quantum Science and Technology}\ }\textbf {\bibinfo {volume} {9}} (\bibinfo {year} {2024})}\BibitemShut {NoStop}\bibitem [{\citenamefont {Huang}\ \emph {et~al.}(2022)\citenamefont {Huang}, \citenamefont {Jochym-O'Connor},\ and\ \citenamefont {Yoder}}]{huang2022homomorphiclogicalmeasurements}\BibitemOpen
	\bibfield  {author} {\bibinfo {author} {S.~Huang}, \bibinfo {author} {T.~Jochym-O'Connor},\ and\ \bibinfo {author} {T.~J. Yoder},\ }\href@noop {} {\emph {Homomorphic logical measurements}} (\bibinfo {year} {2022}),\ \Eprint {https://arxiv.org/abs/2211.03625} {arXiv:2211.03625 [quant-ph]} \BibitemShut {NoStop}\bibitem [{\citenamefont {Cross}\ \emph {et~al.}(2024)\citenamefont {Cross}, \citenamefont {He}, \citenamefont {Rall},\ and\ \citenamefont {Yoder}}]{cross2024linearsizeancillasystemslogical}\BibitemOpen
	\bibfield  {author} {\bibinfo {author} {A.~Cross}, \bibinfo {author} {Z.~He}, \bibinfo {author} {P.~Rall},\ and\ \bibinfo {author} {T.~Yoder},\ }\href@noop {} {\emph {Linear-size ancilla systems for logical measurements in qLDPC codes}} (\bibinfo {year} {2024}),\ \Eprint {https://arxiv.org/abs/2407.18393} {arXiv:2407.18393 [quant-ph]} \BibitemShut {NoStop}\bibitem [{\citenamefont {Xu}\ \emph {et~al.}(2024)\citenamefont {Xu}, \citenamefont {Zhou}, \citenamefont {Zheng}, \citenamefont {Bluvstein}, \citenamefont {Ataides}, \citenamefont {Lukin},\ and\ \citenamefont {Jiang}}]{xu2024fastparallelizablelogicalcomputation}\BibitemOpen
	\bibfield  {author} {\bibinfo {author} {Q.~Xu}, et~al.,\ }\href@noop {} {\emph {Fast and parallelizable logical computation with homological product codes}} (\bibinfo {year} {2024}),\ \Eprint {https://arxiv.org/abs/2407.18490} {arXiv:2407.18490 [quant-ph]} \BibitemShut {NoStop}\bibitem [{\citenamefont {Paetznick}\ and\ \citenamefont {Reichardt}(2013{\natexlab{b}})}]{paetznick2013faulttolerant}\BibitemOpen
	\bibfield  {author} {\bibinfo {author} {A.~Paetznick}\ and\ \bibinfo {author} {B.~W. Reichardt},\ }\href@noop {} {\emph {Fault-tolerant ancilla preparation and noise threshold lower bounds for the 23-qubit Golay code}} (\bibinfo {year} {2013}{\natexlab{b}}),\ \Eprint {https://arxiv.org/abs/1106.2190} {arXiv:1106.2190 [quant-ph]} \BibitemShut {NoStop}\bibitem [{\citenamefont {Kielpinski}\ \emph {et~al.}(2002)\citenamefont {Kielpinski}, \citenamefont {Monroe},\ and\ \citenamefont {Wineland}}]{kielpinski2002architecture}\BibitemOpen
	\bibfield  {author} {\bibinfo {author} {D.~Kielpinski}, \bibinfo {author} {C.~Monroe},\ and\ \bibinfo {author} {D.~J. Wineland},\ }\emph {Architecture for a large-scale ion-trap quantum computer},\ \href {https://doi.org/10.1038/nature00784} {\bibfield  {journal} {\bibinfo  {journal} {Nature}\ }\textbf {\bibinfo {volume} {417}},\ \bibinfo {pages} {709} (\bibinfo {year} {2002})}\BibitemShut {NoStop}\bibitem [{\citenamefont {Durandau}\ \emph {et~al.}(2023)\citenamefont {Durandau}, \citenamefont {Wagner}, \citenamefont {Mailhot}, \citenamefont {Brunet}, \citenamefont {Schmidt-Kaler}, \citenamefont {Poschinger},\ and\ \citenamefont {B{\'{e}}rub{\'{e}}-Lauzi{\`{e}}re}}]{Durandau2023automatedgeneration}\BibitemOpen
	\bibfield  {author} {\bibinfo {author} {J.~Durandau}, et~al.,\ }\emph {Automated generation of shuttling sequences for a linear segmented ion trap quantum computer},\ \href {https://doi.org/10.22331/q-2023-11-08-1175} {\bibfield  {journal} {\bibinfo  {journal} {{Quantum}}\ }\textbf {\bibinfo {volume} {7}},\ \bibinfo {pages} {1175} (\bibinfo {year} {2023})}\BibitemShut {NoStop}\bibitem [{\citenamefont {Schoenberger}\ \emph {et~al.}(2024{\natexlab{b}})\citenamefont {Schoenberger}, \citenamefont {Hillmich}, \citenamefont {Brandl},\ and\ \citenamefont {Wille}}]{schoenberger2024shuttlingscalabletrappedionquantum}\BibitemOpen
	\bibfield  {author} {\bibinfo {author} {D.~Schoenberger}, \bibinfo {author} {S.~Hillmich}, \bibinfo {author} {M.~Brandl},\ and\ \bibinfo {author} {R.~Wille},\ }\href@noop {} {\emph {Shuttling for scalable trapped-ion quantum computers}} (\bibinfo {year} {2024}{\natexlab{b}}),\ \Eprint {https://arxiv.org/abs/2402.14065} {arXiv:2402.14065 [quant-ph]} \BibitemShut {NoStop}\bibitem [{\citenamefont {Kaufmann}\ \emph {et~al.}(2017)\citenamefont {Kaufmann}, \citenamefont {Ruster}, \citenamefont {Schmiegelow}, \citenamefont {Luda}, \citenamefont {Kaushal}, \citenamefont {Schulz}, \citenamefont {von Lindenfels}, \citenamefont {Schmidt-Kaler},\ and\ \citenamefont {Poschinger}}]{PhysRevA.95.052319}\BibitemOpen
	\bibfield  {author} {\bibinfo {author} {H.~Kaufmann}, et~al.,\ }\emph {Fast ion swapping for quantum-information processing},\ \href {https://doi.org/10.1103/PhysRevA.95.052319} {\bibfield  {journal} {\bibinfo  {journal} {Physical Review A}\ }\textbf {\bibinfo {volume} {95}},\ \bibinfo {pages} {052319} (\bibinfo {year} {2017})}\BibitemShut {NoStop}\bibitem [{\citenamefont {Palmero}\ \emph {et~al.}(2015)\citenamefont {Palmero}, \citenamefont {Martínez-Garaot}, \citenamefont {Poschinger}, \citenamefont {Ruschhaupt},\ and\ \citenamefont {Muga}}]{Palmero_2015}\BibitemOpen
	\bibfield  {author} {\bibinfo {author} {M.~Palmero}, \bibinfo {author} {S.~Martínez-Garaot}, \bibinfo {author} {U.~G. Poschinger}, \bibinfo {author} {A.~Ruschhaupt},\ and\ \bibinfo {author} {J.~G. Muga},\ }\emph {Fast separation of two trapped ions},\ \href {https://doi.org/10.1088/1367-2630/17/9/093031} {\bibfield  {journal} {\bibinfo  {journal} {New Journal of Physics}\ }\textbf {\bibinfo {volume} {17}},\ \bibinfo {pages} {093031} (\bibinfo {year} {2015})}\BibitemShut {NoStop}\bibitem [{\citenamefont {Ruster}\ \emph {et~al.}(2014)\citenamefont {Ruster}, \citenamefont {Warschburger}, \citenamefont {Kaufmann}, \citenamefont {Schmiegelow}, \citenamefont {Walther}, \citenamefont {Hettrich}, \citenamefont {Pfister}, \citenamefont {Kaushal}, \citenamefont {Schmidt-Kaler},\ and\ \citenamefont {Poschinger}}]{PhysRevA.90.033410}\BibitemOpen
	\bibfield  {author} {\bibinfo {author} {T.~Ruster}, et~al.,\ }\emph {Experimental realization of fast ion separation in segmented Paul traps},\ \href {https://doi.org/10.1103/PhysRevA.90.033410} {\bibfield  {journal} {\bibinfo  {journal} {Physical Review A}\ }\textbf {\bibinfo {volume} {90}},\ \bibinfo {pages} {033410} (\bibinfo {year} {2014})}\BibitemShut {NoStop}\bibitem [{\citenamefont {Walther}\ \emph {et~al.}(2012)\citenamefont {Walther}, \citenamefont {Ziesel}, \citenamefont {Ruster}, \citenamefont {Dawkins}, \citenamefont {Ott}, \citenamefont {Hettrich}, \citenamefont {Singer}, \citenamefont {Schmidt-Kaler},\ and\ \citenamefont {Poschinger}}]{PhysRevLett.109.080501}\BibitemOpen
	\bibfield  {author} {\bibinfo {author} {A.~Walther}, et~al.,\ }\emph {Controlling fast transport of cold trapped ions},\ \href {https://doi.org/10.1103/PhysRevLett.109.080501} {\bibfield  {journal} {\bibinfo  {journal} {Physical Review Letters}\ }\textbf {\bibinfo {volume} {109}},\ \bibinfo {pages} {080501} (\bibinfo {year} {2012})}\BibitemShut {NoStop}\bibitem [{\citenamefont {Paz-Silva}\ \emph {et~al.}(2010)\citenamefont {Paz-Silva}, \citenamefont {Brennen},\ and\ \citenamefont {Twamley}}]{pazsilva2010fault}\BibitemOpen
	\bibfield  {author} {\bibinfo {author} {G.~A. Paz-Silva}, \bibinfo {author} {G.~K. Brennen},\ and\ \bibinfo {author} {J.~Twamley},\ }\emph {Fault tolerance with noisy and slow measurements and preparation},\ \href {https://doi.org/10.1103/PhysRevLett.105.100501} {\bibfield  {journal} {\bibinfo  {journal} {Physical Review Letters}\ }\textbf {\bibinfo {volume} {105}},\ \bibinfo {pages} {100501} (\bibinfo {year} {2010})}\BibitemShut {NoStop}\bibitem [{\citenamefont {Heu\ss{}en}\ \emph {et~al.}(2024)\citenamefont {Heu\ss{}en}, \citenamefont {Locher},\ and\ \citenamefont {M\"uller}}]{heussen2024measurementfree}\BibitemOpen
	\bibfield  {author} {\bibinfo {author} {S.~Heu\ss{}en}, \bibinfo {author} {D.~F. Locher},\ and\ \bibinfo {author} {M.~M\"uller},\ }\emph {Measurement-free fault-tolerant quantum error correction in near-term devices},\ \href {https://doi.org/10.1103/PRXQuantum.5.010333} {\bibfield  {journal} {\bibinfo  {journal} {PRX Quantum}\ }\textbf {\bibinfo {volume} {5}},\ \bibinfo {pages} {010333} (\bibinfo {year} {2024})}\BibitemShut {NoStop}\bibitem [{\citenamefont {Veroni}\ \emph {et~al.}(2025)\citenamefont {Veroni}, \citenamefont {Paler},\ and\ \citenamefont {Giudice}}]{veroni2025universalquantumcomputationscalable}\BibitemOpen
	\bibfield  {author} {\bibinfo {author} {S.~Veroni}, \bibinfo {author} {A.~Paler},\ and\ \bibinfo {author} {G.~Giudice},\ }\href {https://arxiv.org/abs/2412.15187} {\emph {Universal quantum computation via scalable measurement-free error correction}} (\bibinfo {year} {2025}),\ \Eprint {https://arxiv.org/abs/2412.15187} {arXiv:2412.15187 [quant-ph]} \BibitemShut {NoStop}\bibitem [{\citenamefont {Butt}\ \emph {et~al.}(2025{\natexlab{a}})\citenamefont {Butt}, \citenamefont {Locher}, \citenamefont {Brechtelsbauer}, \citenamefont {B{\"u}chler},\ and\ \citenamefont {M{\"u}ller}}]{butt2025measurement}\BibitemOpen
	\bibfield  {author} {\bibinfo {author} {F.~Butt}, \bibinfo {author} {D.~F. Locher}, \bibinfo {author} {K.~Brechtelsbauer}, \bibinfo {author} {H.~P. B{\"u}chler},\ and\ \bibinfo {author} {M.~M{\"u}ller},\ }\emph {Measurement-free, scalable, and fault-tolerant universal quantum computing},\ \href {https://doi.org/10.1126/sciadv.adv2590} {\bibfield  {journal} {\bibinfo  {journal} {Science Advances}\ }\textbf {\bibinfo {volume} {11}},\ \bibinfo {pages} {eadv2590} (\bibinfo {year} {2025}{\natexlab{a}})}\BibitemShut {NoStop}\bibitem [{\citenamefont {Butt}\ \emph {et~al.}(2025{\natexlab{b}})\citenamefont {Butt}, \citenamefont {Pogorelov}, \citenamefont {Freund}, \citenamefont {Steiner}, \citenamefont {Meyer}, \citenamefont {Monz},\ and\ \citenamefont {Müller}}]{butt2025demonstrationmeasurementfreeuniversalfaulttolerant}\BibitemOpen
	\bibfield  {author} {\bibinfo {author} {F.~Butt}, et~al.,\ }\href {https://arxiv.org/abs/2506.22600} {\emph {Demonstration of measurement-free universal fault-tolerant quantum computation}} (\bibinfo {year} {2025}{\natexlab{b}}),\ \Eprint {https://arxiv.org/abs/2506.22600} {arXiv:2506.22600 [quant-ph]} \BibitemShut {NoStop}\bibitem [{\citenamefont {Brechtelsbauer}\ \emph {et~al.}(2025)\citenamefont {Brechtelsbauer}, \citenamefont {Butt}, \citenamefont {Locher}, \citenamefont {Quintero}, \citenamefont {Weber}, \citenamefont {Müller},\ and\ \citenamefont {Büchler}}]{brechtelsbauer2025measurementfreequantumerrorcorrection}\BibitemOpen
	\bibfield  {author} {\bibinfo {author} {K.~Brechtelsbauer}, et~al.,\ }\href {https://arxiv.org/abs/2505.15669} {\emph {Measurement-free quantum error correction optimized for biased noise}} (\bibinfo {year} {2025}),\ \Eprint {https://arxiv.org/abs/2505.15669} {arXiv:2505.15669 [quant-ph]} \BibitemShut {NoStop}\end{thebibliography}

\appendix

\clearpage
\section{Physical quantum circuits}\label{sec:circs}
\begin{figure}[!h]
    \centering
	\includegraphics[width=0.99\linewidth]{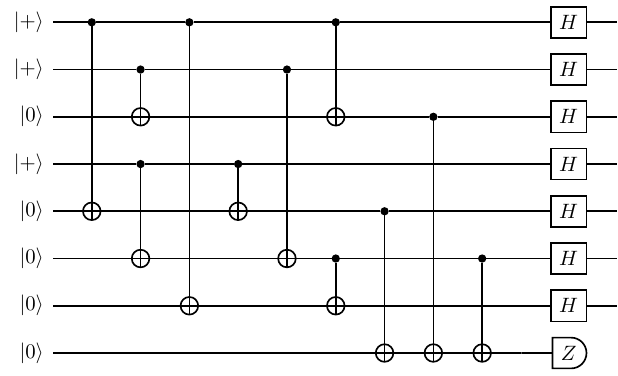}
	\caption{Flag-FT circuit to prepare the logical state $\ket{+}_{\mathrm{S}}$ of the Steane code \cite{goto2016minimizing}. After the first eight CNOT gates, the state $\ket{0}_{\mathrm{S}}$ is prepared non-fault-tolerantly on the upper seven qubits. The last three CNOTs are used to measure the eigenvalue of the logical $Z$-operator $Z_2Z_4Z_5$, with the help of a physical auxiliary qubit, which acts as a flag. If this last qubit is measured as $+1$, the state $\ket{0}_{\mathrm{S}}$ is prepared fault-tolerantly. Otherwise it is discarded and the circuit is repeated. The last layer of physical $H$-gates implements the logical $H$-gate in the Steane code. We assume that these eight qubits can be fit into a single ion trap segment and that entangling gates can be performed between any pair of qubits.}
	\label{fig:init7}
\end{figure}

\begin{figure*}[!h]
	\centering
	\includegraphics[width=0.75\linewidth]{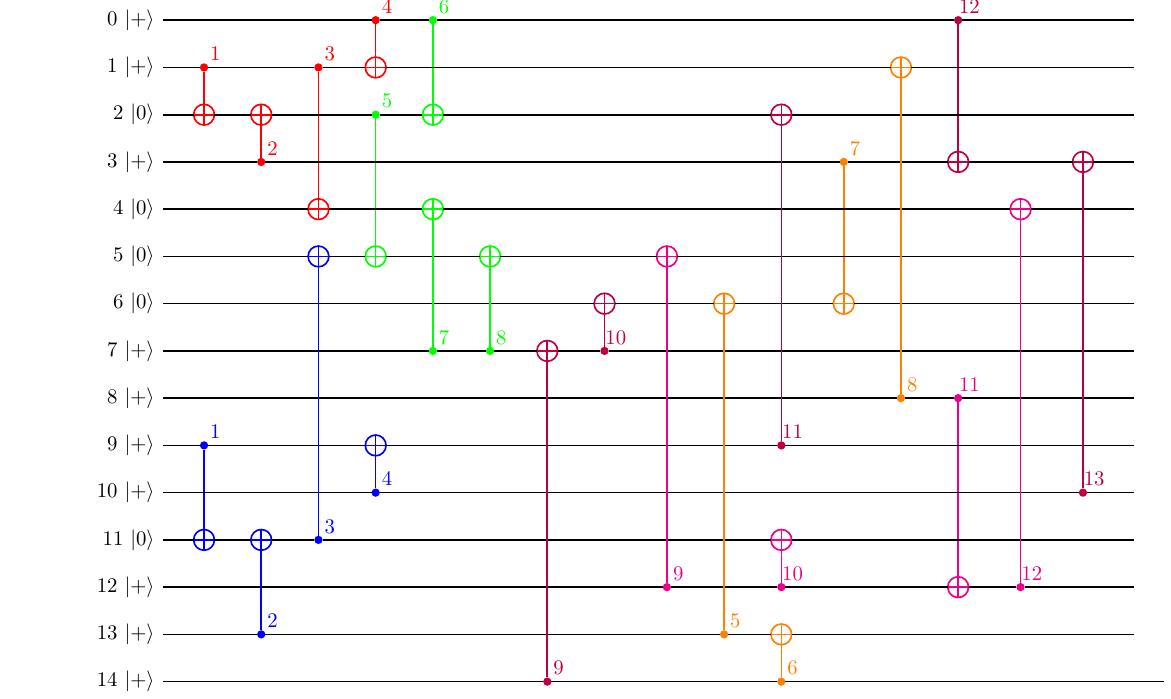}
	\caption{Circuit to non-fault-tolerantly encode the logical zero state of the Tetrahedral code defined by the stabilizer generators in Eqs.~\eqref{eq:tetraxstabs} and \eqref{eq:tetrazstabs} with 25 CNOT gates based on Ref.~\cite{butt2024fault}. Coloring of the CNOT gates indicates parallelizability in two linear ion traps with all-to-all connectivity within each ion crystal. Red and blue gates can be executed in parallel, so can green and orange gates as well as purple and magenta gates (see Fig.~\ref{fig:trap15}). Ion reconfiguration needs to be performed in between. Colored numbers indicate the respective order of CNOT gates in each trap.}
	\label{fig:init15}
\end{figure*}

\begin{figure*}[!h]
	\centering
	\includegraphics[width=0.99\linewidth]{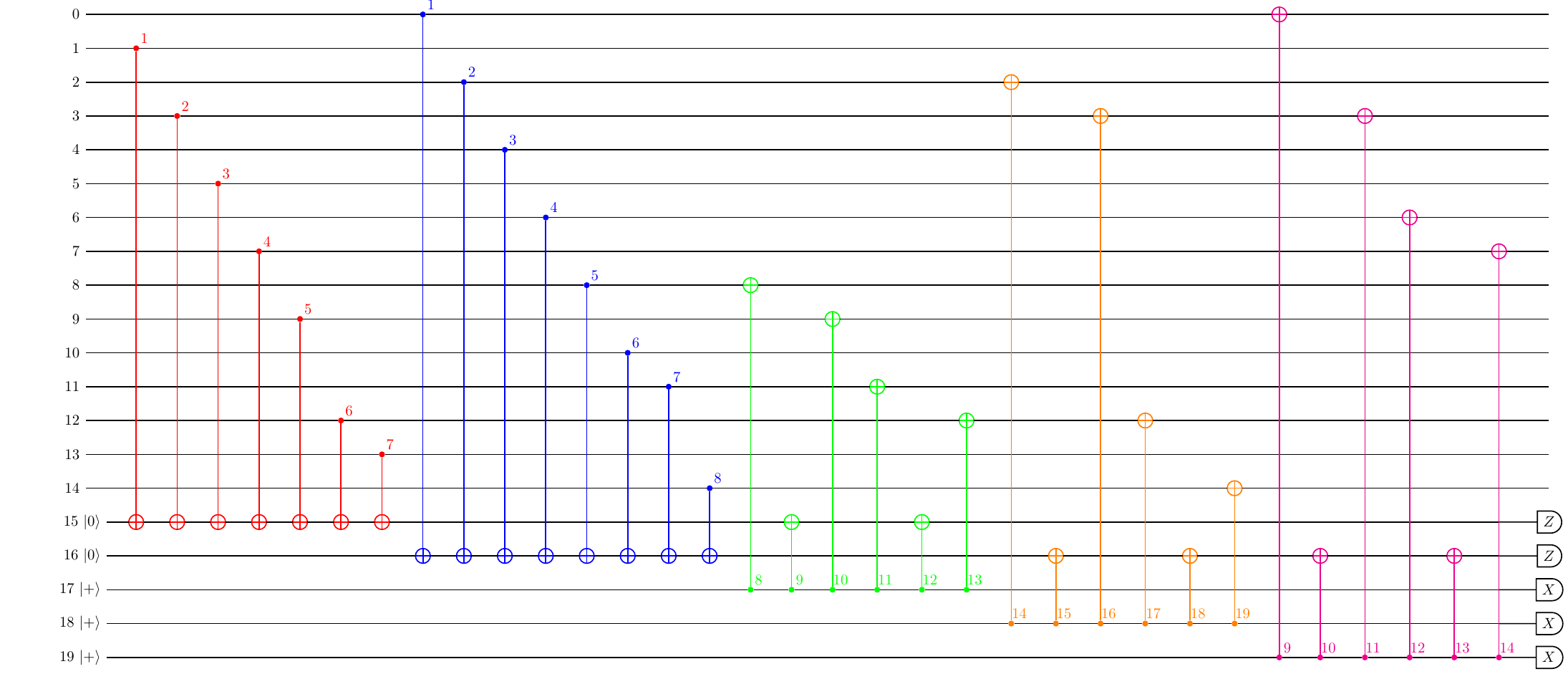}
	\caption{Flag verification circuit consisting of 33 physical CNOT gates that renders the $\ket{0}_{\mathrm{T}}$ state preparation subroutine suitable for fault-tolerant code switching with five auxiliary qubits. This circuit is applied after the one in Fig.~\ref{fig:init15}. A prepared state is accepted if all auxiliary qubits are measured as $+1$ and discarded otherwise. Note that the $Z$-type operator measurements do not require additional flag qubits because only $Z$-errors could propagate from the auxiliary qubit to the data qubits. The $X$-type operator measurements need an additional flag mechanism to sort out propagating high-weight $X$-errors. Colors again indicate parallel execution in two separate ion traps. Red and blue gates can be executed in parallel as well as green and magenta gates. Ion crystals need to be reconfigured in between. Colored numbers indicate the respective order of CNOT gates in each trap. To correctly flag all dangerous errors for the ion mapping given in Figs.~\ref{fig:init15} and \ref{fig:trap15}, the last (magenta) flag that measures the operator $X_0X_3X_6X_7$ needs to be replaced to instead measure $X_2X_3X_5X_6$.}
	\label{fig:flags15}
\end{figure*}

\begin{figure}[!h]
	\centering
	\includegraphics[width=0.99\linewidth]{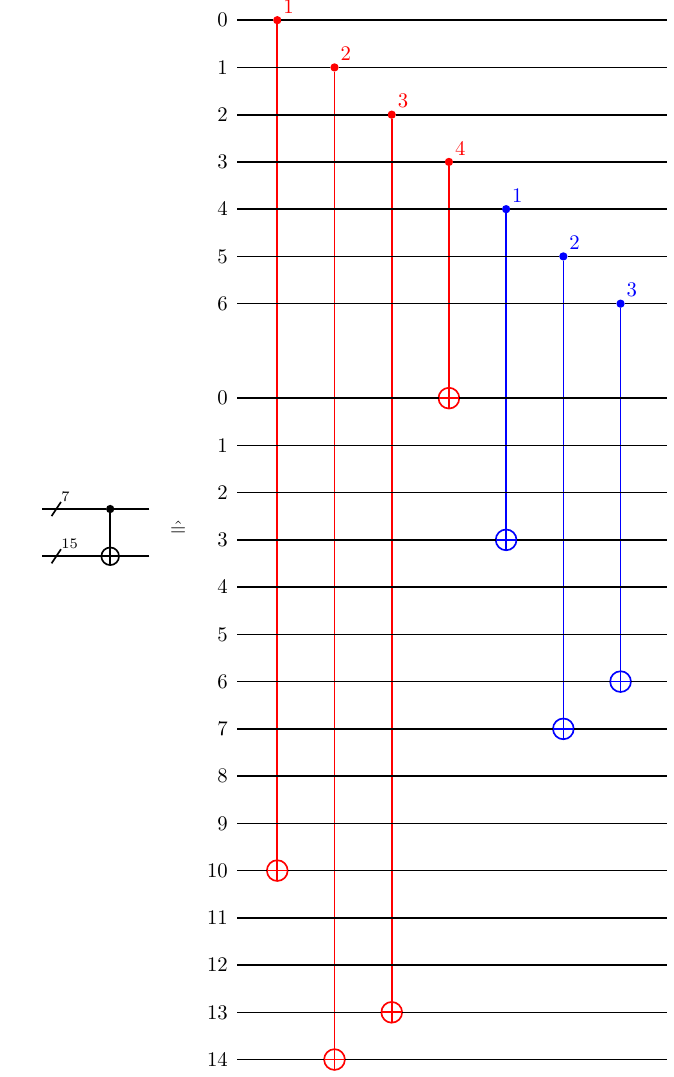}
	\caption{Physical CNOT gates to implement the one-way transversal logical CNOT gate given in Sec.~\ref{sec:colors}. One possible arrangement of physical qubits into two different ion trap segments is indicated by the coloring. Gate sequences as marked by colored numbers can be applied simultaneously within each zone.}
	\label{fig:tv}
\end{figure}

\begin{figure*}[!h]
	\centering
	\includegraphics[width=0.99\linewidth]{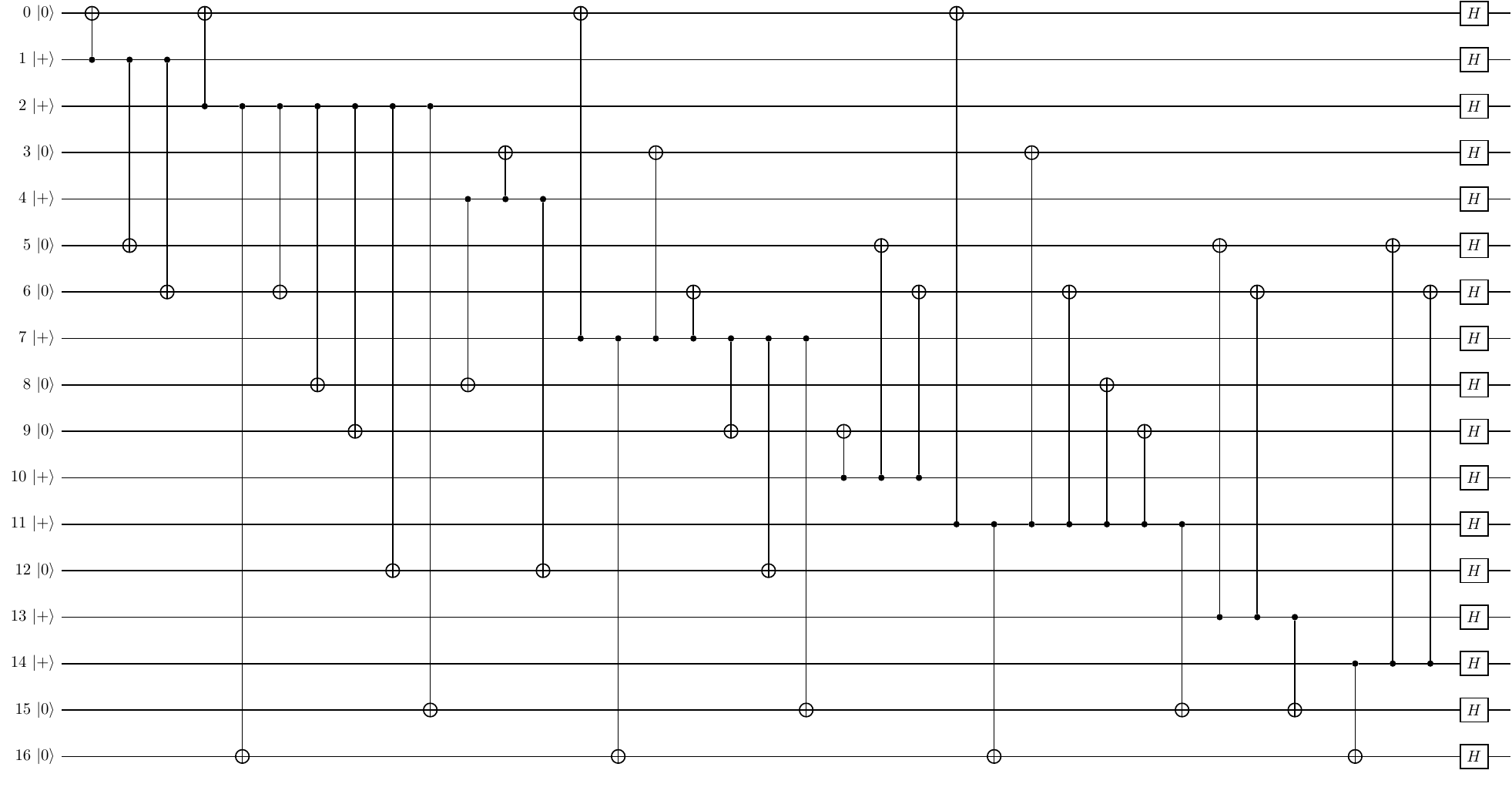}
	\caption{Circuit to non-fault-tolerantly encode the logical plus state of the $[[17,1,5]]$ code, an instance of the 4.8.8 2D color code, with 36 CNOT gates.}
	\label{fig:init17}
\end{figure*}

\begin{figure*}[!h]
	\centering
	\includegraphics[width=0.99\linewidth]{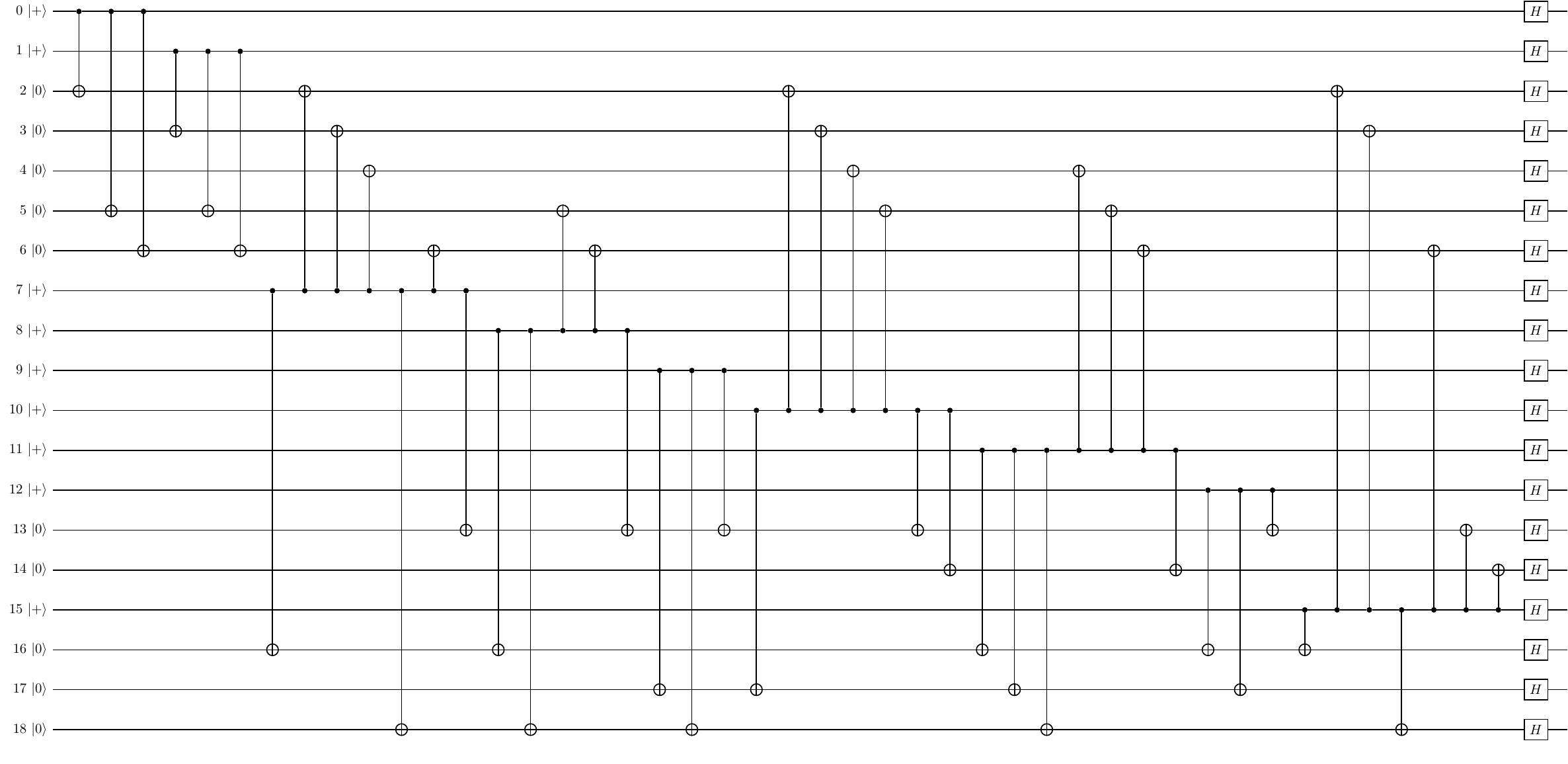}
	\caption{Circuit to non-fault-tolerantly encode the logical plus state of the $[[19,1,5]]$ code, an instance of the 6.6.6 2D color code, with 45 CNOT gates.}
	\label{fig:init19}
\end{figure*}

\begin{figure*}[!h]
	\centering
	\includegraphics[width=0.99\linewidth]{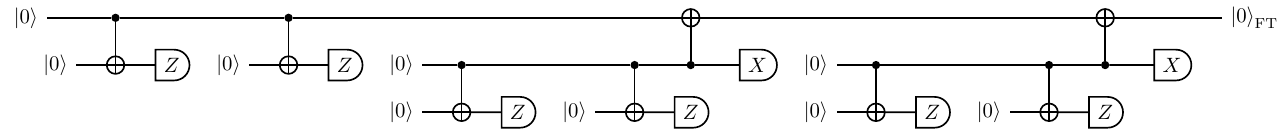}
	\caption{Circuit on the logical level for Steane-type FT state preparation with a distance-5 CSS code that can correct $t=2$ errors. $\ket{0}$ indicates non-FT preparation of the logical zero state. Logical operators of each type $X$ or $Z$ after the first non-FT state preparation propagate to the logical auxiliary qubits. Such errors in the non-FT logical auxiliary qubit state preparations are heralded when propagating to the upper wire by the measurements. Any individual CNOT + measurement is therefore repeated $t$ times. The state is accepted if all $Z$-measurements indicate a trivial $Z$-syndrome and the correct logical state and all $X$-measurements indicate a trivial $X$-syndrome. The state preparation subroutine is aborted and repeated as soon as a single measurement heralds a potential error. Due to transversality of the CNOT gates and measurements, Steane-type state preparation is strictly FT. The corresponding circuit for FT preparation of the logical plus state is obtained by exchanging $X$- and $Z$-type operations and reversing the directions of the CNOT gates. An attempt to optimize the scheme was given, for instance, in Ref.~\cite{paetznick2013faulttolerant}.}
	\label{fig:steane_verification}
\end{figure*}

The subcircuits of fault-tolerant transversal code switching used for logical auxiliary state preparation and implementation of the one-way logical CNOT gate are shown in Figs.~\ref{fig:init7}-\ref{fig:steane_verification}. 

\begin{table*}[!h]    
\begin{center}
\begin{tabular}{c || c | c | c |c}
    $p$ & $10^{-4}$ & $3 \times 10^{-4}$ & $10^{-3}$ & $3 \times 10^{-3}$ \\ \hline
    \rule{0pt}{2.5ex} $[[17,1,5]], \ket{0}$ & 2.2 & 2.5 & 4.0 & 15.9 \\
    \rule{0pt}{2.5ex} $[[49,1,5]], \ket{+}$ & 1.3 & 2.2 & 12.8 & 2098.2 \\ \hline
    \rule{0pt}{2.5ex} $[[19,1,5]], \ket{0}$ & 2.2 & 2.5 & 4.5 & 16.8 \\
    \rule{0pt}{2.5ex} $[[65,1,5]], \ket{+}$ & 1.5 & 3.5 & 64.9 & 2022.2
\end{tabular}
\end{center}
\caption{Average number of repetitions until the respective Pauli states are accepted with Steane-type state preparation in the single-parameter noise model for the distance-5 codes discussed in Sec.~\ref{sec:scaleup}. The required number of repetitions quickly surges for the 3D codes when the depolarizing error rate $p$ is increased. A more efficient FT state preparation subroutine should be employed for a practical application.}
\label{tab:acc_reps}
\end{table*}

The parity check matrix to construct the $[[17,1,5]]$ code is
\setcounter{MaxMatrixCols}{50}
\setlength\arraycolsep{2pt}
\begin{align}
    H_{17} = \begin{bmatrix} 
    1 & 1 & 0 & 0 & 0 & 1 & 1 & 0 & 0 & 0 & 0 & 0 & 0 & 0 & 0 & 0 & 0 \\
    0 & 0 & 0 & 0 & 0 & 1 & 1 & 0 & 0 & 1 & 1 & 0 & 0 & 0 & 0 & 0 & 0 \\
    0 & 0 & 0 & 0 & 0 & 0 & 0 & 0 & 0 & 1 & 1 & 0 & 0 & 1 & 0 & 1 & 0 \\
    0 & 0 & 0 & 0 & 0 & 0 & 0 & 0 & 0 & 0 & 0 & 0 & 0 & 1 & 1 & 1 & 1 \\
    0 & 1 & 1 & 0 & 0 & 0 & 1 & 1 & 0 & 0 & 1 & 1 & 0 & 1 & 1 & 0 & 0 \\
    0 & 0 & 1 & 1 & 0 & 0 & 0 & 1 & 1 & 0 & 0 & 0 & 0 & 0 & 0 & 0 & 0 \\
    0 & 0 & 0 & 0 & 0 & 0 & 0 & 1 & 1 & 0 & 0 & 1 & 1 & 0 & 0 & 0 & 0 \\
    0 & 0 & 0 & 1 & 1 & 0 & 0 & 0 & 1 & 0 & 0 & 0 & 1 & 0 & 0 & 0 & 0
    \end{bmatrix}
\end{align}
and the parity check matrix to construct the $[[19,1,5]]$ code is
\begin{align}
    H_{19} = \begin{bmatrix}
    1 & 1 & 1 & 1 & 0 & 0 & 0 & 0 & 0 & 0 & 0 & 0 & 0 & 0 & 0 & 0 & 0 & 0 & 0 \\
    0 & 0 & 1 & 1 & 1 & 1 & 0 & 1 & 1 & 0 & 0 & 0 & 0 & 0 & 0 & 0 & 0 & 0 & 0 \\
    0 & 1 & 0 & 1 & 0 & 1 & 1 & 0 & 0 & 0 & 0 & 0 & 0 & 0 & 0 & 0 & 0 & 0 & 0 \\
    0 & 0 & 0 & 0 & 0 & 1 & 1 & 0 & 1 & 1 & 0 & 0 & 1 & 1 & 0 & 0 & 0 & 0 & 0 \\
    0 & 0 & 0 & 0 & 1 & 0 & 0 & 1 & 0 & 0 & 1 & 1 & 0 & 0 & 0 & 0 & 0 & 0 & 0 \\
    0 & 0 & 0 & 0 & 0 & 0 & 0 & 1 & 1 & 0 & 0 & 1 & 1 & 0 & 0 & 1 & 1 & 0 & 0 \\
    0 & 0 & 0 & 0 & 0 & 0 & 0 & 0 & 0 & 0 & 1 & 1 & 0 & 0 & 1 & 1 & 0 & 0 & 0 \\
    0 & 0 & 0 & 0 & 0 & 0 & 0 & 0 & 0 & 0 & 0 & 0 & 1 & 1 & 0 & 0 & 1 & 1 & 0 \\
    0 & 0 & 0 & 0 & 0 & 0 & 0 & 0 & 0 & 1 & 0 & 0 & 0 & 1 & 0 & 0 & 0 & 1 & 1
    \end{bmatrix}
\end{align}

Due to their large size, we provide the stabilizers and encoding circuits of the 3D distance-5 codes in a more compact list format below in order to ease reproducing our results. 
\subsection{$[[49,1,5]]$ code} The 13 $X$-stabilizer generators have their non-trivial support on qubits
\texttt{[[ 0,  1,  5,  6, 17, 18, 22, 23],
 [ 5,  6,  9, 10, 22, 23, 26, 27],
 [ 9, 10, 13, 15, 26, 27, 30, 32],
 [13, 14, 15, 16, 30, 31, 32, 33],
 [ 1,  2,  6,  7, 10, 11, 13, 14, 18, 19, 23, 24, 27, 28, 30, 31],
 [ 2,  3,  7,  8, 19, 20, 24, 25],
 [ 7,  8, 11, 12, 24, 25, 28, 29],
 [ 3,  4,  8, 12, 20, 21, 25, 29],
 [17, 18, 19, 20, 21, 22, 23, 24, 25, 26, 27, 28, 29, 30, 31, 32, 33, 34, 35, 36, 37, 38, 39, 40, 41, 42, 43, 44, 45, 46, 47, 48],
 [34, 36, 38, 40, 42, 44, 46, 48],
 [35, 36, 39, 40, 43, 44, 47, 48],
 [37, 38, 39, 40, 45, 46, 47, 48],
 [41, 42, 43, 44, 45, 46, 47, 48]]}
and the 35 $Z$-stabilizer generators are indexed by
\texttt{[[ 0,  1,  5,  6, 17, 18, 22, 23],
 [ 5,  6,  9, 10, 22, 23, 26, 27],
 [ 9, 10, 13, 15, 26, 27, 30, 32],
 [13, 14, 15, 16, 30, 31, 32, 33],
 [ 1,  2,  6,  7, 10, 11, 13, 14, 18, 19, 23, 24, 27, 28, 30, 31],
 [ 2,  3,  7,  8, 19, 20, 24, 25],
 [ 7,  8, 11, 12, 24, 25, 28, 29],
 [ 3,  4,  8, 12, 20, 21, 25, 29],
 [17, 18, 19, 20, 21, 22, 23, 24, 25, 26, 27, 28, 29, 30, 31, 32, 33, 34, 35, 36, 37, 38, 39, 40, 41, 42, 43, 44, 45, 46, 47, 48],
 [34, 36, 38, 40, 42, 44, 46, 48],
 [35, 36, 39, 40, 43, 44, 47, 48],
 [37, 38, 39, 40, 45, 46, 47, 48],
 [41, 42, 43, 44, 45, 46, 47, 48],
 [0, 1, 5, 6],
 [ 5,  6,  9, 10],
 [ 9, 10, 13, 15],
 [13, 14, 15, 16],
 [ 1,  2,  6,  7, 10, 11, 13, 14],
 [2, 3, 7, 8],
 [ 7,  8, 11, 12],
 [ 3,  4,  8, 12],
 [ 5,  6, 22, 23],
 [ 1,  6, 18, 23],
 [ 6, 10, 23, 27],
 [10, 13, 27, 30],
 [13, 14, 30, 31],
 [ 2,  7, 19, 24],
 [ 7, 11, 24, 28],
 [ 7,  8, 24, 25],
 [36, 40, 44, 48],
 [38, 40, 46, 48],
 [42, 44, 46, 48],
 [39, 40, 47, 48],
 [43, 44, 47, 48],
 [45, 46, 47, 48]]}.
 
The logical $\ket{+}$ state of the $[[49,1,5]]$ code can be prepared, starting from the state $\ket{0}^{\otimes 49}$, with the gate sequence \texttt{[{'H': [1, 3, 4, 6, 39, 40, 41, 43, 47, 16, 20, 22, 25, 27]}, {'CNOT': [(1, 0), (3, 0), (20, 0), (6, 0), (22, 0), (3, 2), (4, 2), (6, 2), (39, 2), (22, 2), (41, 2), (47, 2), (3, 5), (20, 5), (22, 5), (3, 7), (4, 7), (6, 7), (39, 7), (41, 7), (47, 7), (20, 7), (22, 7), (25, 7), (25, 8), (3, 8), (20, 8), (27, 9), (6, 9), (22, 9), (3, 10), (27, 10), (20, 10), (20, 11), (6, 11), (39, 11), (22, 11), (41, 11), (25, 11), (47, 11), (25, 12), (4, 12), (20, 12), (16, 13), (1, 13), (6, 13), (39, 13), (41, 13), (27, 13), (47, 13), (16, 14), (3, 14), (20, 14), (6, 14), (22, 14), (1, 15), (3, 15), (39, 15), (41, 15), (47, 15), (16, 15), (20, 15), (22, 15), (27, 15), (1, 17), (6, 17), (22, 17), (1, 18), (3, 18), (20, 18), (4, 19), (20, 19), (6, 19), (39, 19), (22, 19), (41, 19), (47, 19), (3, 21), (4, 21), (20, 21), (3, 23), (20, 23), (6, 23), (4, 24), (6, 24), (39, 24), (22, 24), (41, 24), (25, 24), (47, 24), (3, 26), (20, 26), (6, 26), (22, 26), (27, 26), (3, 28), (6, 28), (39, 28), (22, 28), (41, 28), (25, 28), (47, 28), (25, 29), (3, 29), (4, 29), (1, 30), (3, 30), (6, 30), (39, 30), (41, 30), (47, 30), (16, 30), (20, 30), (27, 30), (16, 31), (6, 31), (22, 31), (16, 32), (1, 32), (22, 32), (39, 32), (41, 32), (27, 32), (47, 32), (16, 33), (3, 33), (20, 33), (40, 34), (41, 34), (47, 34), (43, 35), (39, 35), (47, 35), (40, 36), (43, 36), (47, 36), (41, 37), (43, 37), (39, 37), (40, 38), (41, 38), (43, 38), (40, 42), (41, 42), (39, 42), (40, 44), (43, 44), (39, 44), (41, 45), (43, 45), (47, 45), (39, 46), (40, 46), (41, 46), (43, 46), (47, 46), (40, 48), (39, 48), (47, 48)]}]}. 

The one-way transversal CNOT gate connects qubits 0 to 16 of both codes. 

\subsection{$[[65,1,5]]$ code} 

The 16 $X$-stabilizer generators have their non-trivial support on qubits
\texttt{[[ 0,  1,  2,  3, 19, 20, 33, 50],
 [22, 24, 27, 28, 35, 37, 54, 64],
 [ 4,  7, 10, 11, 21, 23, 25, 26, 52, 57, 58, 59],
 [ 5,  6,  8,  9, 12, 13, 34, 36, 38, 39, 51, 53, 55, 56, 60, 61, 62, 63],
 [ 1,  3,  5,  6, 33, 34, 50, 51],
 [ 9, 13, 17, 18, 39, 42, 49, 63],
 [35, 36, 37, 38, 40, 41, 43, 44, 54, 55, 60, 64],
 [ 7,  8, 11, 12, 15, 16, 45, 46, 47, 48, 52, 54, 56, 57, 58, 59, 61, 62],
 [27, 28, 31, 32, 37, 40, 43, 64],
 [25, 26, 29, 30, 45, 47, 58, 59],
 [38, 39, 41, 42, 44, 46, 48, 49, 60, 61, 62, 63],
 [19, 20, 21, 22, 23, 24, 33, 34, 35, 36, 50, 51, 52, 53, 54, 55, 56, 57],
 [ 2,  3,  4,  5,  7,  8, 19, 21, 50, 51, 52, 53],
 [10, 11, 14, 15, 25, 29, 47, 59],
 [12, 13, 16, 17, 48, 49, 62, 63],
 [23, 24, 26, 27, 30, 31, 43, 44, 45, 46, 54, 55, 56, 57, 58, 60, 61, 64]]}
and the 48 $Z$-stabilizer generators are indexed by
\texttt{[[ 0, 18, 49, 50, 51, 52, 53, 56, 57, 62],
 [ 1, 18, 49, 50, 52, 56, 57, 62],
 [ 2, 18, 42, 50, 51, 53, 56, 61],
 [ 3, 18, 42, 50, 56, 61],
 [ 4, 18, 42, 53, 57, 61],
 [ 5, 18, 42, 51, 56, 61],
 [ 6, 18, 49, 51, 52, 56, 57, 62],
 [ 7, 18, 42, 52, 56, 61],
 [ 8, 18, 42, 52, 57, 61],
 [ 9, 18, 49, 63],
 [10, 18, 42, 52, 53, 58, 59, 61],
 [11, 18, 42, 56, 57, 58, 59, 61],
 [12, 18, 42, 62],
 [13, 18, 42, 63],
 [14, 18, 42, 49, 52, 53, 56, 57, 58, 59, 61, 63],
 [15, 18, 42, 49, 58, 59, 61, 63],
 [16, 18, 42, 49, 62, 63],
 [17, 18, 42, 49],
 [19, 50, 51, 53],
 [20, 42, 49, 50, 51, 52, 53, 57, 61, 62],
 [21, 53, 56, 57],
 [22, 42, 49, 54, 60, 62],
 [23, 52, 53, 56],
 [24, 54, 60, 61],
 [25, 52, 53, 56, 57, 59],
 [26, 52, 53, 56, 57, 58],
 [27, 52, 53, 56, 57, 60, 61, 64],
 [28, 42, 49, 52, 53, 56, 57, 60, 62, 64],
 [29, 49, 52, 53, 59, 63],
 [30, 49, 52, 53, 58, 63],
 [31, 49, 54, 56, 63, 64],
 [32, 42, 54, 56, 61, 62, 63, 64],
 [33, 42, 49, 50, 52, 57, 61, 62],
 [34, 42, 49, 51, 52, 57, 61, 62],
 [35, 42, 49, 52, 53, 54, 56, 57, 61, 62],
 [36, 42, 49, 56, 60, 62],
 [37, 42, 49, 61, 62, 64],
 [38, 42, 49, 60, 61, 62],
 [39, 42, 49, 63],
 [40, 42, 52, 53, 54, 57, 60, 62, 63, 64],
 [41, 42, 60, 61, 62, 63],
 [43, 49, 52, 53, 54, 57, 60, 61, 63, 64],
 [44, 49, 60, 63],
 [45, 49, 56, 57, 58, 63],
 [46, 49, 61, 63],
 [47, 49, 56, 57, 59, 63],
 [48, 49, 62, 63],
 [55, 56, 60, 61]]}.

The logical $\ket{+}$ state of the $[[65,1,5]]$ code can be prepared, starting from the state $\ket{0}^{\otimes 65}$, with the gate sequence \texttt{[{'H': [2, 4, 7, 10, 12, 14, 16, 30, 32, 34, 41, 46, 49, 50, 54, 57, 62]}, {'CNOT': [
(16, 0), (34, 0), (50, 0), (4, 0), (10, 0), (12, 0), (14, 0), (16, 1), (2, 1), (34, 1), (4, 1), (62, 1), (57, 1), (46, 1), (50, 3), (14, 3), (62, 3), (57, 3), (10, 3), (12, 3), (46, 3), (16, 5), (2, 5), (50, 5), (4, 5), (62, 5), (57, 5), (46, 5), (34, 6), (14, 6), (62, 6), (57, 6), (10, 6), (12, 6), (46, 6), (16, 8), (7, 8), (10, 8), (12, 8), (14, 8), (2, 9), (34, 9), (4, 9), (7, 9), (12, 9), (16, 9), (49, 9), (50, 9), (62, 9), (10, 11), (4, 11), (7, 11), (16, 13), (49, 13), (62, 13), (4, 15), (14, 15), (7, 15), (49, 17), (12, 17), (62, 17), (49, 18), (2, 18), (34, 18), (4, 18), (50, 18), (7, 18), (62, 18), (2, 19), (14, 19), (62, 19), (57, 19), (10, 19), (12, 19), (46, 19), (16, 20), (34, 20), (50, 20), (4, 20), (62, 20), (57, 20), (46, 20), (4, 21), (14, 21), (62, 21), (57, 21), (10, 21), (12, 21), (46, 21), (46, 22), (41, 22), (54, 22), (57, 23), (4, 23), (7, 23), (2, 24), (34, 24), (4, 24), (7, 24), (41, 24), (16, 24), (50, 24), (54, 24), (62, 24), (4, 25), (30, 25), (7, 25), (62, 25), (10, 25), (12, 25), (46, 25), (10, 26), (14, 26), (30, 26), (32, 27), (4, 27), (7, 27), (41, 27), (10, 27), (14, 27), (46, 27), (54, 27), (57, 27), (32, 28), (2, 28), (34, 28), (41, 28), (10, 28), (14, 28), (16, 28), (50, 28), (54, 28), (57, 28), (62, 28), (4, 29), (14, 29), (30, 29), (7, 29), (62, 29), (12, 29), (46, 29), (32, 31), (2, 31), (34, 31), (4, 31), (7, 31), (46, 31), (16, 31), (50, 31), (62, 31), (16, 33), (2, 33), (34, 33), (4, 33), (10, 33), (12, 33), (14, 33), (16, 35), (2, 35), (34, 35), (50, 35), (54, 35), (62, 35), (46, 35), (16, 36), (14, 36), (41, 36), (10, 36), (12, 36), (46, 36), (57, 36), (32, 37), (54, 37), (57, 37), (10, 37), (14, 37), (16, 38), (41, 38), (12, 38), (16, 39), (49, 39), (2, 39), (34, 39), (4, 39), (50, 39), (7, 39), (32, 40), (16, 40), (2, 40), (34, 40), (50, 40), (41, 40), (62, 40), (49, 42), (2, 42), (34, 42), (4, 42), (50, 42), (7, 42), (12, 42), (32, 43), (4, 43), (7, 43), (41, 43), (46, 43), (2, 44), (34, 44), (4, 44), (7, 44), (41, 44), (46, 44), (16, 44), (50, 44), (62, 44), (4, 45), (30, 45), (7, 45), (30, 47), (14, 47), (62, 47), (12, 47), (46, 47), (16, 48), (12, 48), (62, 48), (16, 51), (2, 51), (50, 51), (4, 51), (10, 51), (12, 51), (14, 51), (62, 52), (7, 52), (46, 52), (57, 52), (10, 52), (12, 52), (14, 52), (16, 53), (4, 53), (62, 53), (57, 53), (46, 53), (2, 55), (34, 55), (4, 55), (7, 55), (41, 55), (10, 55), (12, 55), (14, 55), (50, 55), (57, 55), (62, 55), (16, 56), (57, 56), (10, 56), (12, 56), (14, 56), (4, 58), (30, 58), (7, 58), (10, 58), (14, 58), (30, 59), (62, 59), (10, 59), (12, 59), (46, 59), (2, 60), (34, 60), (4, 60), (7, 60), (41, 60), (12, 60), (46, 60), (50, 60), (62, 60), (16, 61), (12, 61), (46, 61), (16, 63), (49, 63), (12, 63), (32, 64), (2, 64), (34, 64), (4, 64), (7, 64), (10, 64), (14, 64), (46, 64), (16, 64), (50, 64), (54, 64), (57, 64), (62, 64)
]}]}. 

The one-way transversal CNOT gate connects qubits 0 to 18 of both codes. 

Note that these definitions correspond to the original definition of the 3D color code with $X$-type cells and $Z$-type faces, which achieves a higher protection against $X$-errors than against $Z$-errors, as opposed to the $d=3$-example discussed in the main text. In simulations for the $d=5$-case, we use this convention so that the state $\ket{+}_\mathrm{S}$ as input is analogous to an input state $\ket{0}_\mathrm{S}$ with the triorthogonal code definition where $X$- and $Z$-type stabilizers are interchanged, such that more $Z$- than $X$-errors can be corrected. This interchange does not affect logical failure rate estimations of the $\ket{+\ii}_\mathrm{S}$ state. All circuits can be accessed under \url{https://doi.org/10.5281/zenodo.15674834}.

\section{Large code state preparation in segmented ion trap}\label{sec:ion3d}

\begin{figure}\centering
	\includegraphics[width=0.99\linewidth]{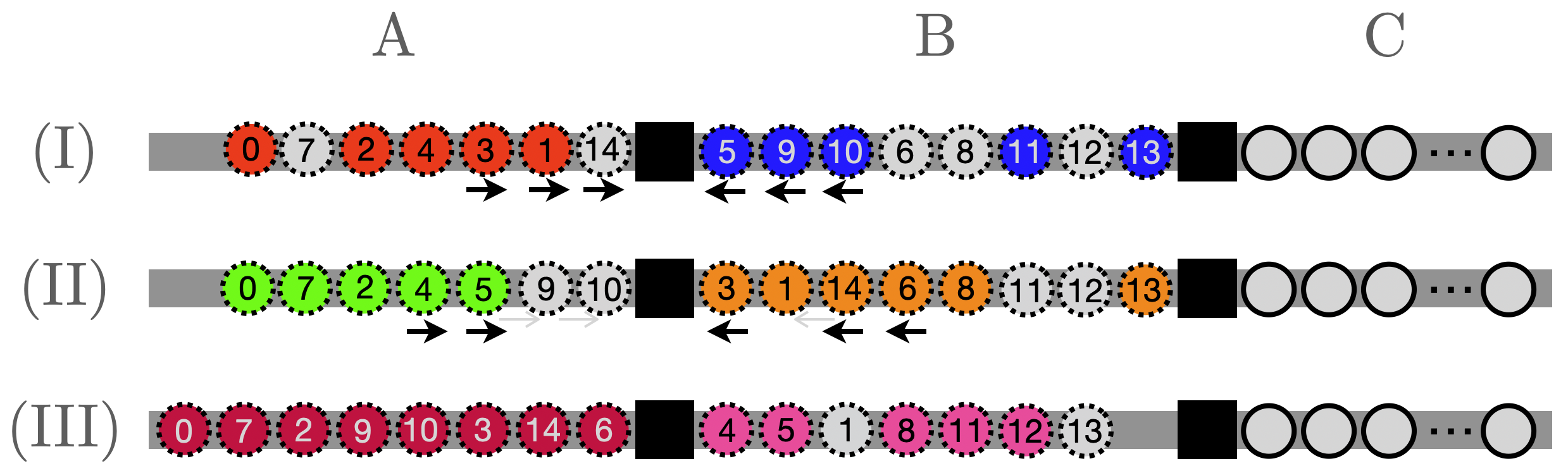}
	\caption{Simplified sketch of a segmented ion trap with three zones (grey bars), which each hold a linear ion crystal with all-to-all connectivity (cf.~Fig.~\ref{fig:yellow_trap}). Ions can be shuttled from one zone to another (separation indicated as black square). The segments A and B are used to prepare the logical state $\ket{0}_{\mathrm{T}}$ on 15 ions marked by the dotted line circles. Zone C holds a Steane code state on seven ions marked as solid line circles. Application of entangling gates is indicated by color in accordance with the coloring in Fig.~\ref{fig:init15}. No entangling gates are applied to ions colored grey. Three ion configurations, labelled (I)-(III), are sufficient for non-FT state preparation. Arrows indicate the direction of ion movement between configurations -- black arrows for shuttling through the junction, grey arrows for moving past another ion within the zone.}
	\label{fig:trap15}
\end{figure}

Entangling gates in ion traps are usually based on coupling of internal electronic degrees of freedom of ions and motional modes of the ion crystal during the gate operation \cite{molmer1999multiparticle, sorensen2000entanglement, ballance2017high}. The number of motional modes in an $N$-ion crystal is given by $3N$. Most gates require cooling all gate and spectator modes at least close to the motional ground state to reach high fidelities, which becomes more challenging for larger ion crystals. Another critical part arising from too large ion crystals can be an increased amount of crosstalk due to the fact that ions reside closer to each other. In this case, one needs to achieve a more tightly focused laser beam and higher beam pointing stability, which can pose considerable technical challenges. A potential solution for this scaling problem is realized in the QCCD architecture \cite{kielpinski2002architecture, pino2021demonstration} -- especially in the combination of 1) addressing a suitably-sized ion crystal with 2) performing ion-shuttling operations in a segmented ion trap.

Let us now discuss the initialization of the logical zero state of the 15-qubit Tetrahedral code in a segmented ion trap architecture where we allow, as an example, up to 8 ions trapped in a single linear crystal. Shuttling of individual ions is employed to perform crystal reconfigurations that restore effective all-to-all connectivity across individual segments. The ion configurations in Fig.~\ref{fig:trap15} match the coloring of CNOT gates in the encoding circuit of Fig.~\ref{fig:init15}. It indicates a possible parallelized gate sequence to initialize $\ket{0}_{\mathrm{T}}$ in two ion crystals labelled A and B with three sequential ion configurations. 

The five subsequent flag measurements that verify the state preparation, shown in Fig.~\ref{fig:flags15}, can be conducted largely independently in separated zones. Only the shared flag qubits need to be shuttled when their respective entangling gates are scheduled for execution. Since the two $Z$-flags and two of the three $X$-flags act on an entirely disjoint set of data qubits it is sensible to use distinct ions for their measurement in order to parallelize the respective gate sequences.

Nowadays, most shuttling compilers focus on reducing the overall amount of ion crystal reconfiguration operations, while some already include consideration of different cost for different types of shuttling operations \cite{Durandau2023automatedgeneration}. It shall be noted that an ``optimal'' shuttling schedule is not necessarily identical with this shortest-sequence approach. For example, moving ions between trap segments can be realized in different ways in a given architecture: One can either use physical SWAP operations of ions directly, where ions are moved out of the otherwise one-dimensional chain to interchange their positions (see Fig.~\ref{fig:yellow_trap}). Another possibility, which retains all ion movement quasi-1D, is to first move some ions out of the way with the help of junctions and then directly transport ions without physical SWAPs \cite{schoenberger2024shuttlingscalabletrappedionquantum} (see Fig.~\ref{fig:embed}).
From the typical reconfiguration operations, usually coined SWAP \cite{PhysRevA.95.052319}, MERGE, SPLIT \cite{Palmero_2015, PhysRevA.90.033410} and MOVE \cite{PhysRevLett.109.080501}, some might be preferred over others, depending on their impact on the ions such as the resulting error rate or induced time overhead for recooling \cite{Kaushal2020shuttlingbasedtiqc}. Compilers could include this behaviour, which can depend very much on the concrete experimental setup, by utilizing customized cost functions for ion reconfiguration instructions.

\section{Measurement-free code switching}\label{sec:mf}
Owing to current experimental limitations of quantum computing hardware, there is a growing interest in running FT quantum circuits without measurements of physical qubits \cite{pazsilva2010fault, heussen2024measurementfree, veroni2024optimized, veroni2025universalquantumcomputationscalable, butt2025measurement, butt2025demonstrationmeasurementfreeuniversalfaulttolerant, brechtelsbauer2025measurementfreequantumerrorcorrection}. 

Note that measurement-free code switching basically amounts to performing a SWAP gate between two different QEC codes and discarding the original qubit. A SWAP gate can be decomposed into three CNOT gates, two of which we can perform in the one-way fashion described in the main text. As of now, we are not aware of a unitary logical CNOT gate between the 2D and 3D color code in the other direction. To obtain this ``reversed'' CNOT gate, we note that one can concatenate each code with the respective other code, e.g., apply encoding circuits of the Tetrahedral code to the Steane code and apply encoding circuits of the Steane code to the Tetrahedral code, which yields a 105-qubit code for both logical qubits. Then, the CNOT gate could be performed transversally between the two logical qubits and decoding circuits of the respective codes might finally be applied to each logical qubit to re-obtain the original code blocks. We conjecture that this could be possible without any additional FT overhead. 

\end{document}